\pdfoutput=1
\documentclass[11pt,twoside,a4paper,cmspaper,final,collab]{cms-tdr}
\def\svnVersion{03624f1}\def\svnDate{2024/06/25}\def\cmsCernNoTag{CERN-EP-2024-032}\def\cmsCernDate{\today}\def\cmsMessage{Published in the Journal of High Energy Physics as \href{http://dx.doi.org/10.1007/JHEP06(2024)123}{\doi{10.1007/JHEP06(2024)123}.}}
\begin{document}\cmsNoteHeader{EXO-22-011}

\providecommand{\cmsTable}[2][1]{\resizebox{#1\textwidth}{!}{#2}}
\newlength\cmsTabSkip\setlength\cmsTabSkip{1ex}

\renewcommand{\ten}[1]{\ensuremath{\times10^{#1}}}
\newcommand{\sqrts}[1][13]{\ensuremath{\sqrt{s}=#1\TeV}\xspace}
\newcommand{\pp}{\ensuremath{\Pp\Pp}\xspace}

\newcommand{\hnl}{\ensuremath{\mathrm{N}}\xspace}
\newcommand{\mhnl}{\ensuremath{m_{\hnl}}\xspace}
\newcommand{\Vhnl}{\ensuremath{V_{\Pell\hnl}}\xspace}
\newcommand{\Vhnlsq}{\ensuremath{\abs{\Vhnl}^2}\xspace}
\newcommand{\Vhnlesq}{\ensuremath{\abs{V_{\Pe\hnl}}^2}\xspace}
\newcommand{\Vhnlmsq}{\ensuremath{\abs{V_{\PGm\hnl}}^2}\xspace}
\newcommand{\Vhnltsq}{\ensuremath{\abs{V_{\PGt\hnl}}^2}\xspace}

\newcommand{\Pellp}{\ensuremath{\HepParticle{\ell}{}{+}}\xspace}
\newcommand{\Pellm}{\ensuremath{\HepParticle{\ell}{}{-}}\xspace}
\newcommand{\Pellpm}{\ensuremath{\HepParticle{\ell}{}{\pm}}\xspace}
\newcommand{\PGth}{\ensuremath{\HepParticle{\PGt}{\mathrm{h}}{}}\xspace}

\newcommand{\sigeta}{\ensuremath{\eta}\xspace}
\newcommand{\abseta}{\ensuremath{\abs{\eta}}\xspace}
\newcommand{\Irel}{\ensuremath{I_{\text{rel}}}\xspace}
\newcommand{\mll}{\ensuremath{m(\Pellp\Pellm)}\xspace}
\newcommand{\ptl}[1]{\ensuremath{\pt(\Pell_{#1})}\xspace}
\newcommand{\ptll}[1]{\ensuremath{\pt(\Pl_{#1})}\xspace}
\newcommand{\pte}[1]{\ensuremath{\pt(\Pe_{#1})}\xspace}
\newcommand{\ptm}[1]{\ensuremath{\pt(\PGm_{#1})}\xspace}
\newcommand{\mthreel}{\ensuremath{m(3\Pell)}\xspace}
\newcommand{\minmllos}{\ensuremath{\min\mll}\xspace}
\newcommand{\mZ}{\ensuremath{m_{\PZ}}\xspace}
\newcommand{\mW}{\ensuremath{m_{\PW}}\xspace}
\newcommand{\LT}{\ensuremath{L_{\mathrm{T}}}\xspace}
\renewcommand{\DR}{\ensuremath{\Delta R}\xspace}
\newcommand{\Deta}{\ensuremath{\Delta\sigeta}\xspace}
\newcommand{\Dphi}{\ensuremath{\Delta\phi}\xspace}
\newcommand{\ttl}{\ensuremath{f}\xspace}
\newcommand{\Ne}{\ensuremath{N_{\Pe}}\xspace}
\newcommand{\Nmu}{\ensuremath{N_{\PGm}}\xspace}
\newcommand{\DRminmllos}{\ensuremath{\DR[\minmllos]}\xspace}

\newcommand{\DeepTau}{\ensuremath{\textsc{DeepTau}}\xspace}

\newcommand{\Zjets}{\ensuremath{\PZ\text{+jets}}\xspace}
\newcommand{\Wjets}{\ensuremath{\PW\text{+jets}}\xspace}
\newcommand{\WW}{\ensuremath{\PW\PW}\xspace}
\newcommand{\WZ}{\ensuremath{\PW\PZ}\xspace}
\newcommand{\ZZ}{\ensuremath{\PZ\PZ}\xspace}
\newcommand{\ZG}{\ensuremath{\PZ\PGg}\xspace}

\newcommand{\EEE}{\ensuremath{\Pe\Pe\Pe}\xspace}
\newcommand{\EEM}{\ensuremath{\Pe\Pe\PGm}\xspace}
\newcommand{\EMM}{\ensuremath{\Pe\PGm\PGm}\xspace}
\newcommand{\MMM}{\ensuremath{\PGm\PGm\PGm}\xspace}
\newcommand{\EET}{\ensuremath{\Pe\Pe\PGth}\xspace}
\newcommand{\EMT}{\ensuremath{\Pe\PGm\PGth}\xspace}
\newcommand{\MMT}{\ensuremath{\PGm\PGm\PGth}\xspace}

\newcommand{\likeli}{\ensuremath{L}\xspace}
\newcommand{\sigstr}{\ensuremath{r}\xspace}
\newcommand{\nuisan}{\ensuremath{\theta}\xspace}

\cmsNoteHeader{EXO-22-011}
\title{Search for heavy neutral leptons in final states with electrons, muons, and hadronically decaying tau leptons in proton-proton collisions at \texorpdfstring{\sqrts}{sqrt(s)=13 TeV}}

\author{The CMS Collaboration}
\date{\today}

\abstract{
A search for heavy neutral leptons (HNLs) of Majorana or Dirac type using proton-proton collision data at \sqrts is presented. The data were collected by the CMS experiment at the CERN LHC and correspond to an integrated luminosity of 138\fbinv. Events with three charged leptons (electrons, muons, and hadronically decaying tau leptons) are selected, corresponding to HNL production in association with a charged lepton and decay of the HNL to two charged leptons and a standard model (SM) neutrino. The search is performed for HNL masses between 10\GeV and 1.5\TeV. No evidence for an HNL signal is observed in data. Upper limits at 95\% confidence level are found for the squared coupling strength of the HNL to SM neutrinos, considering exclusive coupling of the HNL to a single SM neutrino generation, for both Majorana and Dirac HNLs. The limits exceed previously achieved experimental constraints for a wide range of HNL masses, and the limits on tau neutrino coupling scenarios with HNL masses above the \PW boson mass are presented for the first time.
}

\hypersetup{%
    pdfauthor={CMS Collaboration},%
    pdftitle={Search for heavy neutral leptons in decays with electrons, muons, and hadronically decaying tau leptons in proton-proton collisions at sqrt(s)=13 TeV},%
    pdfsubject={CMS},%
    pdfkeywords={CMS, heavy neutral leptons},%
}

\maketitle

\section{Introduction}

The observation of neutrino oscillations~\cite{Super-Kamiokande:1998kpq, SNO:2002tuh, KamLAND:2002uet} implies that neutrinos have a nonzero mass~\cite{Bilenky:2016pep}.
Direct neutrino mass measurements~\cite{Formaggio:2021nfz, KATRIN:2021uub}, as well as constraints from cosmological observations~\cite{Planck:2018vyg, eBOSS:2020yzd, Sakr:2022ans}, indicate that the neutrino masses are much smaller than those of the other fermions in the standard model (SM) of particle physics.
A possible mechanism for the generation of gauge-invariant neutrino mass terms and an explanation of their small scale is the see-saw mechanism~\cite{Minkowski:1977sc, Yanagida:1979as, Gell-Mann:1979vob, Glashow:1979nm, Mohapatra:1979ia, Schechter:1980gr, Shrock:1980ct, Cai:2017mow}, which introduces new heavy neutral leptons (HNLs) with right-handed chirality that are singlets under all SM gauge groups, but mix with the SM neutrinos.
In addition, HNL models can provide a viable dark matter candidate~\cite{Dodelson:1993je, Boyarsky:2018tvu}, and a mechanism to generate the matter-antimatter asymmetry of the universe~\cite{Fukugita:1986hr, Chun:2017spz, Drewes:2021nqr}.

\begin{figure}[!b]
\centering
\includegraphics[width=0.475\textwidth]{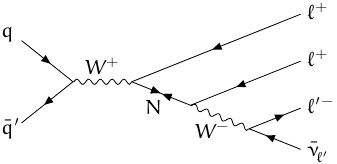}%
\hfill%
\includegraphics[width=0.475\textwidth]{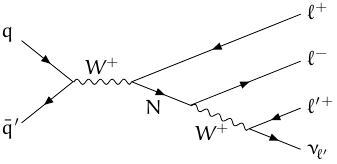} \\[1ex]
\includegraphics[width=0.475\textwidth]{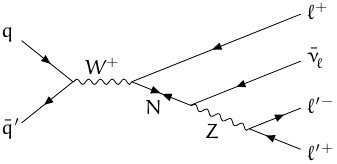}%
\hfill%
\includegraphics[width=0.475\textwidth]{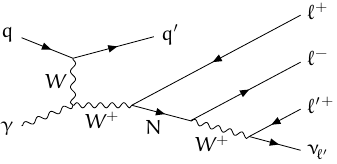}%
\caption{%
    Examples of Feynman diagrams for production and decay of an HNL (indicated with the symbol \hnl) resulting in final states with three charged leptons.
    The production processes DY (upper row and lower left) and VBF (lower right) are shown, with decays mediated by a \PW boson (upper row and lower right) or a \PZ boson (lower left).
    In the left column, HNLs of Majorana type with an LNV decay are shown, whereas the right column has HNLs of Dirac type with an LNC decay.
    The leptons that couple directly to the HNL (indicated with the symbol \Pell) are restricted to the SM generation that couples with the HNL, whereas the leptons from the \PW and \PZ boson decays (indicated with the symbol $\Pell^\prime$) can be from any SM generation.
}
\label{fig:feynman}
\end{figure}

We consider a simplified model with a single HNL (labelled \hnl in diagrams and formulas) of Majorana or Dirac type that couples through the neutrino mixing matrix exclusively to a single generation of SM neutrinos~\cite{Beacham:2019nyx, Drewes:2022akb}.
The signatures of such models in proton-proton (\pp) collisions have been studied extensively~\cite{delAguila:2008cj, Atre:2009rg, Tello:2010am, Das:2012ze, Deppisch:2015qwa, Das:2017nvm, Cai:2017mow, Das:2017gke, Bhardwaj:2018lma, Pascoli:2018heg, Abdullahi:2022jlv, Antel:2023hkf}.
We focus on the production in association with a charged lepton \Pellpm, which proceeds via the charged-current Drell--Yan (DY) process $\PQq\PAQq^\prime\to\PWpm\to\hnl\Pellpm$~\cite{Keung:1983uu, Petcov:1984nf} or via the vector boson fusion (VBF) process $\PQq\PGg\to\hnl\Pellpm\PQq^\prime$~\cite{Datta:1993nm, Dev:2013wba, Alva:2014gxa, Degrande:2016aje}.
Several searches for this production mode have been performed by the ATLAS, CMS, and LHCb experiments at the CERN LHC~\cite{CMS:EXO-11-076, CMS:EXO-12-057, ATLAS:2015gtp, CMS:EXO-14-014, CMS:EXO-17-012, CMS:EXO-17-028, ATLAS:2019kpx, LHCb:2020wxx, CMS:EXO-20-009, ATLAS:2022atq, CMS:EXO-21-013, CMS:EXO-23-006}.
The fully leptonic decay channel $\hnl\to\Pell\Pell\PGn$ results in final states with three charged leptons, as illustrated in Fig.~\ref{fig:feynman}.
Due to the assumption of an exclusive coupling to a single SM generation, the charged lepton originating from the HNL production and the one from the first decay vertex in the case of \PW-boson-mediated decays are necessarily of the same flavour and from the generation to which the HNL couples.
In the case of an HNL of Majorana type, both lepton number violating (LNV) and lepton number conserving (LNC) decays are possible, and as a result these two charged leptons can be of the same or opposite charge.
For an HNL of Dirac type, only LNC decays are possible and these two charged leptons thus always have opposite charge.

In this article, we present a search for HNLs in events with three charged leptons (electrons \Pe, muons \PGm, and hadronically decaying tau leptons \PGth, in the following referred to as ``leptons''), using \pp collision data collected in 2016--2018 at \sqrts and corresponding to an integrated luminosity of 138\fbinv.
We select events with all possible combinations of three light leptons (electrons and muons), resulting in the \EEE, \EEM, \EMM, and \MMM flavour channels, as well as events with one \PGth and any combination of two light leptons, resulting in the \EET, \EMT, and \MMT flavour channels.
Events with an HNL decay mediated by a \PZ boson or with an LNC decay mediated by a \PW boson result in events with an opposite-sign same-flavour (OSSF) lepton pair, whereas events without such a pair are only possible for an LNV decay.
The HNL scenarios with exclusive electron (muon) neutrino coupling are only probed in the \EEE and \EEM (\MMM and \EMM) channels.
In the scenario of exclusive tau neutrino couplings, the two tau leptons can decay leptonically or hadronically, and thus the \EEE, \EEM, \EMM, and \MMM channels provide sensitivity to HNL events where both tau leptons decay leptonically, whereas the \EET, \EMT, and \MMT channels provide sensitivity where one tau lepton each decays leptonically and hadronically.
Two strategies based on event categorization or on machine-learning discriminants are employed to separate the HNL signal from the SM background, where diboson production is the most important contribution.
Our results are interpreted for HNL masses \mhnl between 10\GeV and 1.5\TeV.
To facilitate reinterpretations within more general HNL models~\cite{Abada:2018sfh, Tastet:2021vwp}, we provide tabulated results in the HEPData record for this analysis~\cite{hepdata}.

The CMS Collaboration presented in Ref.~\cite{CMS:EXO-17-012} a search in events with three light leptons using \pp collision data collected in 2016 at \sqrts and corresponding to an integrated luminosity of 35.9\fbinv, constraining the mixing parameter \Vhnlsq between the HNL and the SM neutrino generation for \mhnl between 1\GeV and 1.2\TeV.
This article supersedes those results, and improves them not only because of the larger data set, but also from refined light-lepton identification (ID) criteria, improved background estimation techniques, and signal-to-background discrimination based on machine learning.
Additionally, we include for the first time in HNL searches at the LHC events with \PGth and use state-of-the-art \PGth ID techniques.

The mean lifetime of an HNL is proportional to $\mhnl^{-5}\Vhnl^{-2}$~\cite{Asaka:2005an}.
The HNL events that have a large decay length compared with the impact parameter resolution of the CMS tracker have a reduced selection efficiency in this analysis because we require that leptons originate from the primary interaction vertex (PV).
Two dedicated HNL searches presented by the CMS Collaboration, based on the same \pp collision data set used in this analysis, reconstruct the secondary HNL decay vertex in events with three light leptons~\cite{CMS:EXO-20-009} or apply a displaced jet tagger to events with two light leptons~\cite{CMS:EXO-21-013}, and constrain long-lived HNL scenarios for $1<\mhnl<20\GeV$.
The results of this analysis are complementary since they probe short-lived HNL scenarios with $\mhnl>10\GeV$ not excluded by the dedicated searches for long-lived HNLs.

\section{The CMS detector and event reconstruction}

The central feature of the CMS apparatus is a superconducting solenoid of 6\unit{m} internal diameter, providing a magnetic field of 3.8\unit{T}.
Within the solenoid volume are a silicon pixel and strip tracker, a lead tungstate crystal electromagnetic calorimeter (ECAL), and a brass and scintillator hadron calorimeter (HCAL), each composed of a barrel and two endcap sections.
Forward calorimeters extend the pseudorapidity (\sigeta) coverage provided by the barrel and endcap detectors.
Muons are measured in gas-ionization detectors embedded in the steel flux-return yoke outside the solenoid.
A more detailed description of the CMS detector, together with a definition of the coordinate system used and the relevant kinematic variables, is reported in Refs.~\cite{CMS:Detector-2008, CMS:PRF-21-001}.

Events of interest are selected using a two-tiered trigger system.
The first level, composed of custom hardware processors, uses information from the calorimeters and muon detectors to select events at a rate of around 100\unit{kHz} within a fixed latency of about 4\mus~\cite{CMS:TRG-17-001}.
The second level, known as the high-level trigger, consists of a farm of processors running a version of the full event reconstruction software optimized for fast processing, and reduces the event rate to around 1\unit{kHz} before data storage~\cite{CMS:TRG-12-001}.

The global event reconstruction with the particle-flow (PF) algorithm~\cite{CMS:PRF-14-001} reconstructs and identifies each individual particle in an event, with an optimized combination of all subdetector information.
In this process, the identification of the particle type (photon, electron, muon, charged or neutral hadron) plays an important role in the determination of the particle direction and energy.
Photons are identified as ECAL energy clusters not linked to the extrapolation of any charged-particle trajectory to the ECAL.
Electrons are identified as a charged-particle track and potentially many ECAL energy clusters corresponding to the extrapolation of this track to the ECAL and to possible bremsstrahlung photons emitted along the way through the tracker material.
Muons are identified as tracks in the central tracker consistent with either a track or several hits in the muon system, and associated with calorimeter deposits compatible with the muon hypothesis.
Charged hadrons are identified as charged-particle tracks neither identified as electrons, nor as muons.
Finally, neutral hadrons are identified as HCAL energy clusters not linked to any charged-hadron trajectory, or as a combined ECAL and HCAL energy excess with respect to the expected charged-hadron energy deposit.
The PV is taken to be the vertex corresponding to the hardest scattering in the event, evaluated using tracking information alone, as described in Section 9.4.1 of Ref.~\cite{CMS:TDR-15-02}.

For each event, hadronic jets are clustered from these reconstructed particles using the infrared and collinear safe anti-\kt algorithm~\cite{Cacciari:2008gp, Cacciari:2011ma} with a distance parameter of 0.4.
Jet momentum is determined as the vectorial sum of all particle momenta in the jet, and is found from simulation to be, on average, within 5--10\% of the true momentum over the entire transverse momentum (\pt) spectrum and detector acceptance.
Additional \pp interactions within the same or nearby bunch crossings (pileup) can contribute additional tracks and calorimetric energy depositions to the jet momentum.
To mitigate this effect, charged particles identified to be originating from pileup vertices are discarded and an offset correction is applied to correct for remaining contributions.
Jet energy corrections are derived from simulation to bring the measured response of jets to that of particle-level jets on average.
In situ measurements of the momentum balance in dijet, photon+jet, {\PZ}+jet, and multijet events are used to correct for any residual differences in the jet energy scale between data and simulation~\cite{CMS:JME-13-004}.
Only jets with $\pt>25\GeV$ and $\abseta<2.4$ are considered in this analysis.
Additional selection criteria are applied to each jet to remove jets potentially dominated by anomalous contributions from various subdetector components or reconstruction failures~\cite{CMS:JME-16-003}.

The missing transverse momentum vector \ptvecmiss is computed as the negative \ptvec sum of all PF candidates in an event, and its magnitude is denoted as \ptmiss~\cite{CMS:JME-17-001}.
The \ptvecmiss is modified to account for corrections to the energy scale of the reconstructed jets in the event.
Anomalous high-\ptmiss events can arise from a variety of reconstruction failures, detector malfunctions, or noncollisional backgrounds.
Such events are rejected by event filters that identify more than 85\% of the spurious high-\ptmiss events with a mistagging rate of less than 0.1\%~\cite{CMS:JME-17-001}.

The \textsc{DeepJet} algorithm~\cite{CMS:BTV-16-002, Bols:2020bkb, CMS:DP-2023-005} is applied to identify jets arising from the hadronization of \PQb hadrons.
We use a loose working point to tag jets as ``\PQb jets'' with a selection efficiency for \PQb quark jets of more than 90\%, and a misidentification rate for \PQc quark jets (light quark and gluon jets) of 50 (20)\%.

\section{Event simulation}
\label{sec:simulation}

Event samples simulated with Monte Carlo event generators are used to evaluate the signal selection efficiency, to predict background contributions, to train machine learning discriminators, and to validate background estimation techniques based on control samples in data.
The simulated event samples are processed with a full simulation of the CMS detector based on the \GEANTfour toolkit~\cite{GEANT4:2002zbu}, and are reconstructed with the same software as the data samples.
Additional simulated pileup interactions are added to the simulated events to match the observed pileup distribution as well, with a mean pileup of 23 (32) in 2016 (2017--2018)~\cite{CMS:JME-18-001}.
Separate event samples are generated for each data-taking year, reflecting the differences in the LHC running conditions and the CMS detector performance.

For the signal process, event samples are generated with the \MGvATNLO v2.6.5 program~\cite{Alwall:2014hca, Artoisenet:2012st}, using a model that extends the SM particle content by up to three right-handed neutrinos~\cite{Atre:2009rg, Alva:2014gxa, Degrande:2016aje, Pascoli:2018heg}.
The DY production process is simulated for $\mhnl<80\GeV$ at leading order (LO) in the strong coupling \alpS, whereas the simulation is performed at next-to-LO (NLO) in all other cases.
In the matrix element calculation, the NNPDF3.1~\cite{NNPDF:2017mvq} parton distribution functions (PDFs) are used for the DY production process, and the NNPDF31.luxQED~\cite{Manohar:2016nzj, Manohar:2017eqh, Bertone:2017bme} PDFs for the VBF production process.
Separate samples are generated for HNLs that couple to electron, muon, or tau neutrinos, and for different \mhnl values between 10\GeV and 1.5\TeV.
The VBF samples are generated only for masses of at least 600\GeV, since the contribution from VBF production is only relevant at high masses.
For $\mhnl>30\GeV$, no HNL lifetime effects are included in the simulation and a fixed value of $\Vhnlsq=10^{-4}$ is used.
The HNL production cross section is proportional to \Vhnlsq~\cite{Degrande:2016aje}, and thus the generated samples for a specific \mhnl value can be used to emulate signal scenarios with the same \mhnl and different \Vhnlsq values by applying a corresponding normalization factor.
At 30\GeV and lower, we calculate the HNL mean lifetime analytically~\cite{Bondarenko:2018ptm} and include it in the simulation of the HNL decay.
Samples are generated with one fixed \Vhnlsq value between $10^{-6}$ and $10^{-3}$ for each mass point, and we emulate other \Vhnlsq values by reweighting the HNL decay length distribution as described in Ref.~\cite{CMS:EXO-20-009}.
In all cases, the samples are generated assuming an HNL of Majorana nature, \ie including both LNV and LNC decays.
Samples for a Dirac HNL are obtained by selecting only the subset of simulated events with LNC decays, and the event weights are calculated using the same cross section but half the decay width of the Majorana HNL with the same mass and couplings.

Furthermore, the \MGvATNLO generator is used to simulate background samples at NLO for \WZ and \ZG diboson production, for Higgs boson (\PH) production in association with a vector boson or a top quark pair (\ttbar), for triboson production, for \ttbar production in association with a \PW or \PZ boson, for $s$-channel and $\PQt\PZ$ single top quark production, and for four top quark production.
It is also used at LO for DY vector boson production in association with jets (\Zjets and \Wjets), \PH production in association with a single top quark, \ttbar production in association with two bosons, and three top quark production.
Background samples for \qqbar-initiated \WW and \ZZ diboson production, gluon fusion and VBF \PH production, \ttbar production, and $t$-channel and $\PQt\PW$ single-\PQt production are generated with the \POWHEG2 program~\cite{Nason:2004rx, Frixione:2007nw, Frixione:2007vw, Alioli:2009je, Nason:2009ai, Alioli:2010xd, Re:2010bp, Bagnaschi:2011tu, Nason:2013ydw} at NLO.
The gluon-gluon-initiated \ZZ diboson production is simulated with the \MCFM v7.0.1 generator~\cite{Campbell:1999ah, Campbell:2011bn, Campbell:2015qma} at LO.
In all cases, the NNPDF3.1 PDFs are used.

The generators are interfaced with the \PYTHIA v8.230 program~\cite{Sjostrand:2014zea} for the underlying event description with the CP5 tune~\cite{CMS:GEN-17-001}, the parton shower simulation, and hadronization.
For \MGvATNLO samples simulated at LO (NLO), jets from matrix element calculations are merged with those from the parton shower using the MLM~\cite{Alwall:2007fs} (FxFx~\cite{Frederix:2012ps}) matching scheme.
In \POWHEG samples for \PH production, the decay to four leptons is simulated with the \textsc{JHUgen} v5.2.5 program~\cite{Bolognesi:2012mm}.

\section{Lepton selection}

Electrons are measured in the range $\abseta<2.5$, and their momentum is estimated by combining the energy measurement in the ECAL with the momentum measurement in the tracker~\cite{CMS:EGM-17-001, CMS:DP-2020-021}.
Electrons with $1.44<\abseta<1.57$ in the transition region between the barrel and endcap are not considered in the analysis because of performance limitations of the electron reconstruction in this region.
Muons are measured in the range $\abseta<2.4$, with detection planes made using three technologies: drift tubes, cathode strip chambers, and resistive plate chambers~\cite{CMS:MUO-16-001}.

We select reconstructed electrons and muons with $\pt>10\GeV$ that are compatible with originating from the PV and isolated from other particles in the event.
The relative isolation variable \Irel is defined as the scalar \pt sum of all PF particles reconstructed within a cone around the lepton direction divided by the lepton \pt, with the cone size defined in terms of $\DR=\sqrt{\smash[b]{(\Deta)^2+(\Dphi)^2}}$, where \Deta and \Dphi are the \sigeta and azimuthal angle difference between the particle and the lepton.
We use a variable cone size of 0.2 for leptons with $\pt<50\GeV$, of $10\GeV/\pt$ for $50<\pt<200\GeV$, and of 0.05 for $\pt>200\GeV$, which improves the efficiency for high-\pt leptons by removing the accidental overlap with other particles~\cite{Rehermann:2010vq}.
Additionally, corrections for pileup contributions to \Irel are applied.
All reconstructed electrons and muons are required to have $\Irel<0.4$.
For electrons, we additionally require that there be at most one tracker layer that contributes no hit on the track, to reduce contributions from photon conversions~\cite{CMS:EGM-17-001}.
For muons, we additionally apply the ``medium'' set of ID criteria defined in Ref.~\cite{CMS:MUO-16-001}.

Electrons and muons produced directly from the prompt decay of HNLs, \PW and \PZ bosons, or tau leptons are referred to as ``prompt'' leptons.
Background contributions with ``nonprompt'' leptons arise from events with genuine leptons produced in hadron decays and photon conversions, as well as from evens with jet constituents misidentified as leptons.
To distinguish between prompt and nonprompt electrons and muons, the two additional sets of ID criteria defined in Ref.~\cite{CMS:TOP-22-013} are applied, labelled ``loose'' and ``tight''.
The tight ID is based on a multivariate analysis (MVA) discriminant using the methods developed for various CMS measurements and searches with multilepton signatures~\cite{CMS:HIG-17-018, CMS:TOP-18-008, CMS:HIG-19-008, CMS:SUS-19-012, CMS:SMP-20-012, CMS:TOP-20-010}, described in more detail for the case of muons in Ref.~\cite{CMS:MUO-22-001}.
Tight electrons and muons are required to have the MVA discriminant exceed certain thresholds, resulting in a prompt electron (muon) selection efficiency of about 85 (92)\%.
The misidentification rate for nonprompt electrons (muons) is less than 0.6\% (about 1\%).
The loose ID is defined by requiring that electrons and muons either pass the tight ID, or pass selection requirements on some properties that are also used as inputs to the MVA discriminant.

Jets are used to reconstruct \PGth candidates with the hadrons-plus-strips algorithm~\cite{CMS:TAU-16-003}, which combines one or three tracks with energy deposits in the calorimeters, to identify the \PGth decay modes.
Neutral pions are reconstructed as ECAL energy deposition ``strips'' with dynamic size in \sigeta--$\phi$ from reconstructed electrons and photons, where the strip size varies as a function of the \pt of the electron or photon candidate.
The \PGth decay mode is then obtained by combining the charged hadrons with the strips.
We consider decay modes with one or three charged hadrons, with or without neutral pions, and require the \PGth candidate to have $\pt>20\GeV$ and $\abseta<2.3$.

To distinguish genuine \PGth decays from jets originating from the hadronization of quarks or gluons, and from electrons and muons, the \DeepTau algorithm is used~\cite{CMS:TAU-20-001}.
Information from all individual reconstructed particles near the \PGth axis is combined with properties of the \PGth candidate and the event to provide a multiclassification output equivalent to a Bayesian probability of the \PGth to originate from a genuine tau lepton, the hadronization of a quark or gluon, an electron, or a muon.
We define a ``loose'' and a ``tight'' ID for \PGth by choosing different working points of the \DeepTau discriminant for genuine tau leptons.
The rate of a jet to be misidentified as \PGth by the \DeepTau algorithm depends on the \pt and quark flavour of the jet.
We estimate it in simulated events from \PW boson production in association with jets to be 0.43\% for a genuine \PGth identification efficiency of 70\%.
The misidentification rate for electrons (muons) is 2.6 (0.03)\% for a genuine \PGth identification efficiency of 80 ($>$99)\%.

To avoid double counting of charged-particle candidates that pass both the electron and muon reconstruction, we remove reconstructed electrons that are within $\DR<0.05$ of any reconstructed muon.
Furthermore, we require that \PGth candidates be separated from any reconstructed electron or muon passing the tight working point by $\DR>0.5$.
Any jet that is within $\DR<0.4$ of any reconstructed electron or muon that passes the loose ID or within $\DR<0.5$ of a \PGth candidate that passes the tight ID is removed as well.

\section{Event selection and search strategy}

We analyse events that were collected with various triggers that require the presence of one, two, or three light leptons, with \pt thresholds that depend on the data-taking year and the flavour combination of the reconstructed leptons, as listed in Table~\ref{tab:ptthresholds}.
The efficiency of the trigger selection is larger than 90\% for three-lepton events everywhere, approaching 100\% for events with large lepton \pt.

\begin{table}[!ht]
\centering\renewcommand\arraystretch{1.3}
\topcaption{%
    Requirements on the light-lepton \pt values in the online and offline selections.
    The first two columns give the numbers of electrons and muons in the event (\Ne and \Nmu).
    The third column lists the \pt thresholds on the reconstructed electrons and muons in the online trigger selection, where the indices 1, 2, and 3 refer to the highest \pt, second-highest \pt, and third-highest \pt lepton, respectively.
    The fourth column lists the offline event selection requirements applied in addition to the baseline requirements of $\ptl1>15\GeV$ and $\ptl{2,3}>10\GeV$, where \Pell refers to reconstructed leptons of any flavour.
    For the $\Pe\PGm$ trigger, the requirements are given for the highest and second-highest \pt light lepton, referred to as $\Pl_1$ and $\Pl_2$ to indicate that a \PGth present in the event is not considered for the ordering.
    The values in parentheses give the thresholds applied in 2017 and 2018, where they are different from 2016.
    All events are required to pass the conditions of at least one of the rows.
}
\label{tab:ptthresholds}
\cmsTable{\begin{tabular}{cccc}
    \Ne & \Nmu & Online selection & Offline selection \\ \hline
    $\geq$1 & \NA & $\pte1>27$ (32)\GeV & $\pte1>30$ (35)\GeV \\
    \NA & $\geq$1 & $\ptm1>24\GeV$ & $\ptm1>25\GeV$ \\
    $\geq$2 & \NA & $\pte1>23\GeV$, $\pte2>12\GeV$ & $\pte1>25\GeV$, $\pte2>15\GeV$ \\
    \NA & $\geq$2 & $\ptm1>17\GeV$, $\ptm2>8\GeV$ & $\ptm1>20\GeV$ \\
    $\geq$1 & $\geq$1 & $\ptll1>23\GeV$, $\ptll2>8$ (12)\GeV & $\ptll1>25\GeV$, $\ptll2>10$ (15)\GeV \\
    $\geq$3 & \NA & $\pte1>16\GeV$, $\pte2>12\GeV$, $\pte3>8\GeV$ & $\pte1>25\GeV$, $\pte2>15\GeV$ \\
    $\geq$2 & $\geq$1 & $\pte{1,2}>12\GeV$, $\ptm1>8\GeV$ & $\pte1>25\GeV$, $\pte2>15\GeV$ \\
    $\geq$1 & $\geq$2 & $\pte1>9\GeV$, $\ptm{1,2}>9\GeV$ & \NA \\
    \NA & $\geq$3 & $\ptm1>12$ (10)\GeV, $\ptm2>10$ (5)\GeV, $\ptm3>5\GeV$ & \NA \\
\end{tabular}}
\end{table}

We select events with exactly three leptons that pass the tight ID criteria.
For a sideband enriched in events with nonprompt leptons, we retain events where at least one lepton fails the tight but passes the loose ID criteria.
Events with additional loose leptons or with at least one \PQb jet are removed, as well as events where all leptons have the same charge.
The highest \pt (leading) lepton, referred to as $\Pell_1$, is required to have $\ptl1>15\GeV$.
To ensure a high trigger efficiency, higher \pt thresholds are applied to the leading and second-highest \pt (subleading) light lepton depending on the lepton flavours present in the event, as summarized in Table~\ref{tab:ptthresholds}.
If OSSF lepton pairs are present in an event, they are required to have an invariant mass $\mll>5\GeV$ to remove contributions from low-mass resonances.
Additionally, we require $\abs{\mll-\mZ}>15\GeV$ for any OSSF lepton pair, where $\mZ=91.2\GeV$ is the \PZ boson mass~\cite{ParticleDataGroup:2022pth}, to remove background events with \PZ bosons.
While this removes HNL signal events with decays mediated by an on-shell \PZ boson, the SM \PZ boson background is overwhelming in this phase space and thus the loss of sensitivity incurred by this requirement is negligible.
Finally, events with \PQb jets are removed to suppress background contributions with top quarks.

Events are categorized by the flavour of the selected leptons.
For events with only electrons and muons, this results in the four categories \EEE, \EEM, \EMM, and \MMM.
For events with exactly one \PGth, we distinguish the three categories \EET, \EMT, and \MMT.
We do not select events with more than one \PGth, since the typically smaller signal efficiency, higher background contamination, and lower resolution of the \PGth reconstruction compared with those for electrons and muons result in a low signal acceptance and a significant background yield for these events.

For HNL models with \mhnl below the \PW boson mass $\mW=80.4\GeV$~\cite{ParticleDataGroup:2022pth}, the HNL decay proceeds via a virtual \PW or \PZ boson, resulting in typically low-\pt leptons.
In the case of $\mhnl>\mW$, the decay will first proceed to an on-shell \PW boson and a lepton, with a subsequent leptonic decay of the \PW boson, resulting in typically higher \pt for at least one of the leptons.
Similarly for $\mhnl>\mZ$, the decay via an on-shell \PZ boson also results in events with leptons of typically higher \pt.
Other kinematic properties of the final-state leptons will be significantly different as well for the cases of $\mhnl<\mW$ (``low mass'') and $>$\mW (``high mass'').
Thus, we define two orthogonal event selections to target the two separate mass ranges, by categorizing events with $\ptl1<55\GeV$ as low-mass and $>$55\GeV as high-mass events.

In the low-mass selection, events are further required to have $\ptmiss<75\GeV$ to remove background contributions with SM neutrinos, such as \ttbar and diboson production, and to have a trilepton invariant mass $\mthreel<80\GeV$ to remove \ZG photon conversion events.
In the high-mass selection, the subleading lepton is required to have $\ptl2>15\GeV$ to reduce background contributions with nonprompt leptons, and events with an OSSF lepton pair and $\abs{\mthreel-\mZ}<15\GeV$ are removed to reduce \ZG photon conversion backgrounds.
To remove background contributions with charge-misidentified electrons, events in the high-mass selection with two same-sign electrons and a muon are required to have consistent results between three independent charge measurements~\cite{CMS:EGM-13-001} for the two electrons and to have a dielectron mass more than 15\GeV away from \mZ.
Although charge-misidentified electrons can also play a role in other final states, their contribution to the background is small in other flavour channels.
Charge mismeasurement for muons is negligible~\cite{CMS:CFT-09-014, CMS:MUO-17-001}.

\begin{table}[!p]
\centering\renewcommand\arraystretch{1.1}
\topcaption{%
    Definitions of the search regions (SRs) for events in the low-mass (upper part) and high-mass (lower part) selections.
}
\cmsTable[0.67]{\begin{tabular}{cccccc}
    \multirow{2}{*}{OSSF pair} & \ptl1 & \mthreel & \minmllos & \mT & \multirow{2}{*}{SR name} \\
    & ({\GeVns}) & ({\GeVns}) & ({\GeVns}) & ({\GeVns}) & \\
    \hline
    \multicolumn{6}{c}{\textit{Low-mass selection}} \\[0.5\cmsTabSkip]
    No & $<$30 & $<$80 & $<$10 & any & La1 \\
    & & & 10--20 & & La2 \\
    & & & 20--30 & & La3 \\
    & & & $>$30 & & La4 \\[0.5\cmsTabSkip]
    & 30--55 & $<$80 & $<$10 & any & La5 \\
    & & & 10--20 & & La6 \\
    & & & 20--30 & & La7 \\
    & & & $>$30 & & La8 \\[\cmsTabSkip]
    Yes & $<$30 & $<$80 & $<$10 & any & Lb1 \\
    & & & 10--20 & & Lb2 \\
    & & & 20--30 & & Lb3 \\
    & & & $>$30 & & Lb4 \\[0.5\cmsTabSkip]
    & 30--55 & $<$80 & $<$10 & any & Lb5 \\
    & & & 10--20 & & Lb6 \\
    & & & 20--30 & & Lb7 \\
    & & & $>$30 & & Lb8 \\[\cmsTabSkip]
    \multicolumn{6}{c}{\textit{High-mass selection}} \\[0.5\cmsTabSkip]
    No & $>$55 & $<$100 & any & $<$100 & Ha1 \\
    & & & & $>$100 & Ha2 \\[0.5\cmsTabSkip]
    & & $>$100 & $<$100 & $<$100 & Ha3 \\
    & & & & 100--150 & Ha4 \\
    & & & & 150--250 & Ha5 \\
    & & & & $>$250 & Ha6 \\[0.5\cmsTabSkip]
    & & & 100--200 & $<$100 & Ha7 \\
    & & & & $>$100 & Ha8 \\[0.5\cmsTabSkip]
    & & & $>$200 & any & Ha9 \\[\cmsTabSkip]
    Yes & $>$55 & $<$75 & any & $<$100 & Hb1 \\
    & & & & 100--200 & Hb2 \\
    & & & & $>$200 & Hb3 \\[\cmsTabSkip]
    & & $>$105 & $<$100 & $<$100 & Hb4 \\
    & & & & 100--200 & Hb5 \\
    & & & & 200--300 & Hb6 \\
    & & & & 300--400 & Hb7 \\
    & & & & $>$400 & Hb8 \\[0.5\cmsTabSkip]
    & & & 100--200 & $<$100 & Hb9 \\
    & & & & 100--200 & Hb10 \\
    & & & & 200--300 & Hb11 \\
    & & & & $>$300 & Hb12 \\[0.5\cmsTabSkip]
    & & & $>$200 & $<$100 & Hb13 \\
    & & & & 100--200 & Hb14 \\
    & & & & 200--300 & Hb15 \\
    & & & & $>$300 & Hb16 \\
\end{tabular}}
\label{tab:searchregions}
\end{table}

Following the strategy applied in Ref.~\cite{CMS:EXO-17-012}, we define a number of orthogonal search regions (SRs) by classifying the events according to several kinematic variables that provide a good discrimination between signal and background contributions, as summarized in Table~\ref{tab:searchregions}.
Events are first sorted based on whether they have an OSSF lepton pair or not, since background processes with a \PZ boson contribute primarily to the OSSF events.
In the low-mass region, the SRs are then defined in bins of \ptl1 and the smallest invariant mass of any opposite-sign (OS) lepton pair, \minmllos.
The SRs with $\ptl1<30\GeV$ target HNL scenarios where \mhnl is close to \mW, such that both the lepton from the HNL production and the leptons from the HNL decay have low \pt.
The SRs with $30<\ptl1<55\GeV$, on the other hand, target HNL scenarios with a smaller \mhnl, such that the lepton from the HNL production can have a large \pt.
The \minmllos variable is bounded to be smaller than \mhnl for HNL signal events, and thus provides sensitivity to distinguish different HNL signal scenarios.

In the high-mass region, the SRs are defined in bins of \minmllos, \mthreel, and the transverse mass \mT calculated with \ptvecmiss and the lepton not used for \minmllos, defined as $\mT=\sqrt{\smash[b]{2\ptmiss\pt(\Pell)(1-\cos\Dphi)}}$, where \Dphi is the $\phi$ angle between the lepton and \ptvecmiss.
The variable \minmllos again provides discriminating power between different HNL masses, whereas \mT is targeted at the reconstruction of a resonance decaying into the third lepton and an SM neutrino that causes the \ptvecmiss in the events.
Although the HNL signal events have no such resonance and thus are distributed towards high \mT values, SM background processes like \WZ production with $\PW\to\Pell\PGn$ decays will have a distribution of \mT mostly below \mW.
Finally, \mthreel measures a lower bound on the energy of the $s$-channel resonance that produced the leptons in the event, with larger values as \mhnl increases due to the large partonic centre-of-mass energy required to produce high-mass HNLs, and a distribution at lower values for the SM backgrounds that generally have a lower production threshold.

{\tolerance=800
To further improve the separation of signal and background events, we employ machine learning classifiers based on boosted decision trees (BDTs) as implemented in the \textsc{tmva} package~\cite{Voss:2007jxm}.
The classifiers are trained to distinguish HNL signal events from background events taken from simulated samples for the \Zjets, \ttbar, and \WZ background processes, using both selected and sideband events to also train against nonprompt-lepton background events.
We train separate BDTs for different HNL coupling scenarios and \mhnl ranges, using different event selections and categories, and label these trainings as ``BDT(\mhnl, \Pell, $i$\PGth)''.
The first argument specifies one of the five \mhnl ranges (in \GeVns), where we use the low-mass (high-mass) selection for the ranges 10--40 and 50--75\GeV (85--150, 200--250, and 300--400\GeV).
The second argument specifies the lepton flavour of the neutrino generation to which the HNL couples exclusively.
The third argument specifies the event categories used in the training: for electron and muon neutrino couplings, the event categories without \PGth are used, \ie \EEE, \EEM, \EMM, and \MMM, labelled as ``0\PGth''.
For tau neutrino couplings, separate trainings are performed for final states with no \PGth at generator level, using the 0\PGth event categories, and for final states with at least one \PGth at generator level, using the \EET, \EMT, and \MMT event categories, labelled as ``1\PGth''.
\par}

\begin{figure}[!p]
\centering
\includegraphics[width=0.42\textwidth]{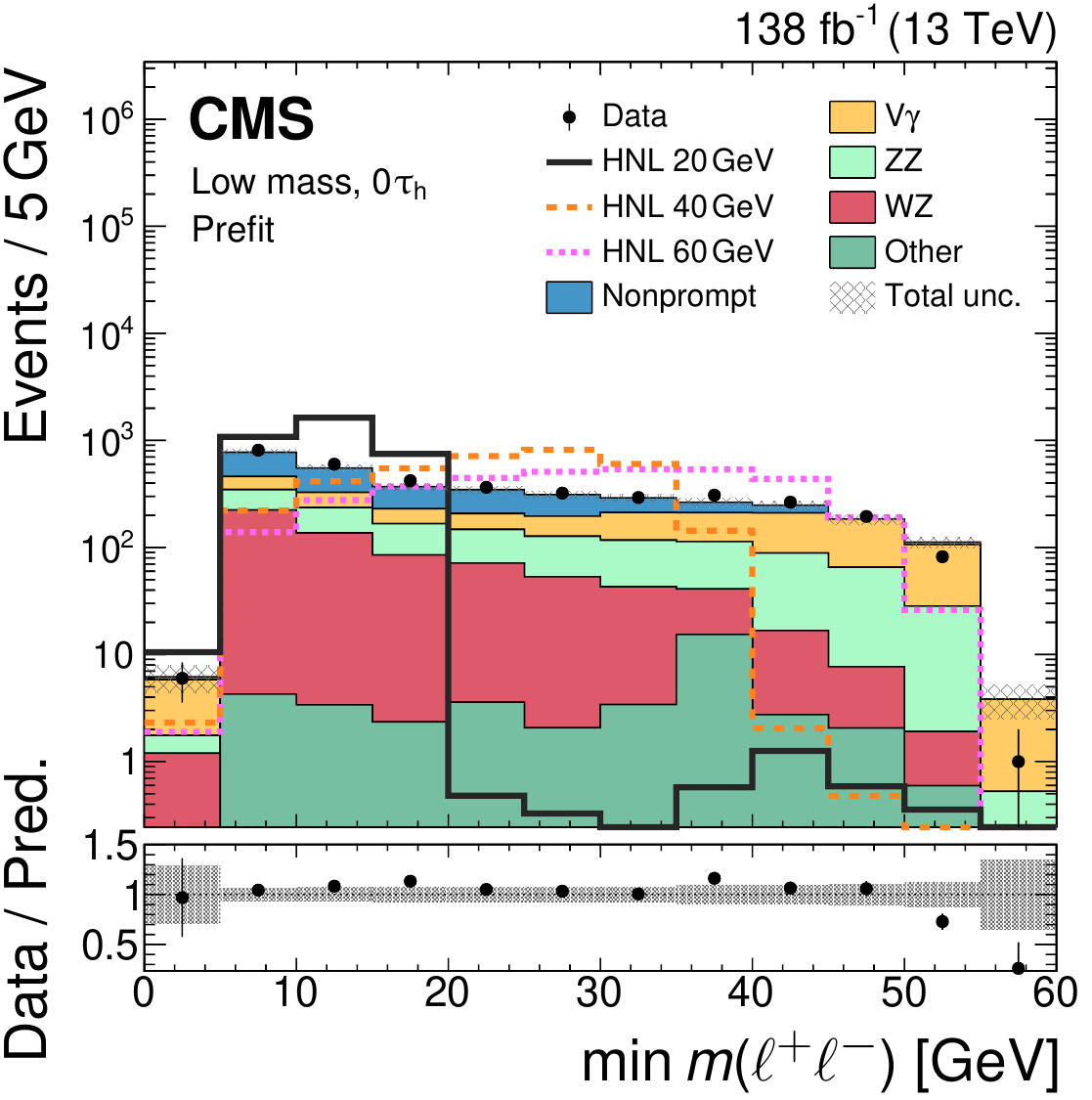}%
\hspace*{0.05\textwidth}%
\includegraphics[width=0.42\textwidth]{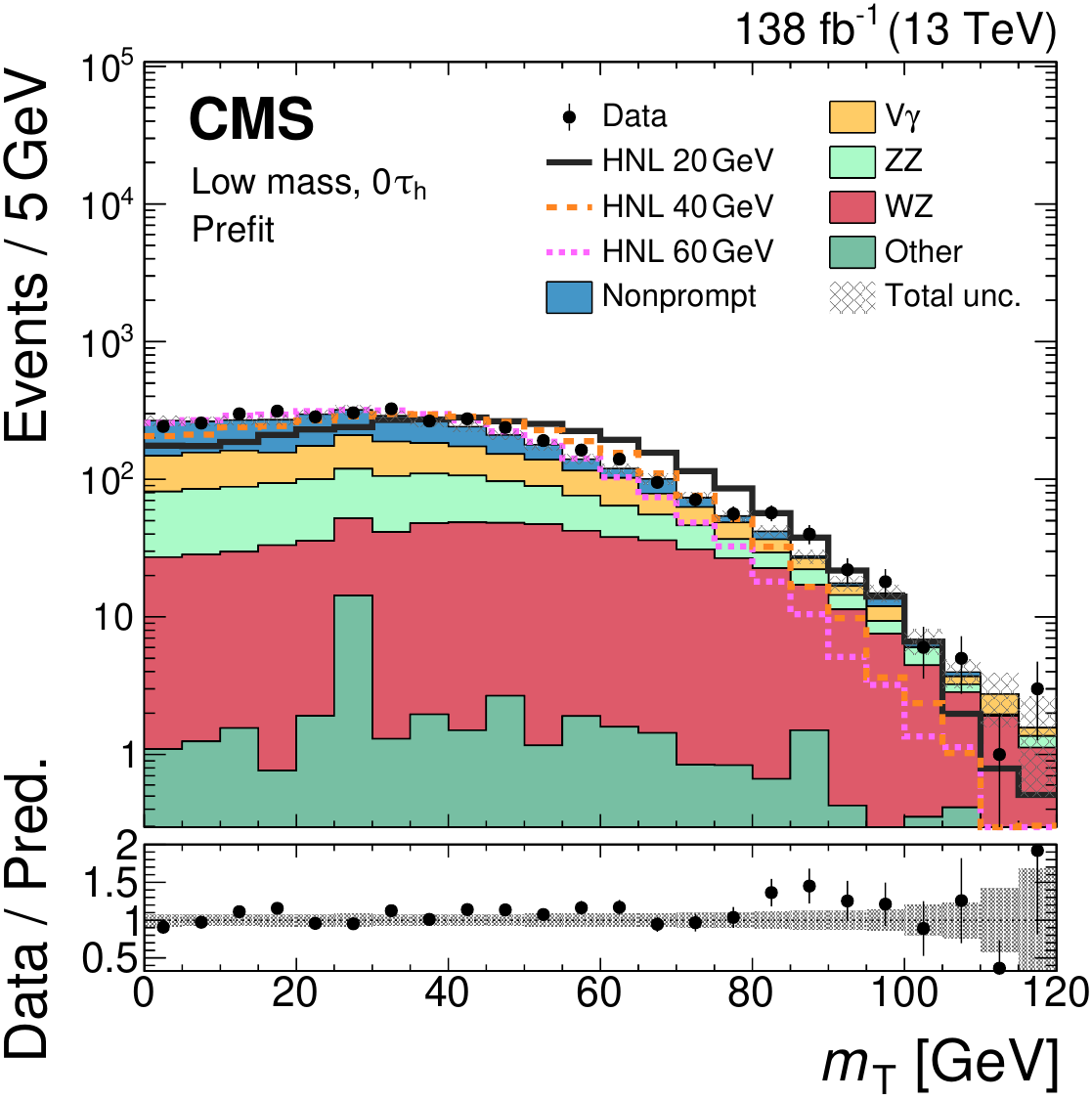} \\[1ex]
\includegraphics[width=0.42\textwidth]{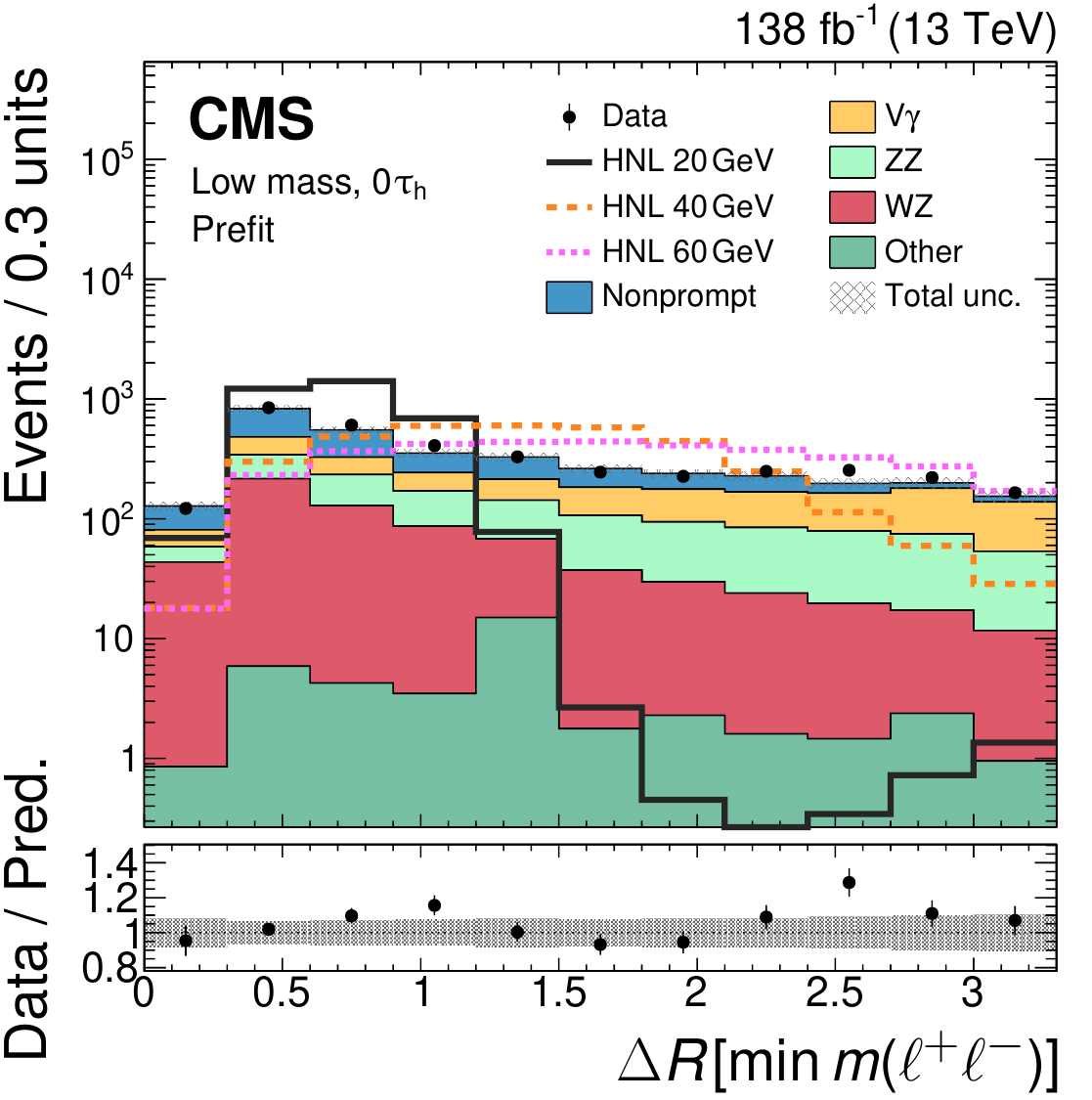}%
\hspace*{0.05\textwidth}%
\includegraphics[width=0.42\textwidth]{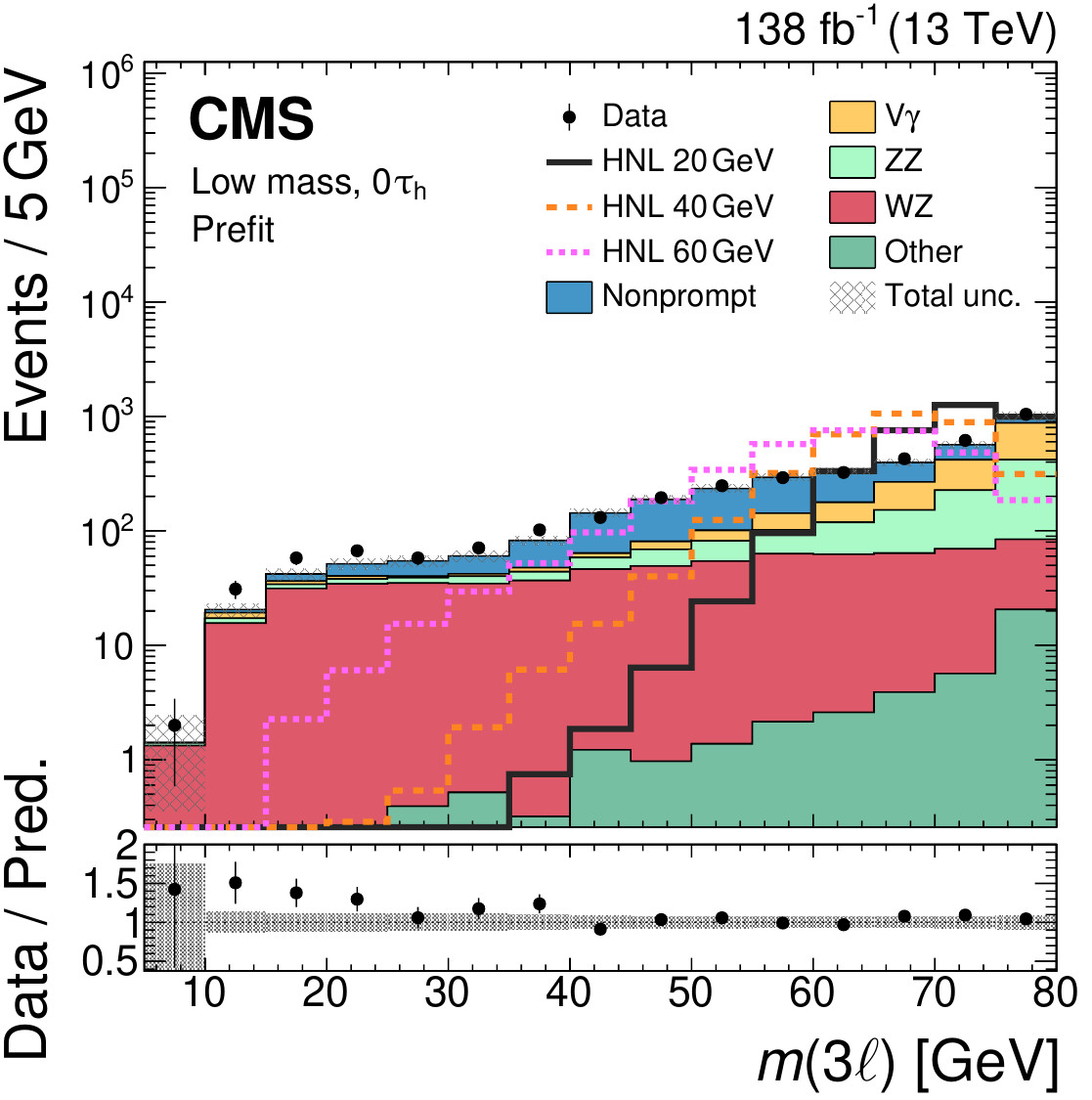}%
\caption{%
    Comparison of observed (points) and predicted (coloured histograms) distributions in the low-mass selection for the 0\PGth categories combined.
    Important input variables to the BDT training are shown:
    \minmllos (upper left),
    \mT (upper right),
    \DR between the two leptons used for \minmllos (\DRminmllos, lower left),
    \mthreel (lower right).
    The predicted background yields are shown before the fit to the data (``prefit'').
    The HNL predictions for three different \mhnl values with exclusive coupling to tau neutrinos are shown with coloured lines, and are normalized to the total background yield.
    The vertical bars on the points represent the statistical uncertainties in the data, and the hatched bands the total uncertainties in the predictions.
    The last bins include the overflow contributions.
    In the lower panels, the ratios of the event yield in data to the overall sum of the predictions are shown.
}
\label{fig:inputs_low_notau}
\end{figure}

\begin{figure}[!p]
\centering
\includegraphics[width=0.42\textwidth]{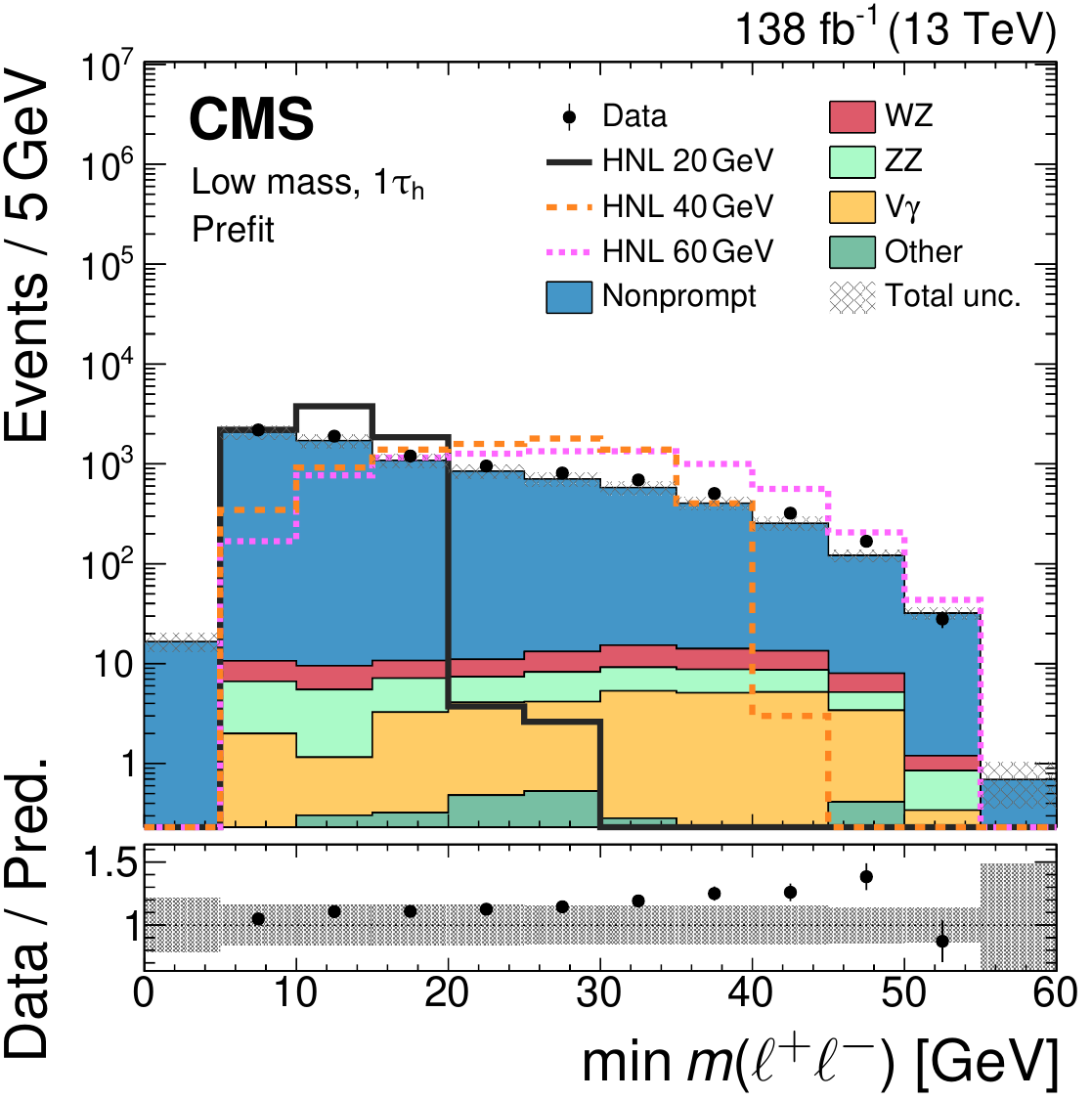}%
\hspace*{0.05\textwidth}%
\includegraphics[width=0.42\textwidth]{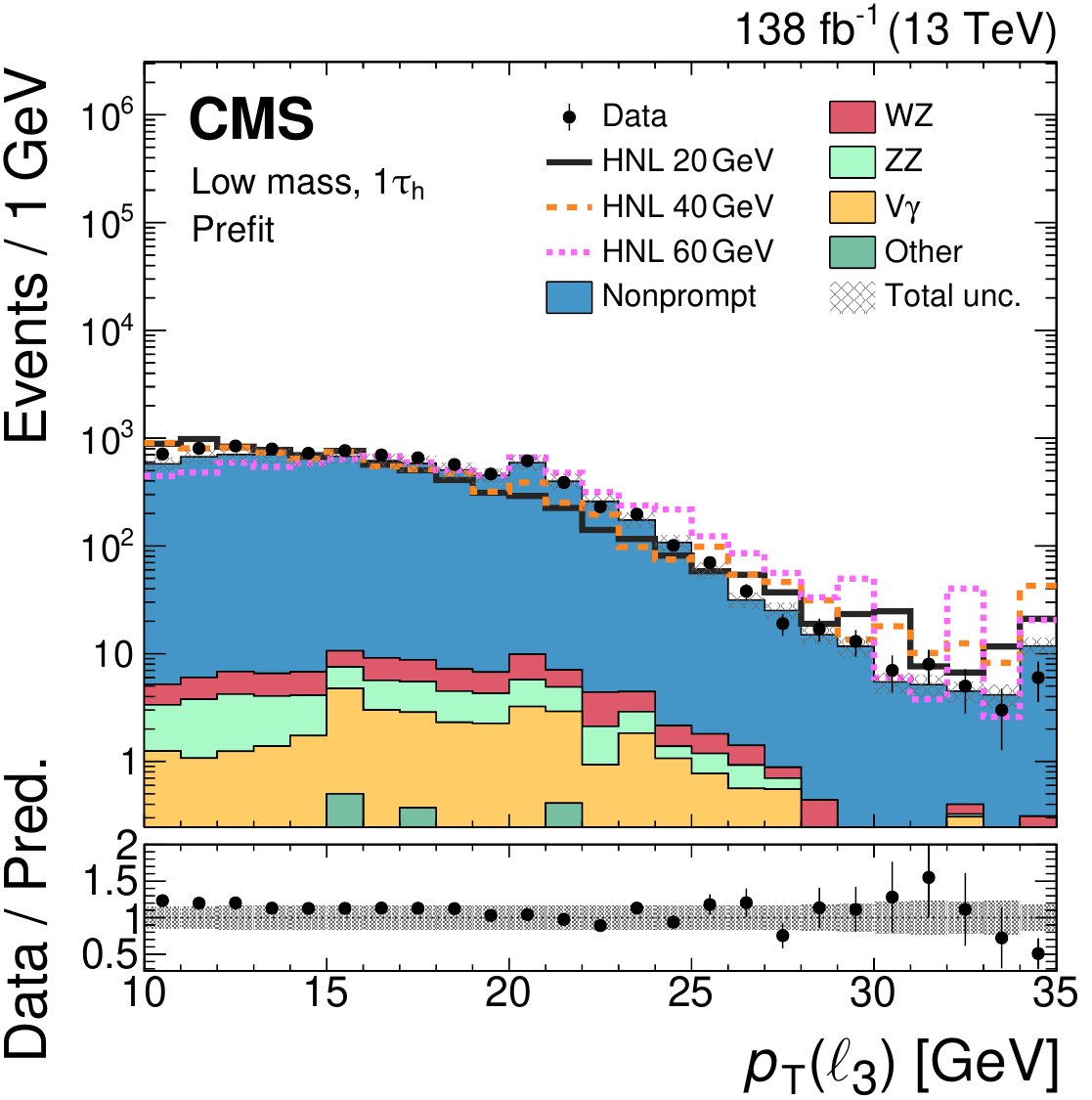} \\[1ex]
\includegraphics[width=0.42\textwidth]{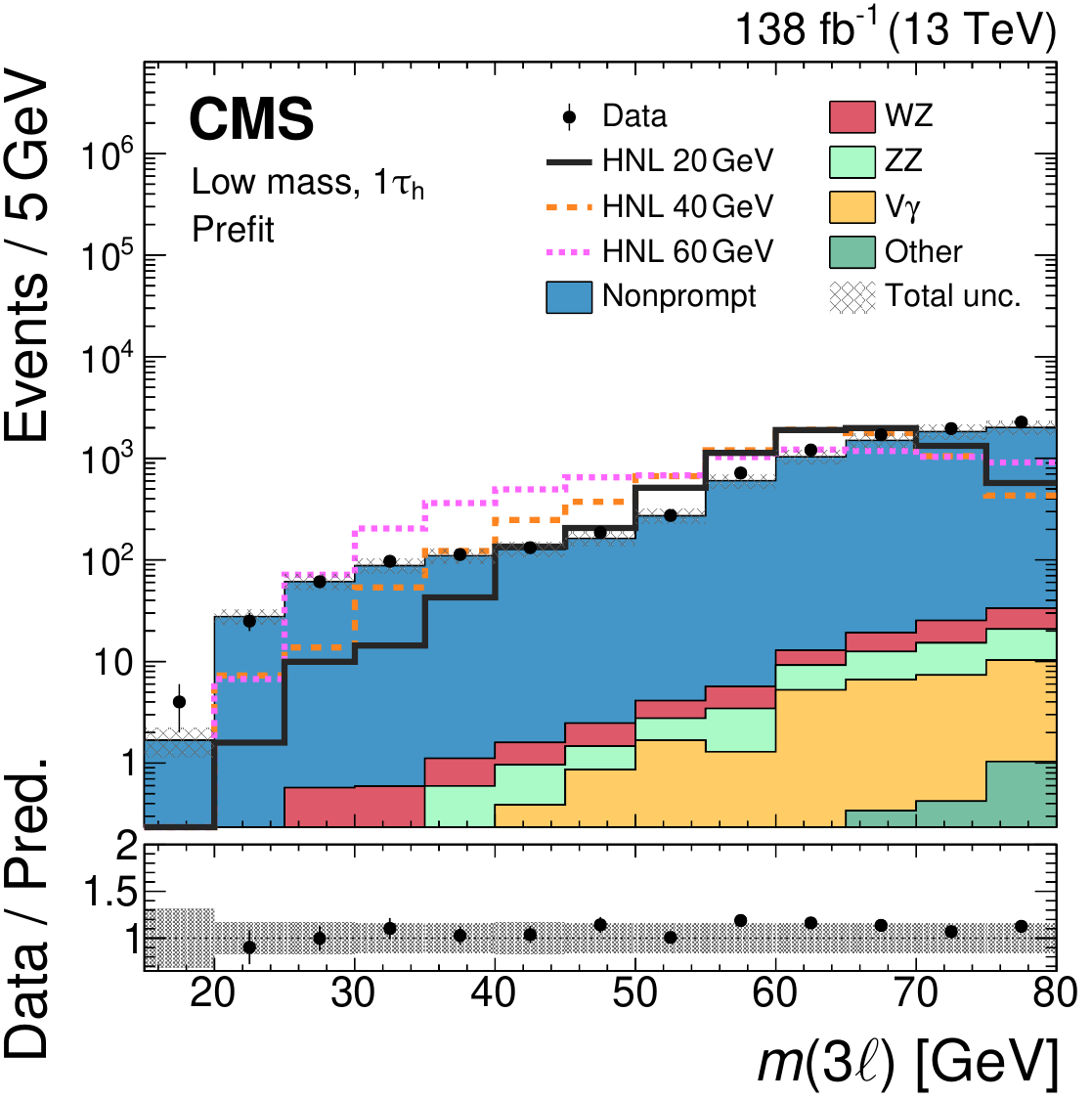}%
\hspace*{0.05\textwidth}%
\includegraphics[width=0.42\textwidth]{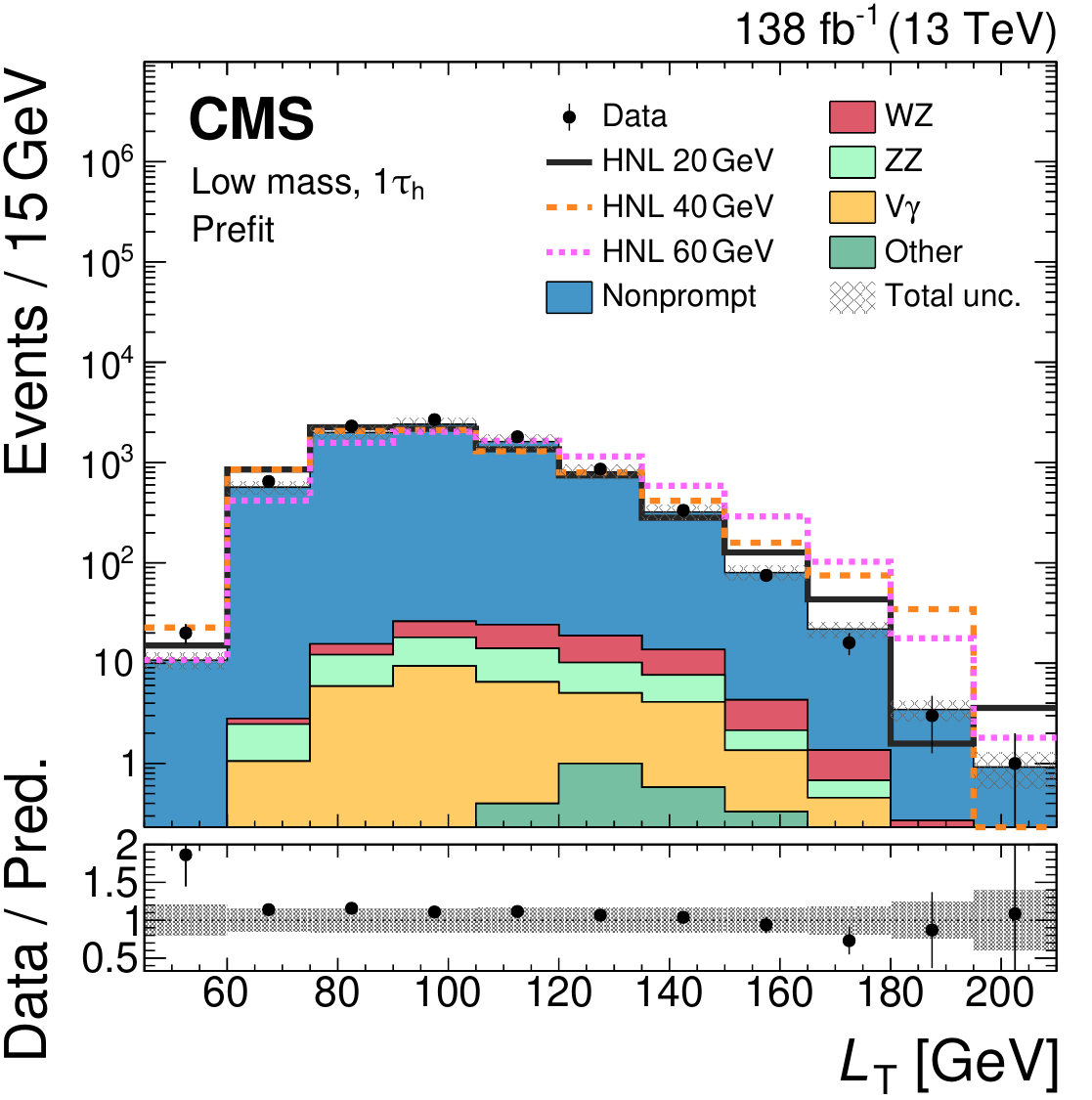}%
\caption{%
    Comparison of observed (points) and predicted (coloured histograms) distributions in the low-mass selection for the 1\PGth categories combined.
    Important input variables to the BDT training are shown:
    \minmllos (upper left),
    \ptl3 (upper right),
    \mthreel (lower left),
    \LT (lower right).
    The predicted background yields are shown before the fit to the data (``prefit'').
    The HNL predictions for three different \mhnl values with exclusive coupling to tau neutrinos are shown with coloured lines, and are normalized to the total background yield.
    The vertical bars on the points represent the statistical uncertainties in the data, and the hatched bands the total uncertainties in the predictions.
    The last bins include the overflow contributions.
    In the lower panels, the ratios of the event yield in data to the overall sum of the predictions are shown.
}
\label{fig:inputs_low_tau}
\end{figure}

\begin{figure}[!t]
\centering
\includegraphics[width=0.42\textwidth]{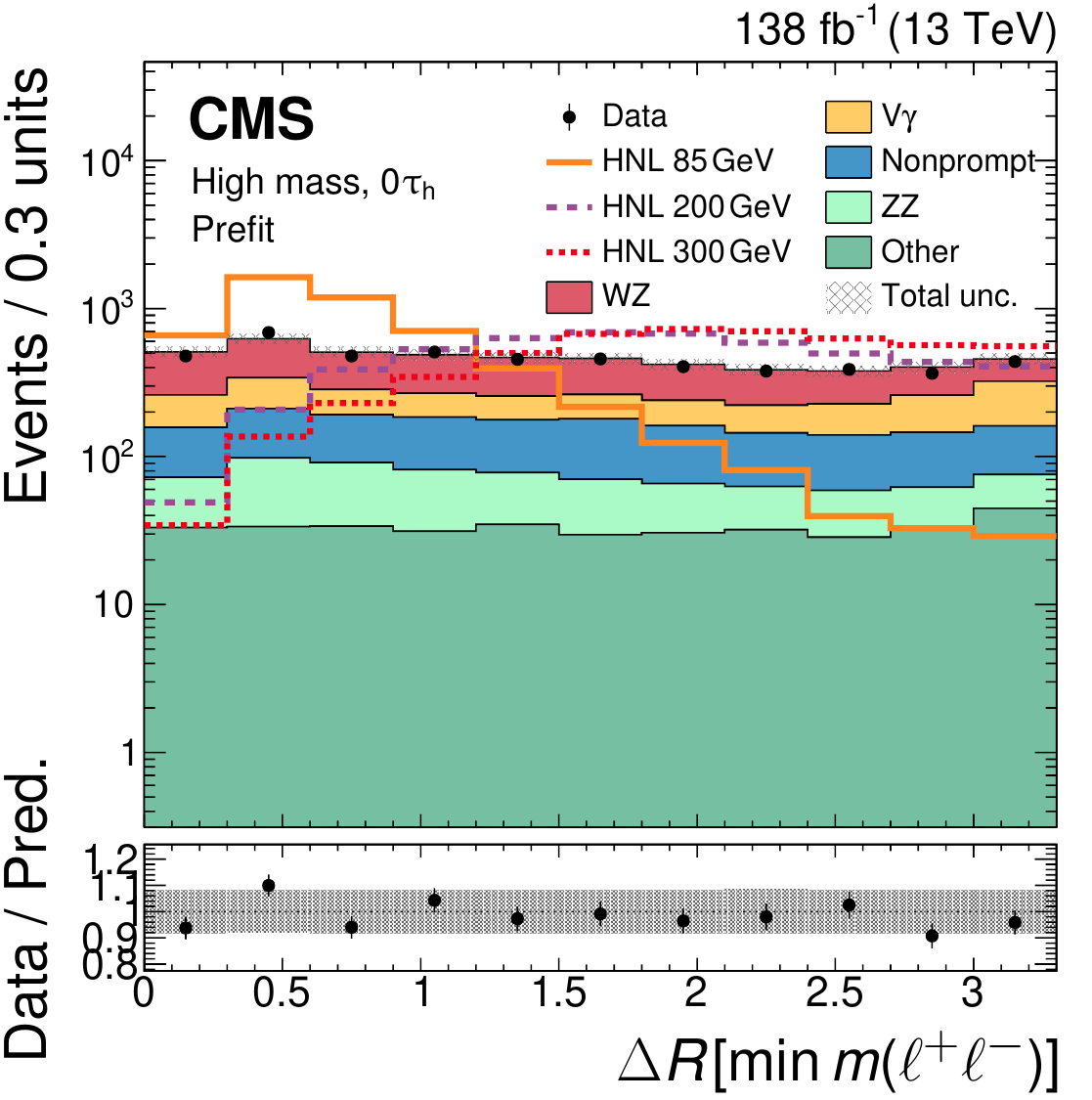}%
\hspace*{0.05\textwidth}%
\includegraphics[width=0.42\textwidth]{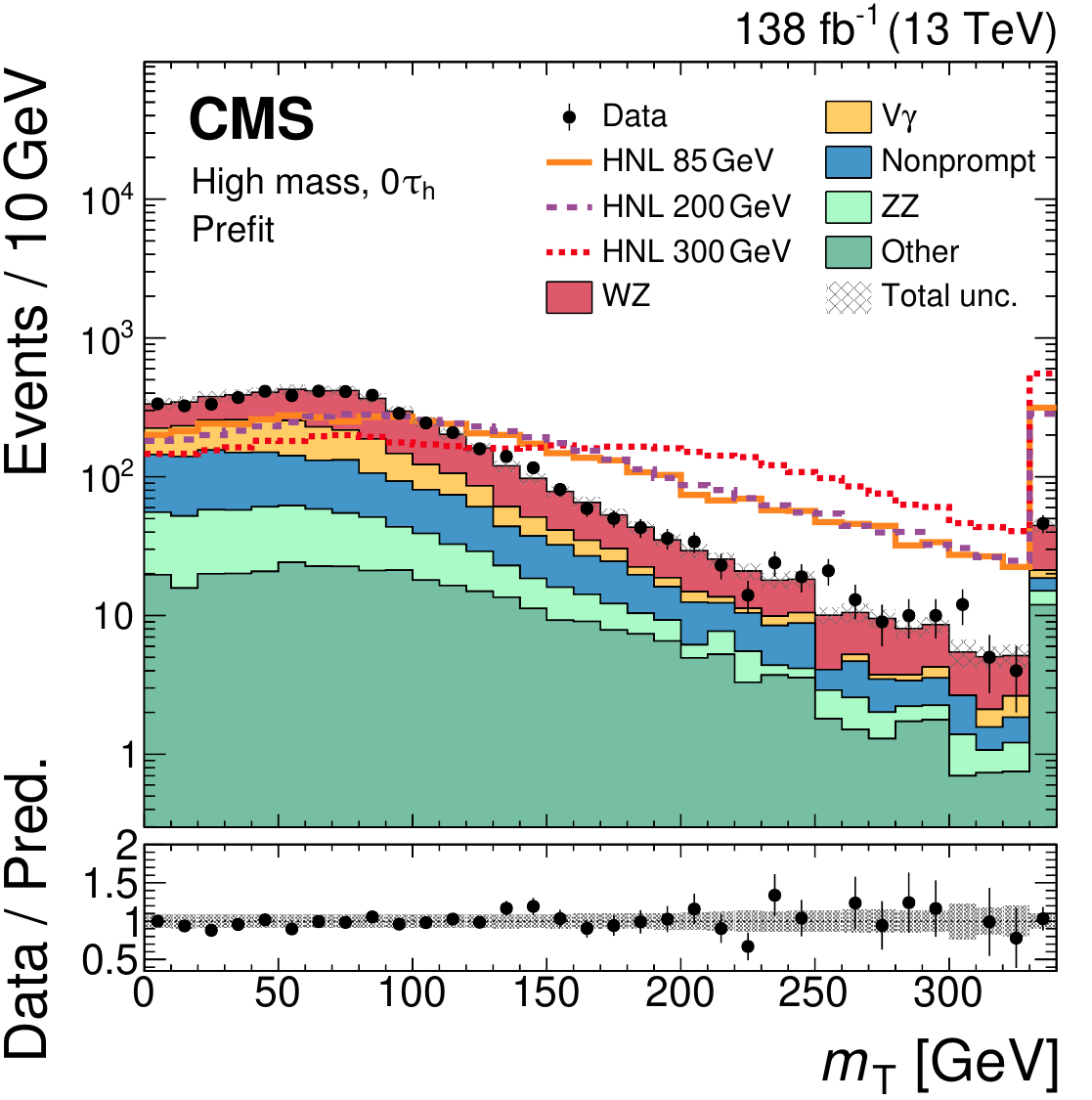} \\[1ex]
\includegraphics[width=0.42\textwidth]{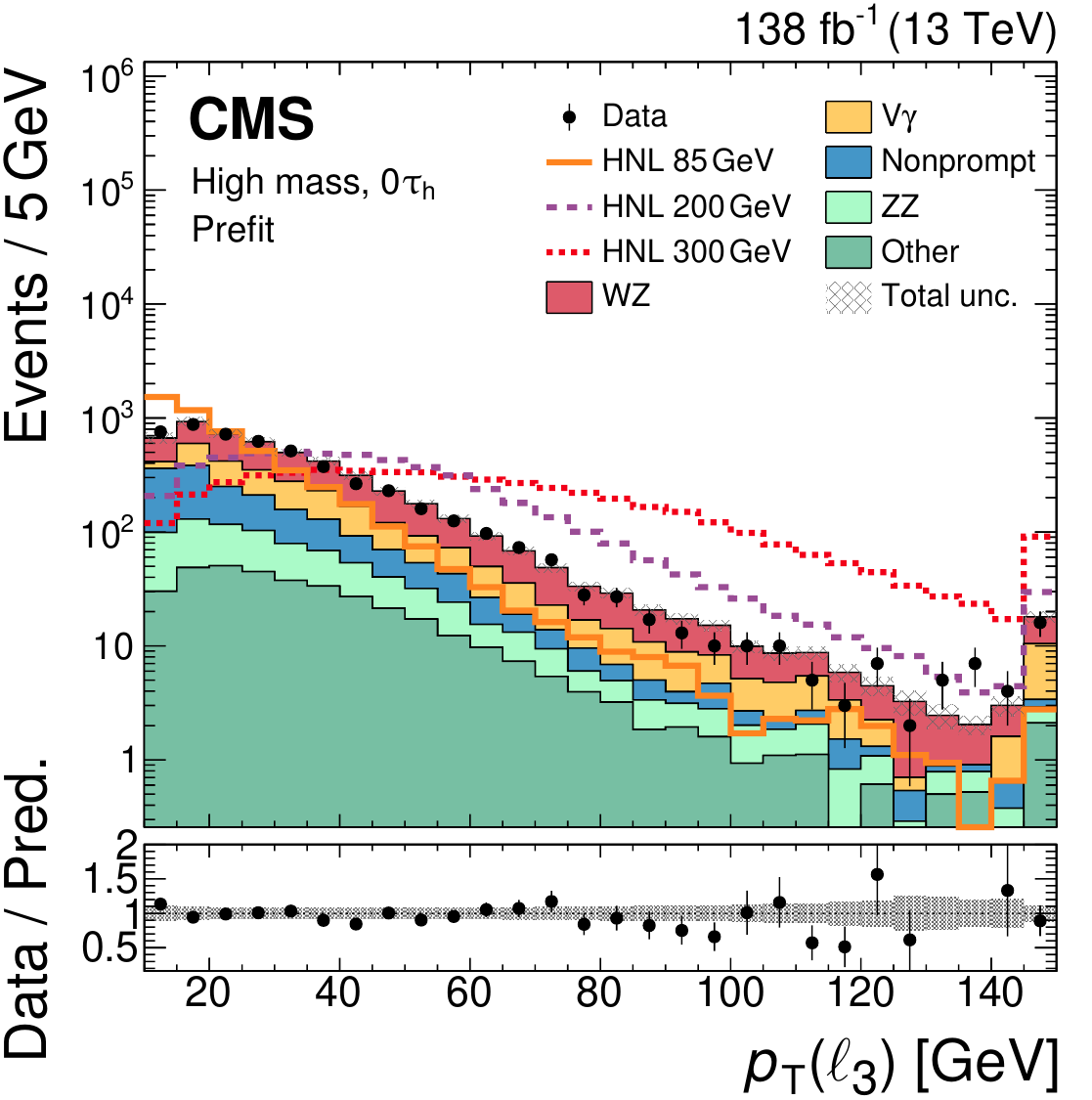}%
\hspace*{0.05\textwidth}%
\includegraphics[width=0.42\textwidth]{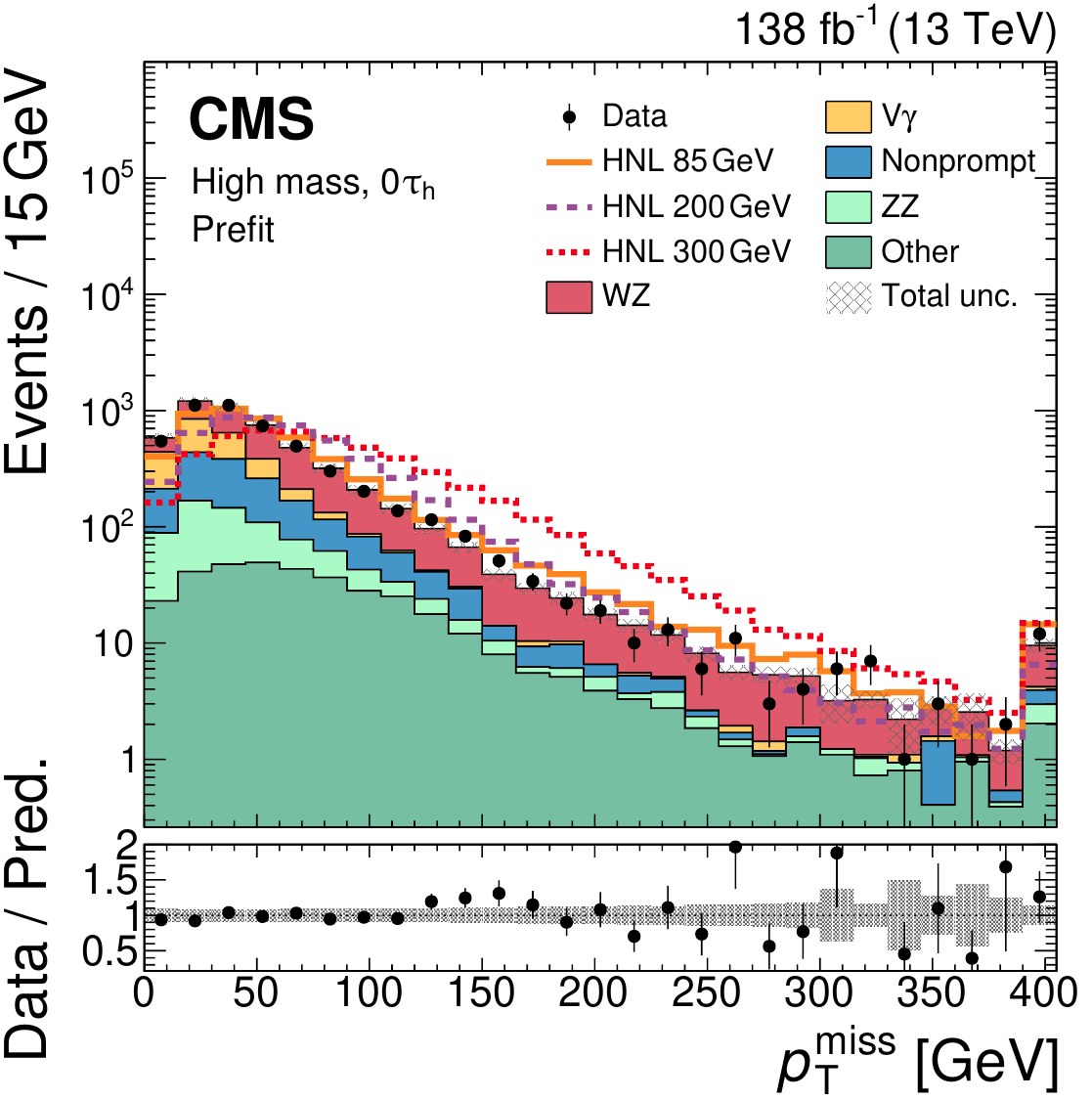}%
\caption{%
    Comparison of observed (points) and predicted (coloured histograms) distributions in the high-mass selection for the 0\PGth categories combined.
    Important input variables to the BDT training are shown:
    \DRminmllos (upper left),
    \mT (upper right),
    \ptl3 (lower left),
    \ptmiss (lower right).
    The predicted background yields are shown before the fit to the data (``prefit'').
    The HNL predictions for three different \mhnl values with exclusive coupling to tau neutrinos are shown with coloured lines, and are normalized to the total background yield.
    The vertical bars on the points represent the statistical uncertainties in the data, and the hatched bands the total uncertainties in the predictions.
    The last bins include the overflow contributions.
    In the lower panels, the ratios of the event yield in data to the overall sum of the predictions are shown.
}
\label{fig:inputs_high}
\end{figure}

The input variables to the BDTs are the kinematic properties of the reconstructed leptons; invariant and transverse masses of different dilepton and trilepton systems; the number of jets; kinematic properties of the reconstructed jets; \DR between different lepton and lepton-jet pairs; \ptmiss; and the sum of the \pt of all reconstructed jets (leptons), referred to as \HT (\LT).
Additional input variables to the BDTs trained with the low-mass selection are various \Dphi between \ptvecmiss and leptons or jets.
Furthermore, the BDTs trained with the low-mass selection in the 0\PGth categories have the flavours and charges of the reconstructed electrons and muons as additional input variables.
A selection of the most discriminating variables used in the different trainings is shown in Figs.~\ref{fig:inputs_low_notau}--\ref{fig:inputs_high}.
It can be seen that the different distributions provide sensitivity to distinguish between HNL signal and background distributions, but also show differences in the expected distributions for different \mhnl values, which is the reason to train separate BDTs for different mass ranges.
Generally good agreement is observed between data and background prediction.

The BDTs calculate event scores based on these input variables, and the score can be interpreted as a measure of how likely an event is to originate from the HNL signals used in the respective BDT training.
The agreement between data and background prediction is validated for all BDT input variables and output scores in the control regions (CRs) defined in Section~\ref{sec:backgrounds}, and good agreement is found.

For the final results, we combine the approaches based on SRs and BDTs.
In the low-mass selection, the background events used in the BDT training predominantly have an OSSF lepton pair, whereas the SRs with no OSSF pair have only small background yields and no significant separation between signal and background contributions is provided by the BDTs for these events.
Thus, we analyse together the BDT score distributions for the combined SRs with an OSSF pair (\ie Lb1--8, as defined in Table~\ref{tab:searchregions}) and the yields in the SRs without an OSSF pair (\ie La1--8).
In the high-mass selection, good sensitivity to the HNL signal is provided by the BDT scores for \mhnl up to 400\GeV for electron and muon neutrino couplings, whereas the low expected signal yields for larger \mhnl or exclusive couplings to tau neutrinos result in a loss of sensitivity for the BDT approach.
Consequentially, we use the BDT score distributions for the combined SRs Ha1--9 and Hb1--16 when targeting \mhnl up to 400\GeV for electron or muon neutrino couplings, and the yields in these SRs otherwise.

\section{Background estimation}
\label{sec:backgrounds}

Background contributions from SM processes are estimated with a combination of methods based on CRs in data and simulated event samples.
We distinguish between background contributions where all three selected leptons are prompt and those that have at least one nonprompt lepton.
For prompt-lepton backgrounds, we additionally treat processes separately that have a charge-misidentified electron (referred to as ``charge misID'') or at least one lepton originating from the conversion of a prompt photon produced at the interaction point.
Prompt-lepton backgrounds are estimated from the simulated event samples discussed in Section~\ref{sec:simulation}, and are dominated by \WZ and \ZZ diboson production, where the latter includes resonant contributions from $\PH\to\ZZ$ production.
All other prompt-lepton background contributions, of which the largest is from associated top quark and triboson production, are grouped together with charge-misID contributions as ``Other'' in the figures.

To validate the modelling of the dominant diboson background contributions in the simulated event samples, we define three CRs that are orthogonal to the SRs by the requirement of an OSSF lepton pair consistent with \mZ.
These CRs target \WZ and \ZZ production, as well as \ZG production with photon conversion.
The total yield predicted by the SM backgrounds is compared with the observed data yield, and a correction factor is derived where necessary.
Additionally, the distributions for several observables relevant in the SR definitions and the BDTs are compared to ensure that these background contributions are well modelled, and to derive uncertainties in the background normalizations.

\begin{figure}[!ht]
\centering
\includegraphics[width=0.42\textwidth]{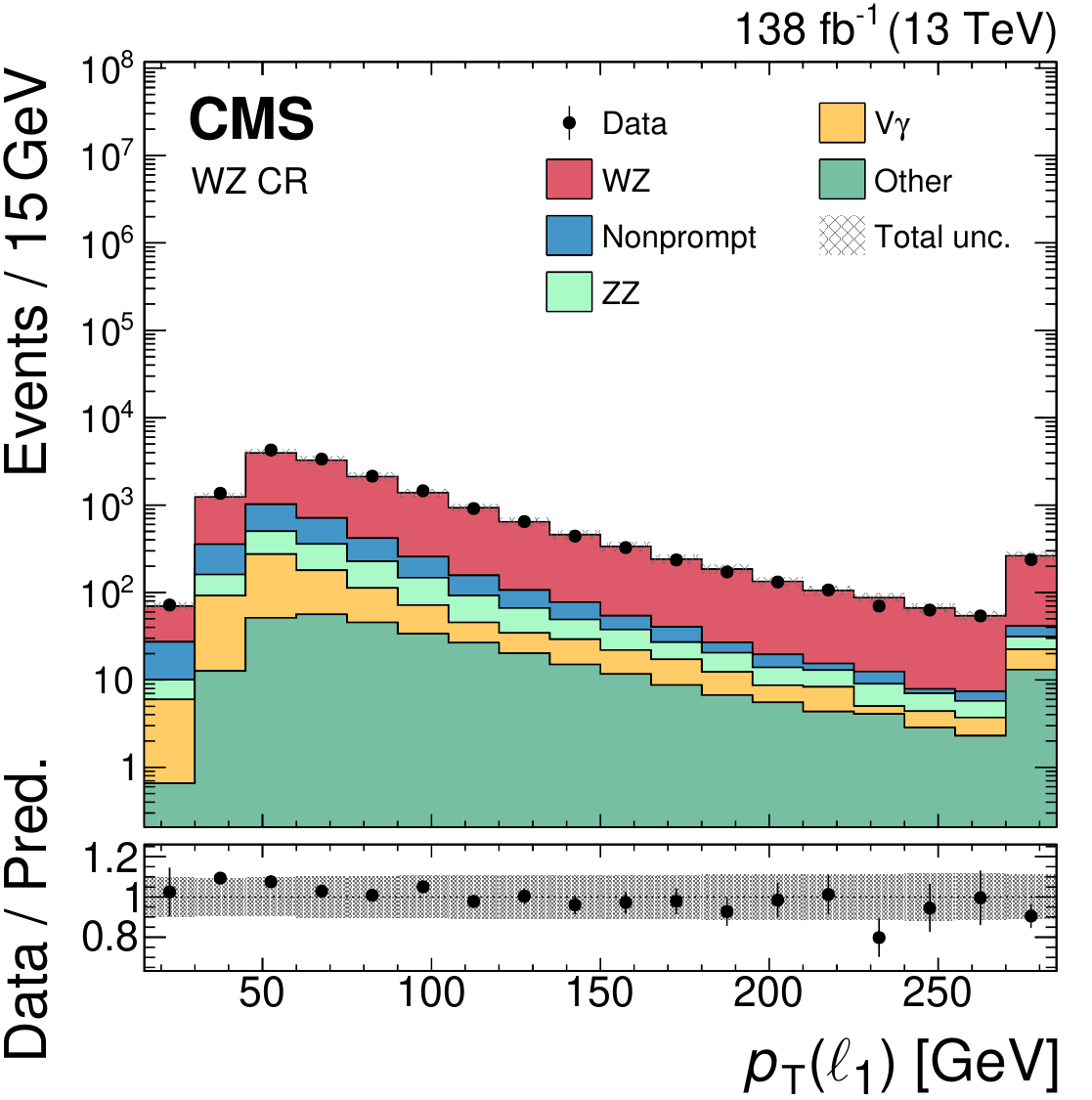}%
\hspace*{0.05\textwidth}%
\includegraphics[width=0.42\textwidth]{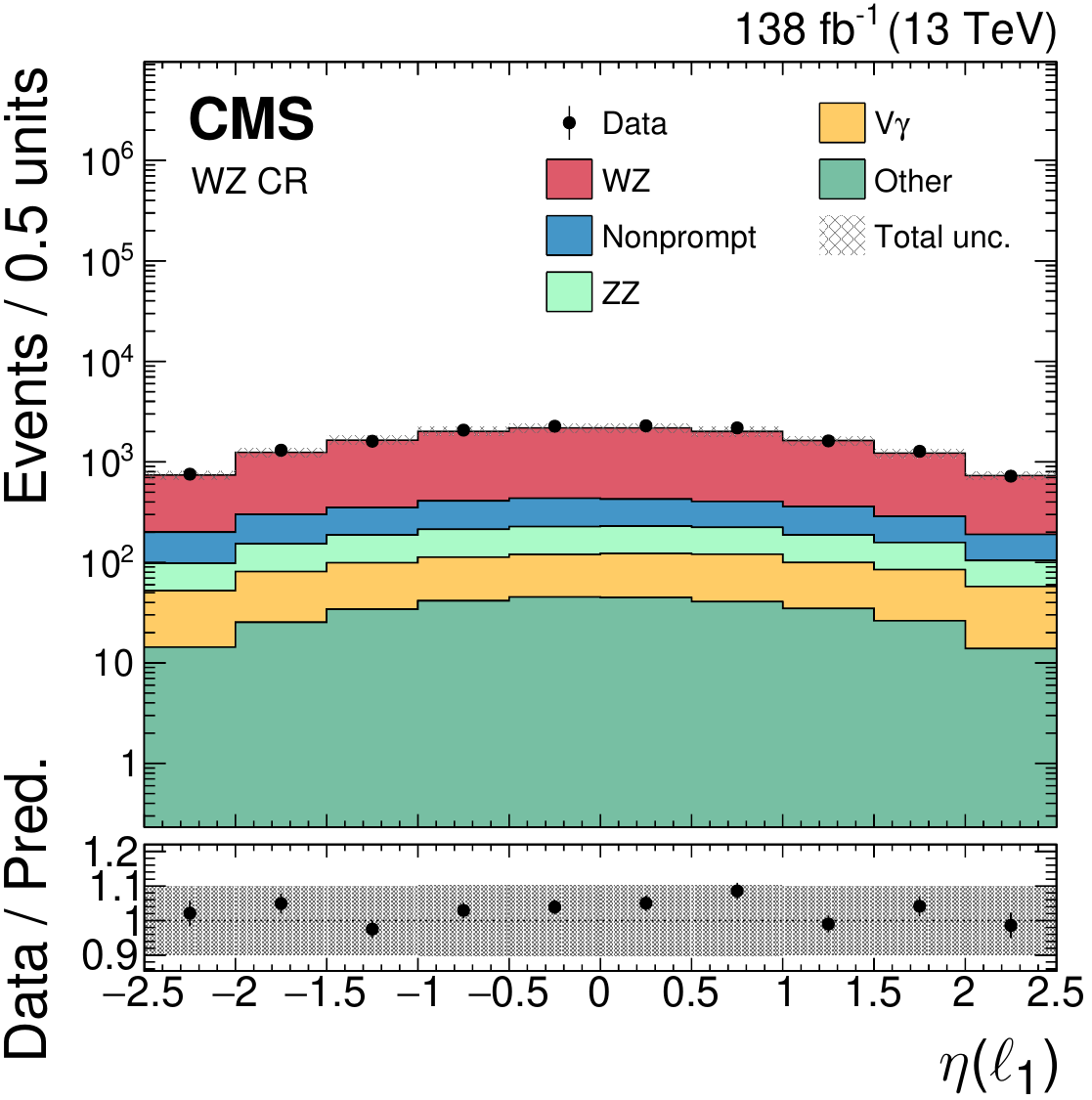} \\[1ex]
\includegraphics[width=0.42\textwidth]{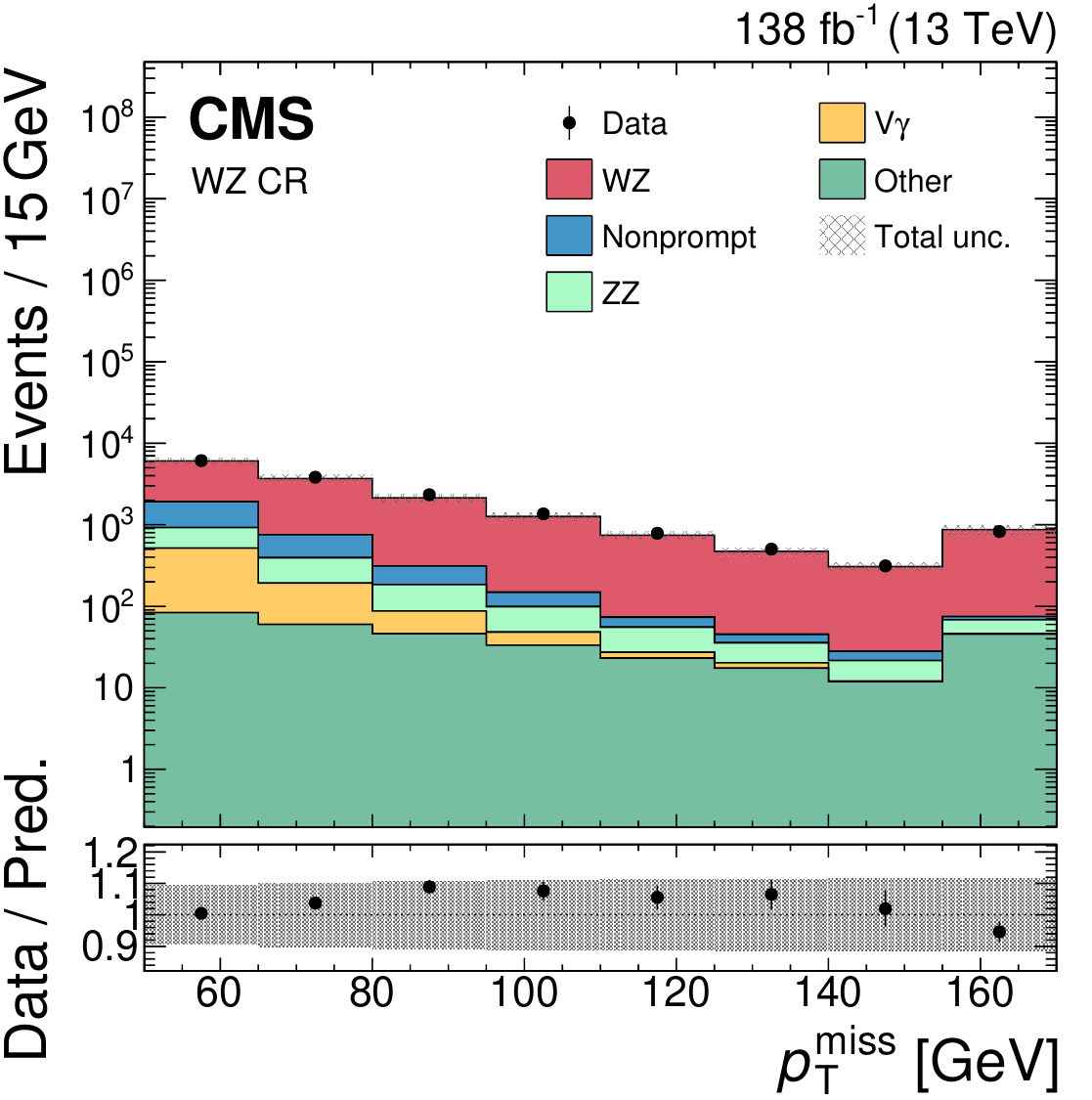}%
\hspace*{0.05\textwidth}%
\includegraphics[width=0.42\textwidth]{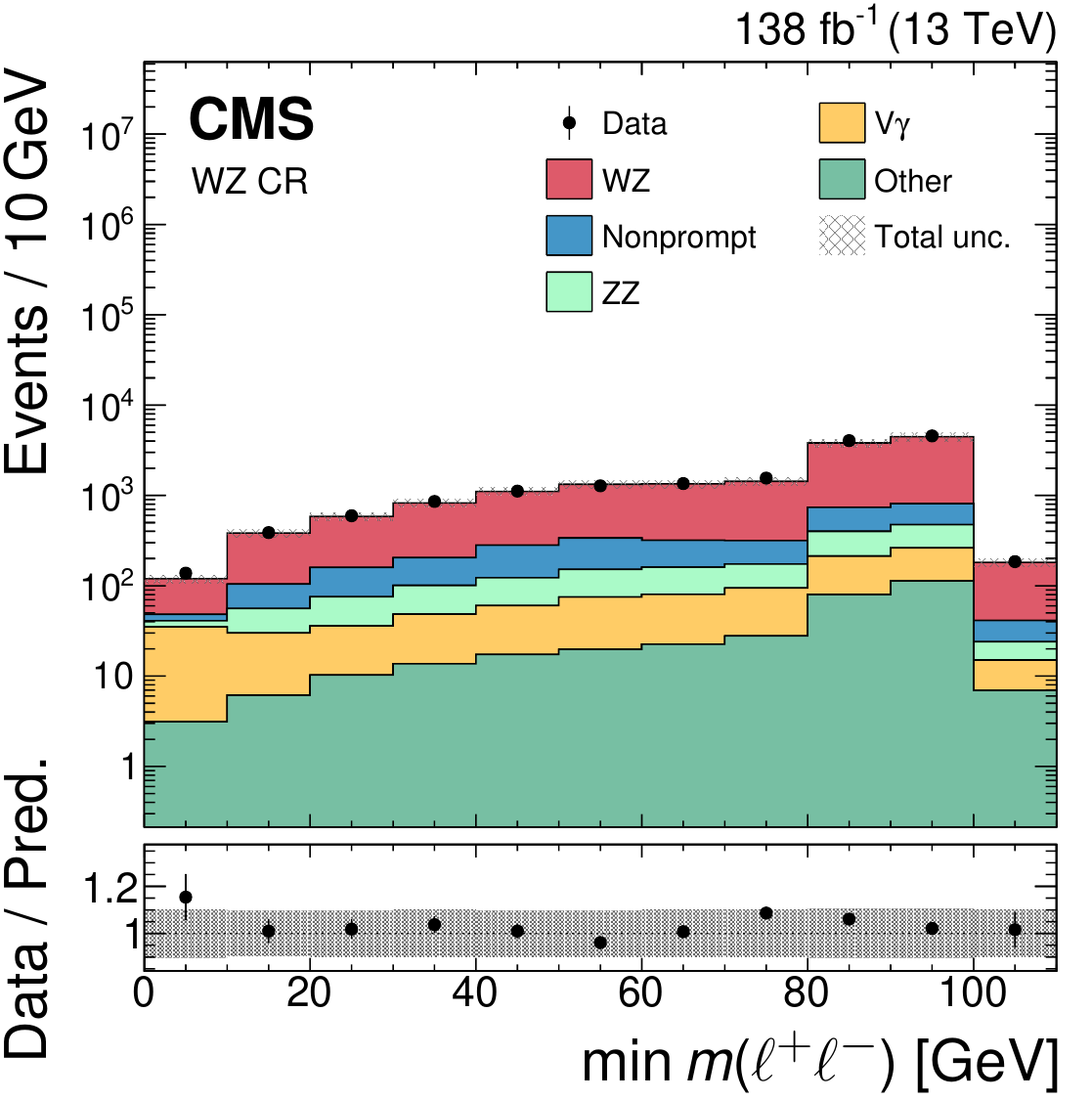}%
\caption{%
    Comparison of observed (points) and predicted (coloured histograms) distributions in the \WZ CR.
    The leading lepton \pt (upper left) and \sigeta (upper right), as well as \ptmiss (lower left) and \minmllos (lower right) are shown.
    The vertical bars on the points represent the statistical uncertainties in the data, and the hatched bands the total uncertainties in the predictions.
    The last bins include the overflow contributions.
    In the lower panels, the ratios of the event yield in data to the overall sum of the predictions are shown.
}
\label{fig:cr_wz}
\end{figure}

The \WZ CR is defined by selecting events with exactly three tight light leptons with $\ptl1>25\GeV$ and $\ptl2>15\GeV$, where two leptons form an OSSF pair with $\abs{\mll-\mZ}<15\GeV$.
Events with \PQb jets are excluded to reduce contributions from associated top quark production, $\ptmiss>50\GeV$ is required to account for the SM neutrino from the \PW boson decay, and the requirement of $\abs{\mthreel-\mZ}>15\GeV$ removes contributions with photon conversions.
More than 80\% of the events in the CR originate from \WZ production.
Signal contributions to this CR are negligible, with predicted yields of signal processes not excluded by Ref.~\cite{CMS:EXO-17-012} less than 0.5\% of the total background yield for all mass points in this analysis.
The total yields observed in the data agree with the prediction.
The comparison of the distributions for several observables in Fig.~\ref{fig:cr_wz} demonstrates good agreement as well, with deviations smaller than 10\% in most bins.

\begin{figure}[!ht]
\centering
\includegraphics[width=0.42\textwidth]{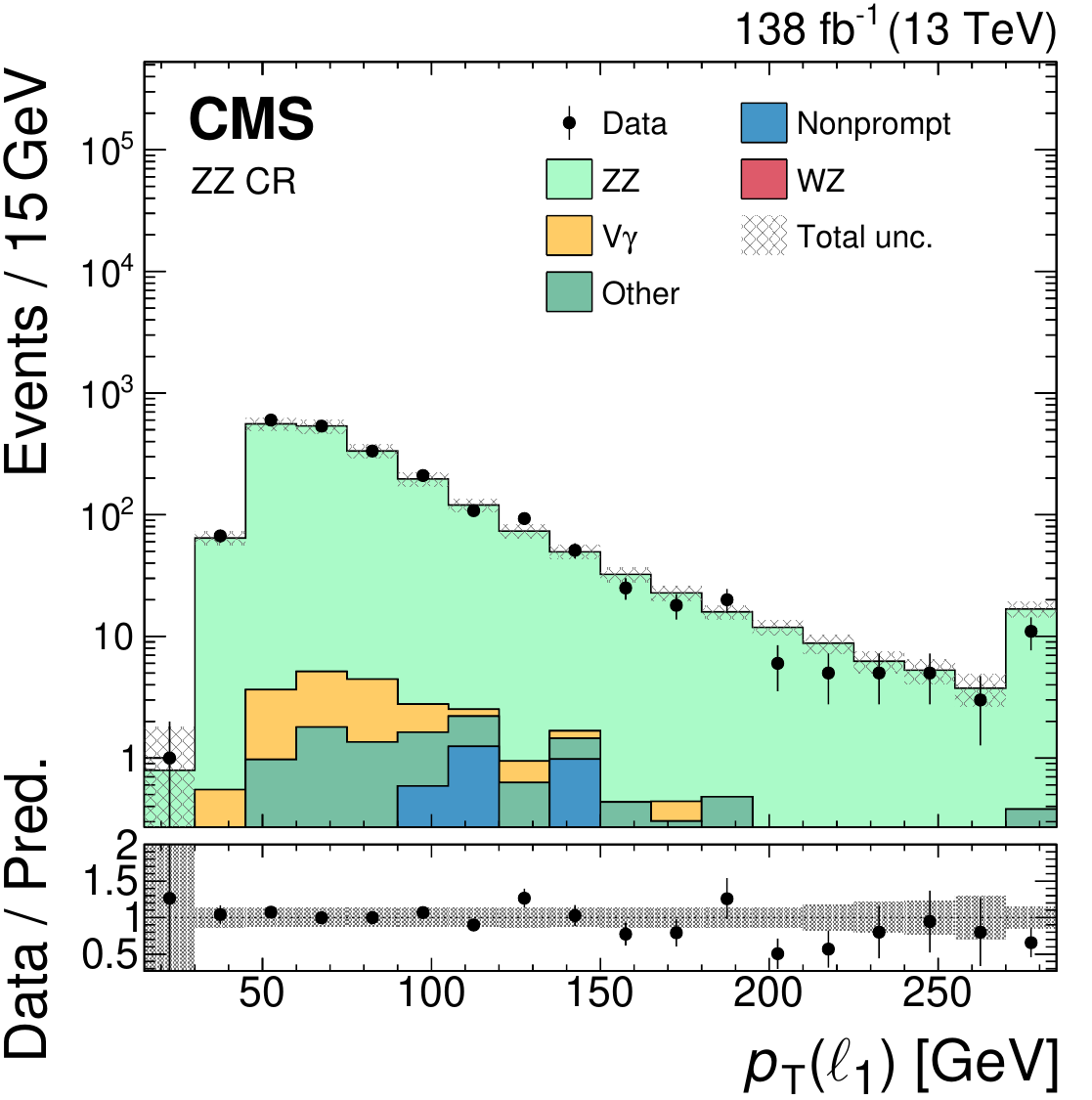}%
\hspace*{0.05\textwidth}%
\includegraphics[width=0.42\textwidth]{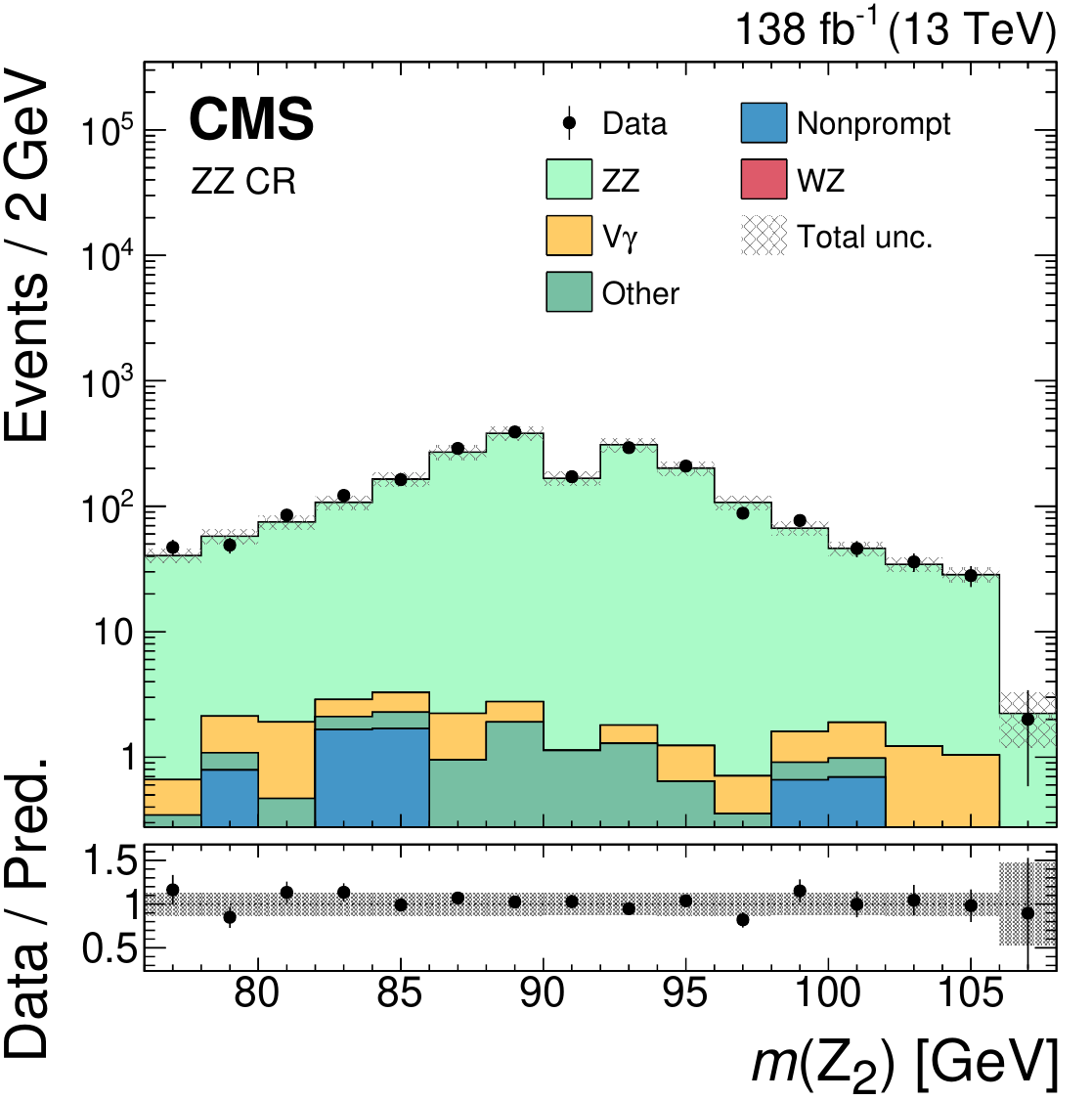} \\[1ex]
\includegraphics[width=0.42\textwidth]{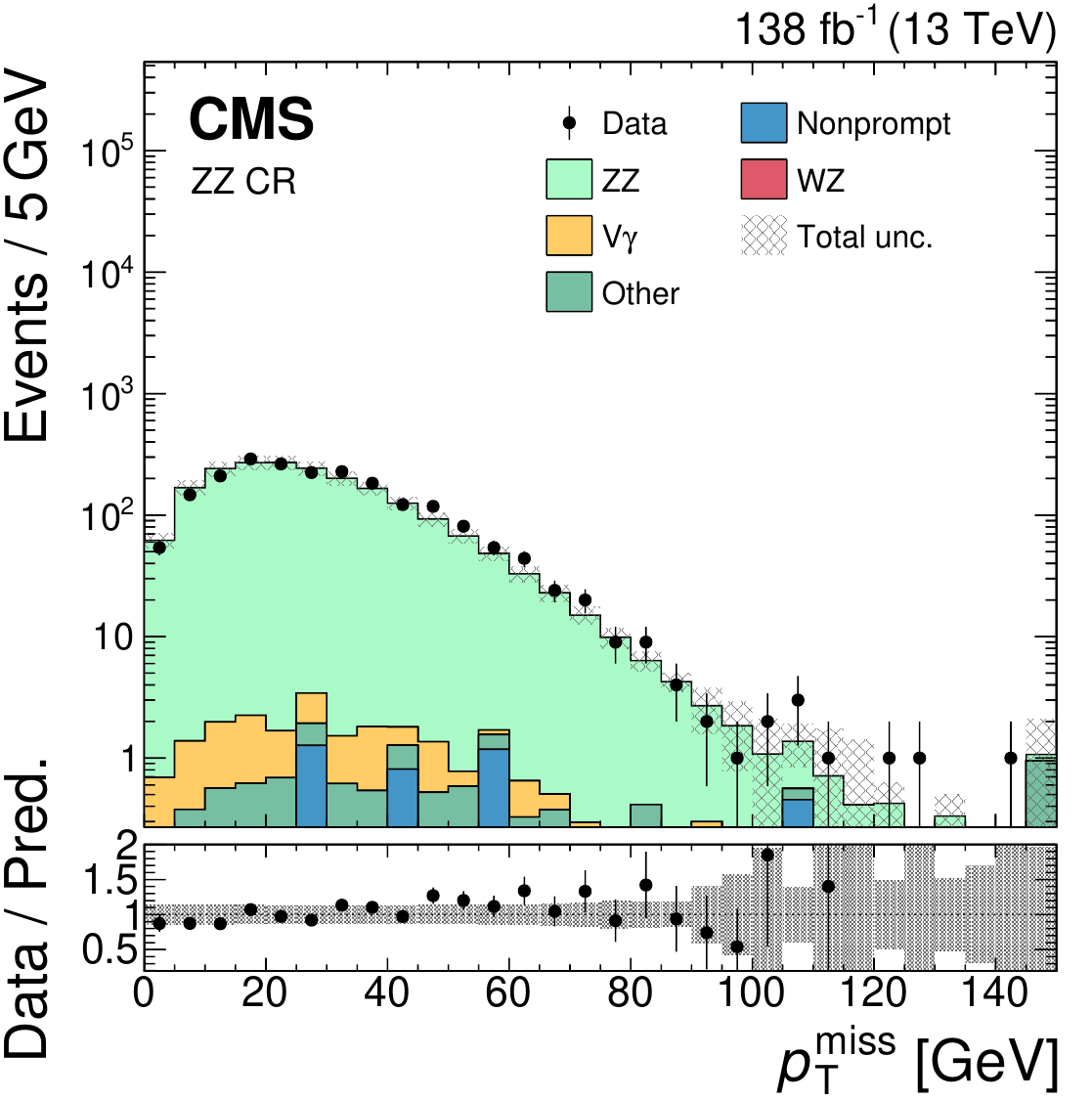}%
\hspace*{0.05\textwidth}%
\includegraphics[width=0.42\textwidth]{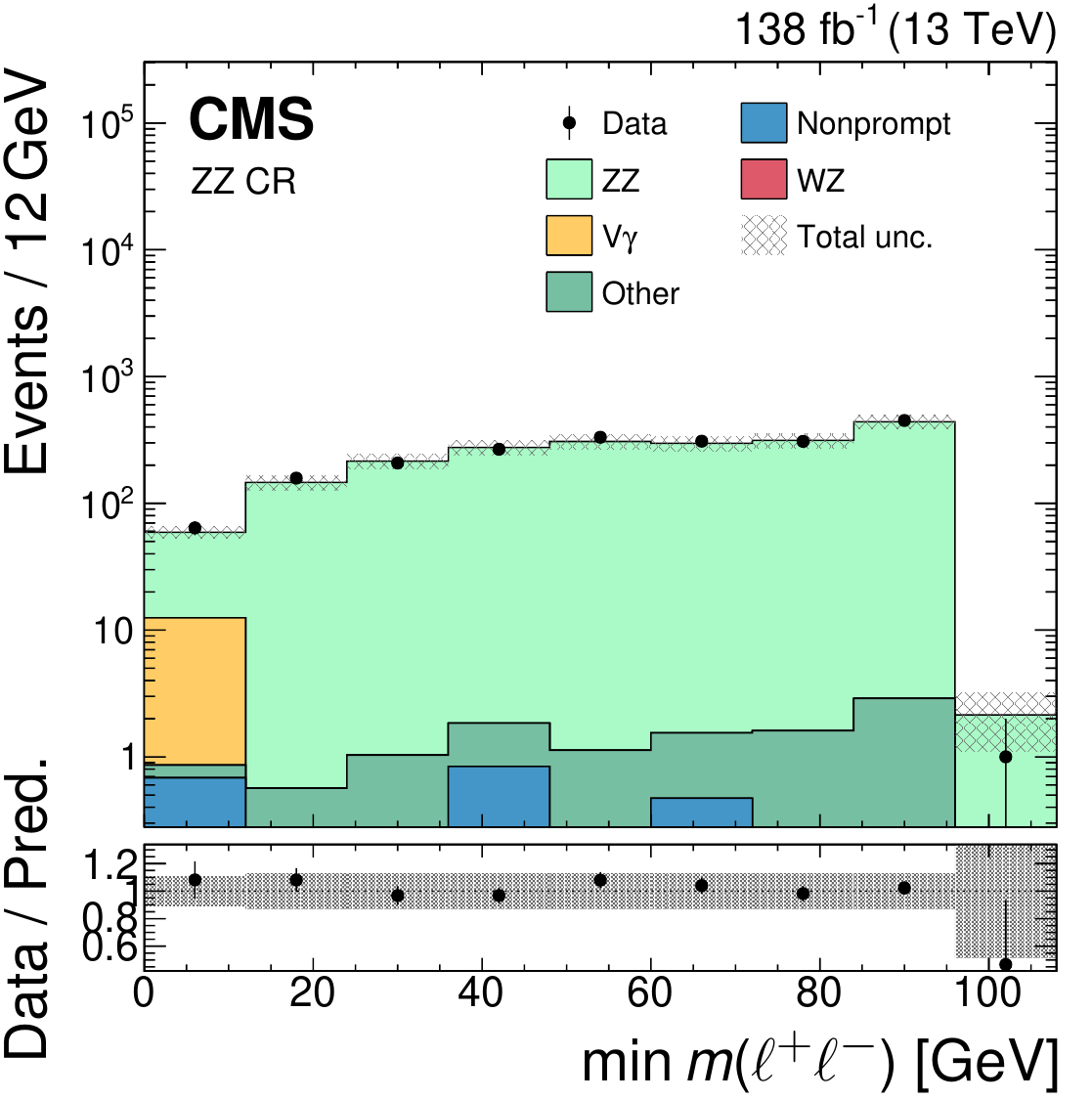}%
\caption{%
    Comparison of observed (points) and predicted (coloured histograms) distributions in the \ZZ CR.
    The leading lepton \pt (upper left), \mll of $\PZ_2$ ($m(\PZ_2)$, upper right), \ptmiss (lower left), and \minmllos (lower right) are shown.
    The \ZZ prediction is scaled with a normalization factor of 1.12, as discussed in the text.
    The vertical bars on the points represent the statistical uncertainties in the data, and the hatched bands the total uncertainties in the predictions.
    The last bins include the overflow contributions.
    In the lower panels, the ratios of the event yield in data to the overall sum of the predictions are shown.
}
\label{fig:cr_zz}
\end{figure}

The \ZZ CR is defined by selecting events with exactly four tight light leptons with $\ptl1>15\GeV$, where the four leptons form two OSSF pairs with $\abs{\mll-\mZ}<15\GeV$ each.
In the case of four leptons of the same flavour, the pairs are chosen such that the sum of the mass differences with respect to \mZ is minimized.
The OSSF pair with the invariant mass further away from \mZ is labelled ``$\PZ_2$''.
Events with \PQb jets are removed to reduce contributions from associated top quark production, and $\mll>12\GeV$ is required for every OS lepton pair to remove contributions from low-mass resonances.
Contributions from background processes other than \ZZ production or from signal contamination are negligible in this CR.
We find the observed yields to be larger than the prediction, and assign a scale factor of 1.12 to the simulated \ZZ samples to correct for the difference in the total yield.
After applying the scale factor, good agreement between the prediction and the observation is found across several observables, some of which are shown in Fig.~\ref{fig:cr_zz}.
Except for a few bins with lower statistical precision, the agreement is generally better than 10\%.

\begin{figure}[!ht]
\centering
\includegraphics[width=0.42\textwidth]{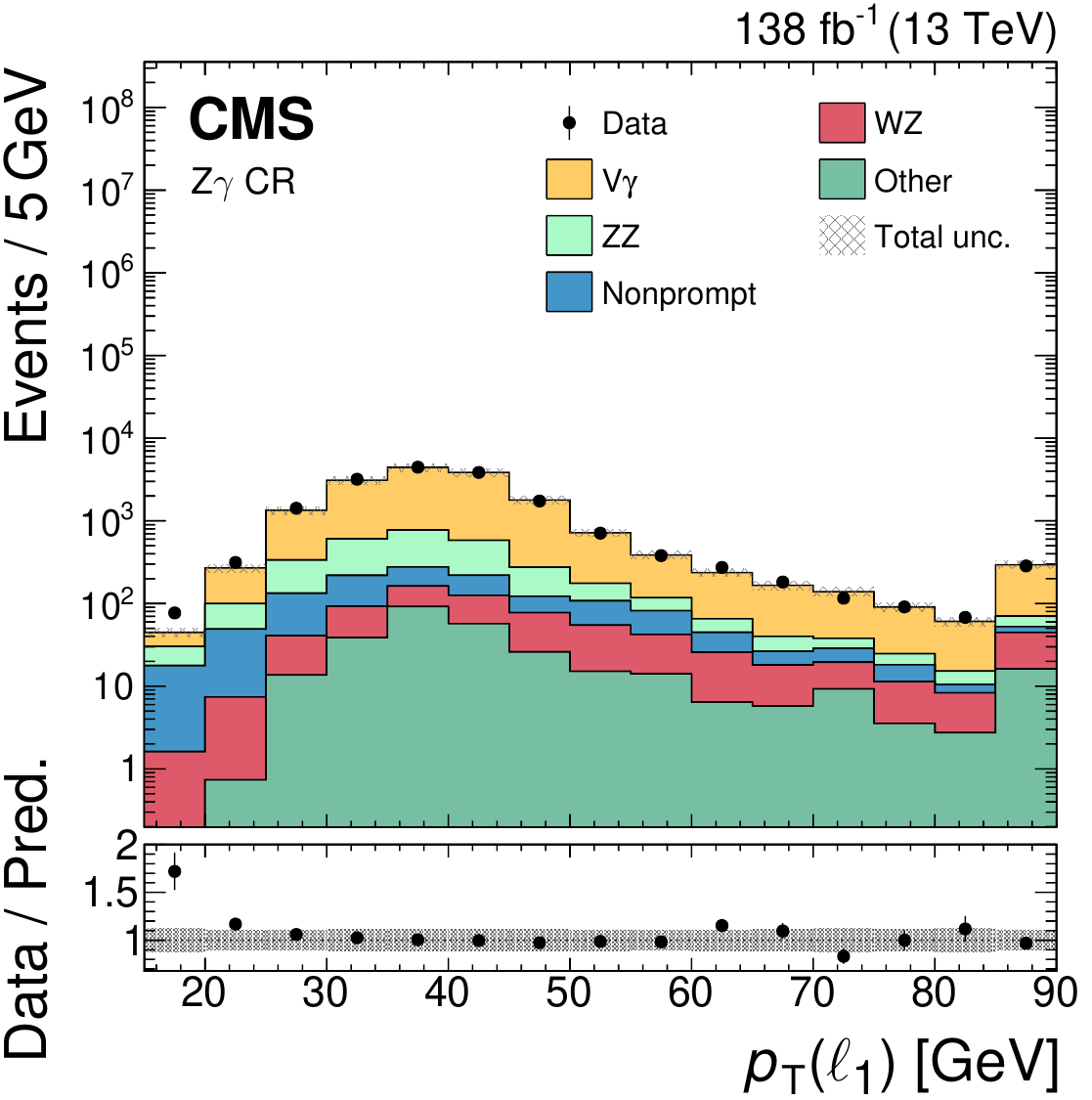}%
\hspace*{0.05\textwidth}%
\includegraphics[width=0.42\textwidth]{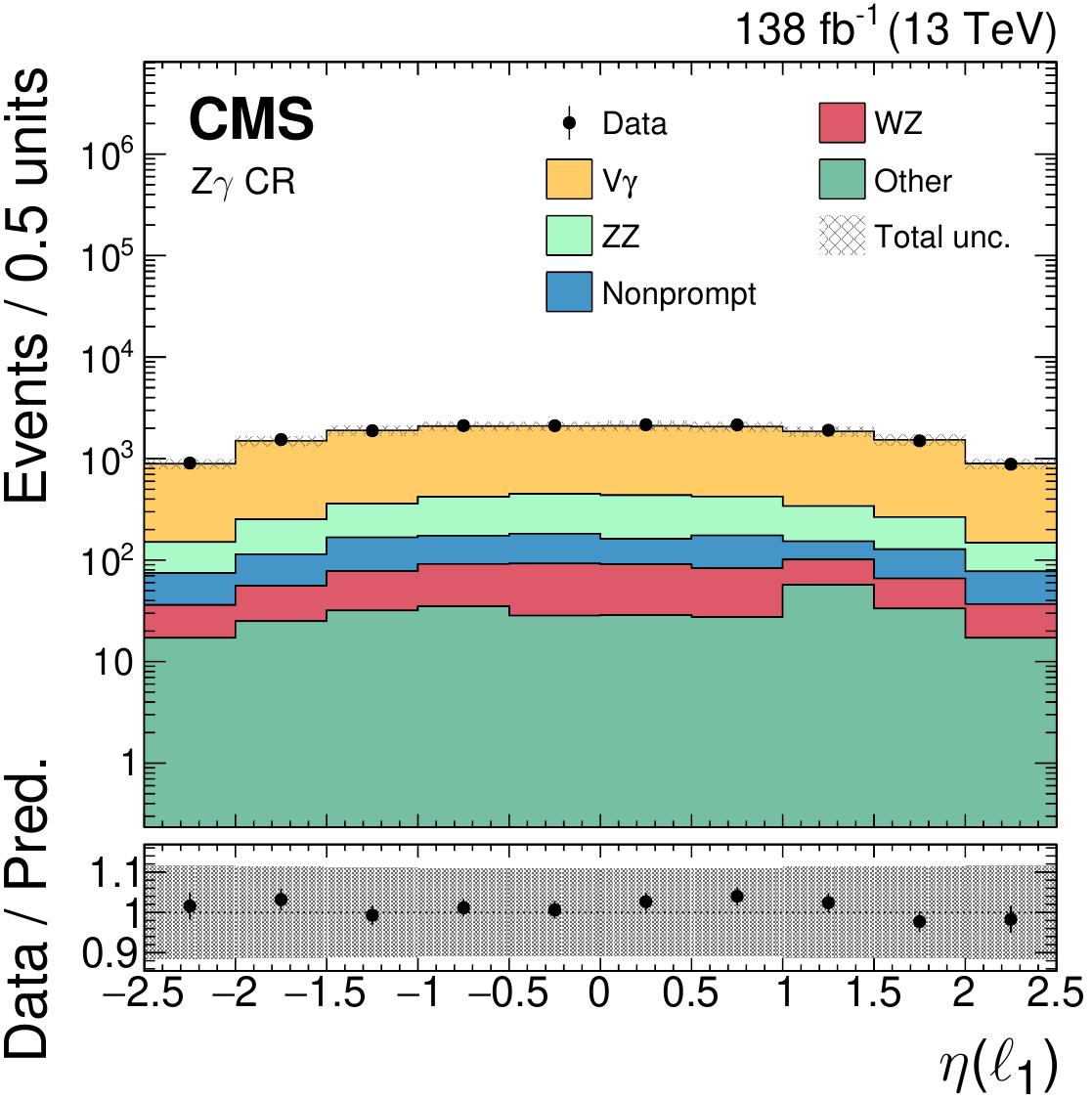} \\[1ex]
\includegraphics[width=0.42\textwidth]{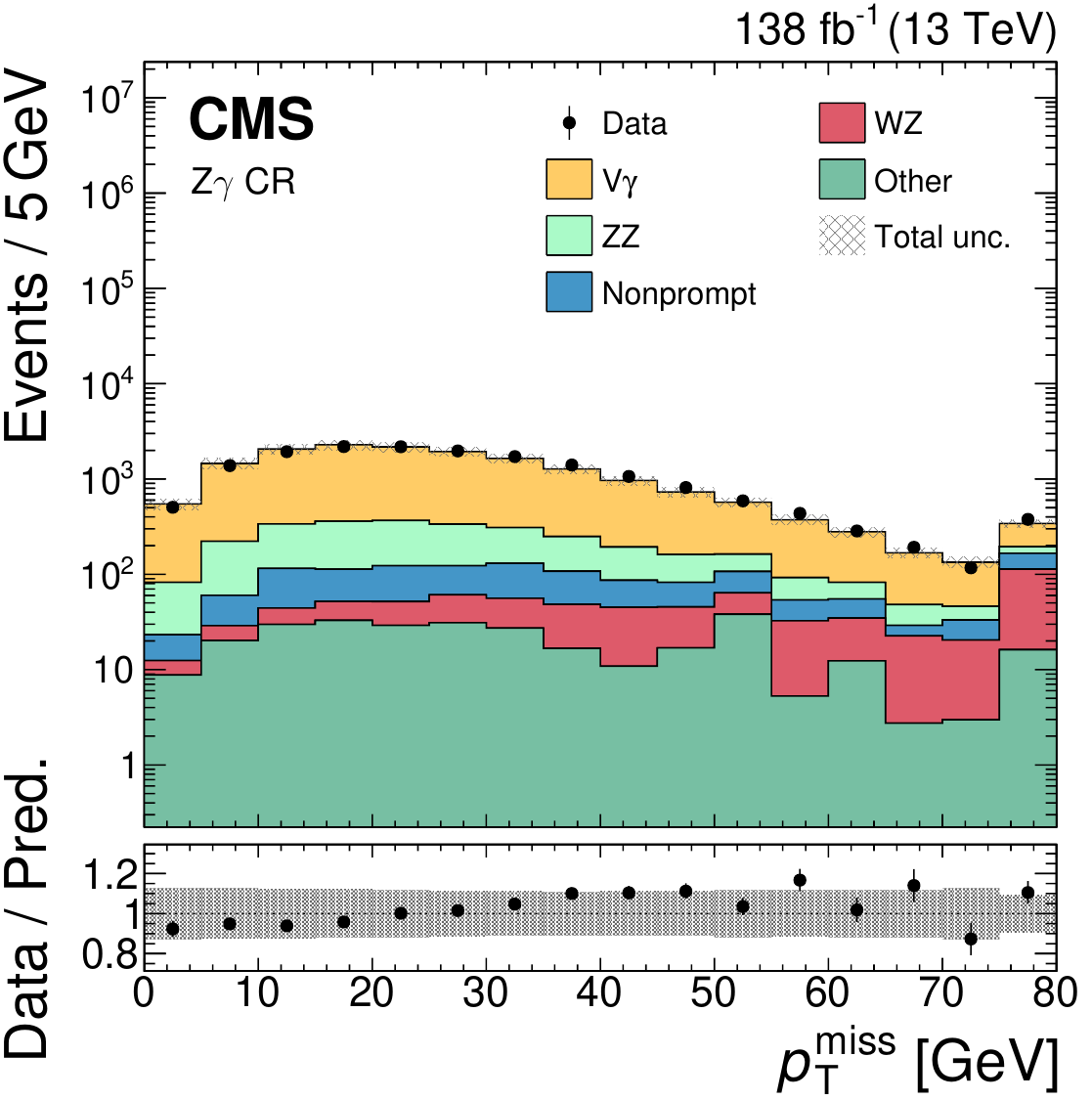}%
\hspace*{0.05\textwidth}%
\includegraphics[width=0.42\textwidth]{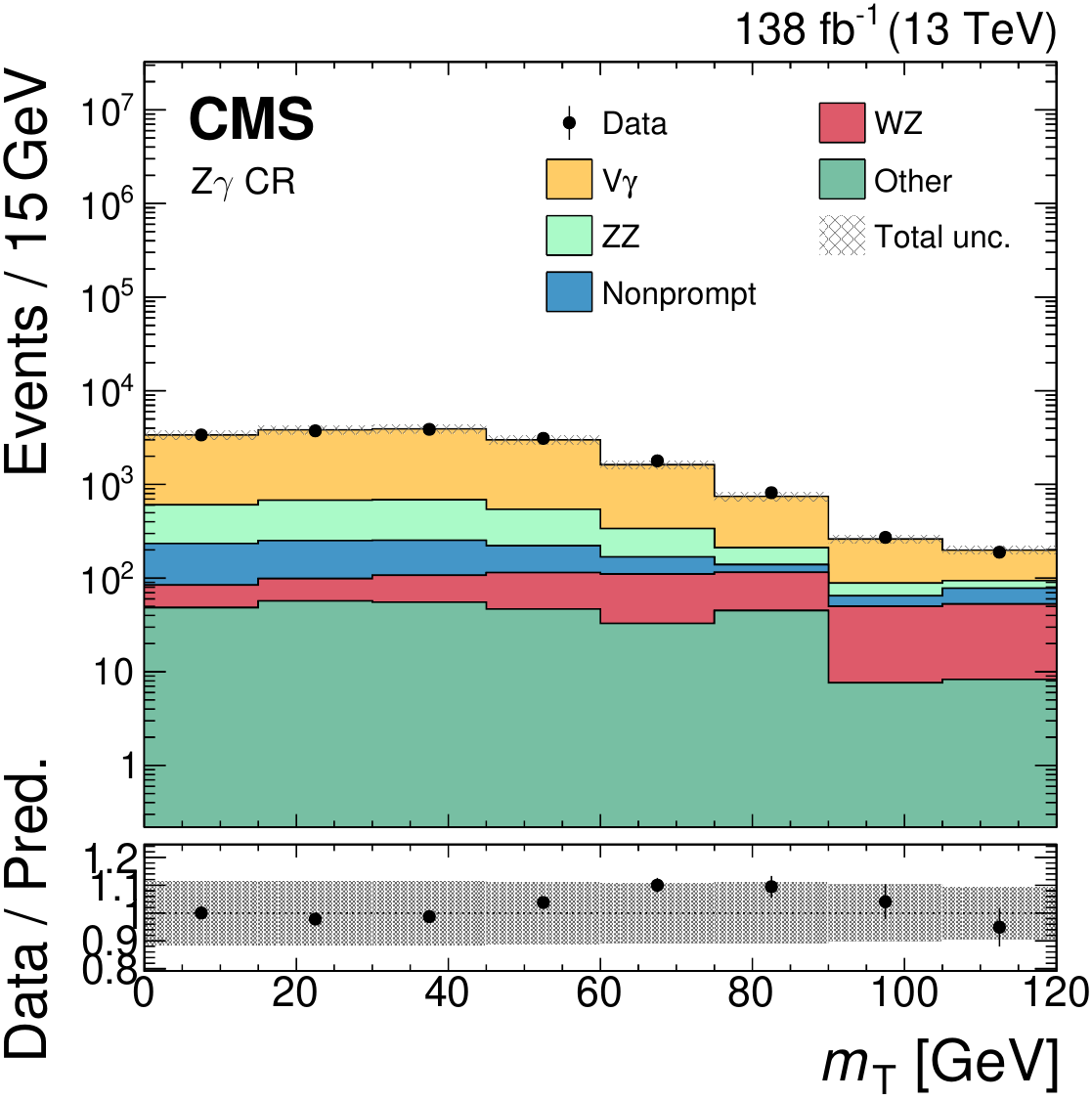}%
\caption{%
    Comparison of observed (points) and predicted (coloured histograms) distributions in the \ZG CR.
    The leading lepton \pt (upper left) and \sigeta (upper right), \ptmiss (lower left), and \mT (lower right) are shown.
    The \ZG prediction is scaled with a normalization factor of 1.11, as discussed in the text.
    The vertical bars on the points represent the statistical uncertainties in the data, and the hatched bands the total uncertainties in the predictions.
    The last bins include the overflow contributions.
    In the lower panels, the ratios of the event yield in data to the overall sum of the predictions are shown.
}
\label{fig:cr_zg}
\end{figure}

Background contributions from processes with photon conversions are also estimated from simulated samples.
The main photon conversion background arises from \ZG production, where the photon undergoes an asymmetric conversion into two leptons of which one has very low \pt and is not reconstructed.
For the \ZG CR, events with exactly three tight light leptons with $\ptl1>15\GeV$ are selected.
To select \ZG events where the photon is radiated from one of the leptons from the \PZ boson decay but at the same time remove contributions from \Zjets and \WZ production, we require $\abs{\mthreel-\mZ}<10\GeV$, as well as that two leptons form an OSSF pair with $\abs{\mll-\mZ}>15\GeV$.
Events with \PQb jets are removed.
In this CR, about 70\% of the events originate from photon conversions.
Expected signal yields for processes not excluded by Ref.~\cite{CMS:EXO-17-012} are below 0.5\% of the total yield.
To correct for differences in the total yield between data and prediction, we apply a scale factor of 1.11 to simulated \ZG samples.
Figure~\ref{fig:cr_zg} shows the data and predicted distributions for several observables, with the scale factor applied, and exhibits agreement that is typically better than 10\%.

Simulated event samples are used to predict background contributions with charge-misID electrons.
The misID rate in simulation depends strongly on the material included in the detector model, and is validated by dedicated measurements in data by comparing event yields with same- and opposite-sign electron pairs~\cite{CMS:TOP-22-013}.
It is found that the misidentification rate is overestimated (underestimated) in 2016 (2017--2018) samples by about 10 (50)\%, and we apply correction factors to the normalization of the charge-misID background correspondingly.

Nonprompt-lepton background contributions arise mostly from \ttbar and \Zjets production with an additional nonprompt lepton.
They are especially relevant in the 1\PGth categories.
A ``tight-to-loose'' ratio method~\cite{CMS:SUS-15-008, CMS:HIG-19-008, CMS:TOP-21-011} is applied to estimate the nonprompt-lepton background contributions from control samples in data.
The tight-to-loose ratio is defined as the probability \ttl for a loose lepton to also satisfy the tight ID selection.
It is evaluated separately for the different lepton flavours, and is measured as a function of \pt and \abseta.
For electrons and muons, \ttl is measured in a sample enriched in SM events composed uniquely of jets produced through the strong interaction selected with nonisolated single-lepton triggers.
For \PGth, \ttl is measured in samples enriched in \Zjets and \ttbar events.
The measured values of \ttl are applied as weights to events that pass the SR selection but have one or more leptons that pass the loose and fail the tight selection.
Both simulated events and data samples enriched in nonprompt leptons are used to validate the tight-to-loose ratio method for all lepton flavours.
Good agreement of better than 30\% is found in these tests in the most relevant kinematic distributions, with larger deviations up to 50\% only for nonprompt electrons with $\pt>55\GeV$.

\section{Systematic uncertainties}

Multiple sources of systematic uncertainty affect the signal prediction, the background event yields, and the distributions of the observables used for the signal extraction.
The sources and their correlations between the data-taking years are described below, and their impact on the fits described in Section~\ref{sec:interpretation} is summarized in Table~\ref{tab:systematics}.

The integrated luminosities for the three data-taking years have individual uncertainties between 1.2 and 2.5\%~\cite{CMS:LUM-17-003, CMS:LUM-17-004, CMS:LUM-18-002}, and the overall uncertainty for the 2016--2018 period is 1.6\%.
This uncertainty affects the normalization of the background contributions from simulated event samples, as well as the extraction of cross section limits from the final estimate of the limit on the number of signal events.

The distribution of the number of additional \pp interactions per event in simulation is matched to data by reweighting the profile of the true number of interactions to the one inferred from the instantaneous luminosity profile in data.
The systematic uncertainty is estimated from a variation of the total inelastic cross section used for this reweighting by $\pm$4.6\%, which is treated as correlated among the data-taking years.

\begin{table}[!b]
\centering\renewcommand{\arraystretch}{1.1}
\topcaption{%
    Relative impacts of the uncertainty sources in fits for six different fit models specified with \mhnl value and coupling scenario, where the relative impact is defined as the ratio between the uncertainty from the respective source and the total uncertainty in the HNL signal strength.
    The symbol ``\NA'' indicates that the corresponding uncertainty source is not applicable.
}
\label{tab:systematics}
\renewcommand{\arraystretch}{1.2}
\cmsTable{\begin{tabular}{lcccccc}
    \multirow{2}{*}{Uncertainty source} & \multicolumn{3}{c}{$\mhnl=40\GeV$} & \multicolumn{3}{c}{$\mhnl=200\GeV$} \\
    & \Pe & \PGm & \PGt & \Pe & \PGm & \PGt \\ \hline
    Luminosity, pileup reweighting & 5.0\% & 2.6\% & 11.3\% & 5.8\% & 7.6\% & 4.4\% \\
    Trigger efficiency & 2.4\% & 10.6\% & 26.4\% & 2.9\% & 6.9\% & 5.1\% \\
    Light-lepton selection efficiency \& energy calibration & 8.7\% & 15.5\% & 7.9\% & 10.4\% & 18.3\% & 1.2\% \\
    \PGth selection efficiency & \NA & \NA & 2.7\% & \NA & \NA & 14.2\% \\
    Jet energy calibration, \ptmiss, \PQb tagging efficiency & 10.6\% & 8.4\% & 34.4\% & 8.6\% & 12.4\% & 24.0\% \\
    \WZ background normalization & 1.1\% & 9.6\% & 7.4\% & 2.8\% & 5.4\% & 4.0\% \\
    \ZZ background normalization & 3.9\% & 9.9\% & 8.0\% & 4.1\% & 6.1\% & 6.5\% \\
    \ZG background normalization & 7.4\% & 8.3\% & 5.1\% & 1.9\% & 1.5\% & 12.1\% \\
    Other background normalization & 1.4\% & 5.6\% & 1.6\% & 13.8\% & 6.3\% & 10.5\% \\
    Nonprompt light-lepton background & 10.8\% & 16.0\% & 20.9\% & 15.6\% & 26.4\% & 9.1\% \\
    Nonprompt \PGth background & \NA & \NA & 14.3\% & \NA & \NA & 66.7\% \\
    HNL cross section prediction & 4.7\% & 3.5\% & 3.8\% & 3.7\% & 3.0\% & 2.1\% \\[\cmsTabSkip]
    Total systematic & 23.3\% & 25.5\% & 55.1\% & 27.7\% & 35.8\% & 75.5\% \\[\cmsTabSkip]
    Statistical & 96.8\% & 96.5\% & 83.4\% & 96.1\% & 93.3\% & 65.5\% \\
\end{tabular}}
\end{table}

The trigger selection efficiency is measured in data with independent trigger paths based on hadronic activity and \ptmiss signatures, and agrees with the efficiency estimated in simulation within 3\%.
Thus, no correction is applied to the simulated event samples and a systematic uncertainty of 3\% is assigned that is correlated between the data-taking years.
Additionally, the statistical uncertainty in the measured trigger efficiencies in data is considered, separately for each data-taking year.

During the 2016 and 2017 data-taking periods, a gradual shift in the timing of the inputs of the ECAL level-1 trigger in the region $\abseta>2.0$ caused a specific trigger inefficiency~\cite{CMS:TRG-17-001}.
For events containing an electron (a jet) with $\pt>50$ (100)\GeV in the region $2.5<\abseta<3.0$ the efficiency loss is $\approx$10--20\%, depending on \pt, $\eta$, and time.
Correction factors are derived from data and applied to the acceptance evaluated by simulation, and the impact on our results is small.

The efficiency of the tight ID selection of light leptons is measured in data and simulation using a ``tag-and-probe'' method applied to $\PZ\to\Pellp\Pellm$ events~\cite{CMS:EWK-10-002}.
Per-lepton corrections are derived separately for electrons, muons, and \PGth.
Statistical and systematic uncertainties in the correction factors are included, with the former (latter) treated as uncorrelated (correlated) between the data-taking years.
Corrections for the differences in the electron energy scale and resolution between data and simulation are derived from $\PZ\to\Pe\Pe$ events using only ECAL information~\cite{CMS:EGM-17-001}, and systematic uncertainties are considered that are correlated between the data-taking years.

Uncertainties in the jet energy scale and resolution are evaluated from the \pt variations of the reconstructed jets in simulated events~\cite{CMS:JME-13-004}.
The variation due to the jet energy scale, as well as an additional variation to account for the uncertainty in the contribution from unclustered PF particles~\cite{CMS:JME-17-001}, is propagated to \ptmiss.
The jet energy scale (jet energy resolution and unclustered energy) variation is treated as correlated (uncorrelated) between the data-taking years.
Differences in the \PQb tagging efficiency between data and simulation are corrected by applying scale factors to simulated events.
Uncertainties in the scale factors are evaluated by separate variations for light- and heavy-flavour jets, where both correlated and uncorrelated variations between the three data-taking years are considered~\cite{CMS:BTV-16-002}.

Several uncertainties are considered for the normalization of the background processes.
For the dominant \WZ, \ZZ, and \ZG contributions, we assign an uncertainty of 10\% each, corresponding to the level of agreement in the CRs described in Section~\ref{sec:backgrounds}.
The SM predictions for triboson production have a precision of about 10\%~\cite{Lazopoulos:2007ix, Binoth:2008kt, Hankele:2007sb, Campanario:2008yg, Dittmaier:2017bnh}, which is assigned as a normalization uncertainty to the triboson background.
For associated top quark production, a normalization uncertainty of 10\% is assigned, matching the experimental precision of the latest CMS measurements of the most important contributions~\cite{CMS:TOP-20-010, CMS:TOP-21-011}.
The uncertainty in the charge-misID contribution originates from the correction factors and is taken to be 15\%~\cite{CMS:TOP-22-013}.
For all remaining contributions in the ``Other'' category, we assign a normalization uncertainty of 20\% corresponding to the experimental precision in the signal strength of \PH production in association with a vector boson~\cite{CMS:HIG-22-001}.

The uncertainties in the nonprompt background contributions stem from the tight-to-loose ratio method.
For nonprompt light leptons, a normalization uncertainty of 30\% is applied when the leading nonprompt lepton is a muon, and a \pt-dependent uncertainty when it is an electron.
In the latter case, the uncertainty is 15\% for $\pt<35\GeV$ of the electron, 30\% for $35<\pt<55\GeV$, and 50\% for $\pt>55\GeV$.
In the case of nonprompt \PGth leptons, a normalization uncertainty of 30\% is assigned to account for observed differences in the validation, separately for events with and without an OSSF lepton pair because of the different composition of sources of nonprompt \PGth leptons in these two event selections.

The HNL signal samples for $\mhnl<\mW$ are simulated at LO accuracy, and their normalization is scaled with the ratio between the SM cross sections of \PW boson production evaluated at next-to-NLO with the \FEWZ v3.1 program~\cite{Melnikov:2006kv, Gavin:2010az, Gavin:2012sy, Li:2012wna} and at LO with \MGvATNLO, using settings identical to those of the signal samples.
The uncertainty in the signal cross section is then evaluated from the variations of the renormalization and factorization scale and of the PDFs in the next-to-NLO calculation and amounts to 4\% in total~\cite{CMS:EXO-20-009}.
For the HNL signal samples with $\mhnl>\mW$, the simulation is done at NLO accuracy, and no additional scale factor is applied.
From the evaluation of scale variations and the PDF choice, we find a signal cross section uncertainty of 3 (15)\% for the DY (VBF) production mode.

\section{Results}

\begin{figure}[!p]
\centering
\includegraphics[width=0.40\textwidth]{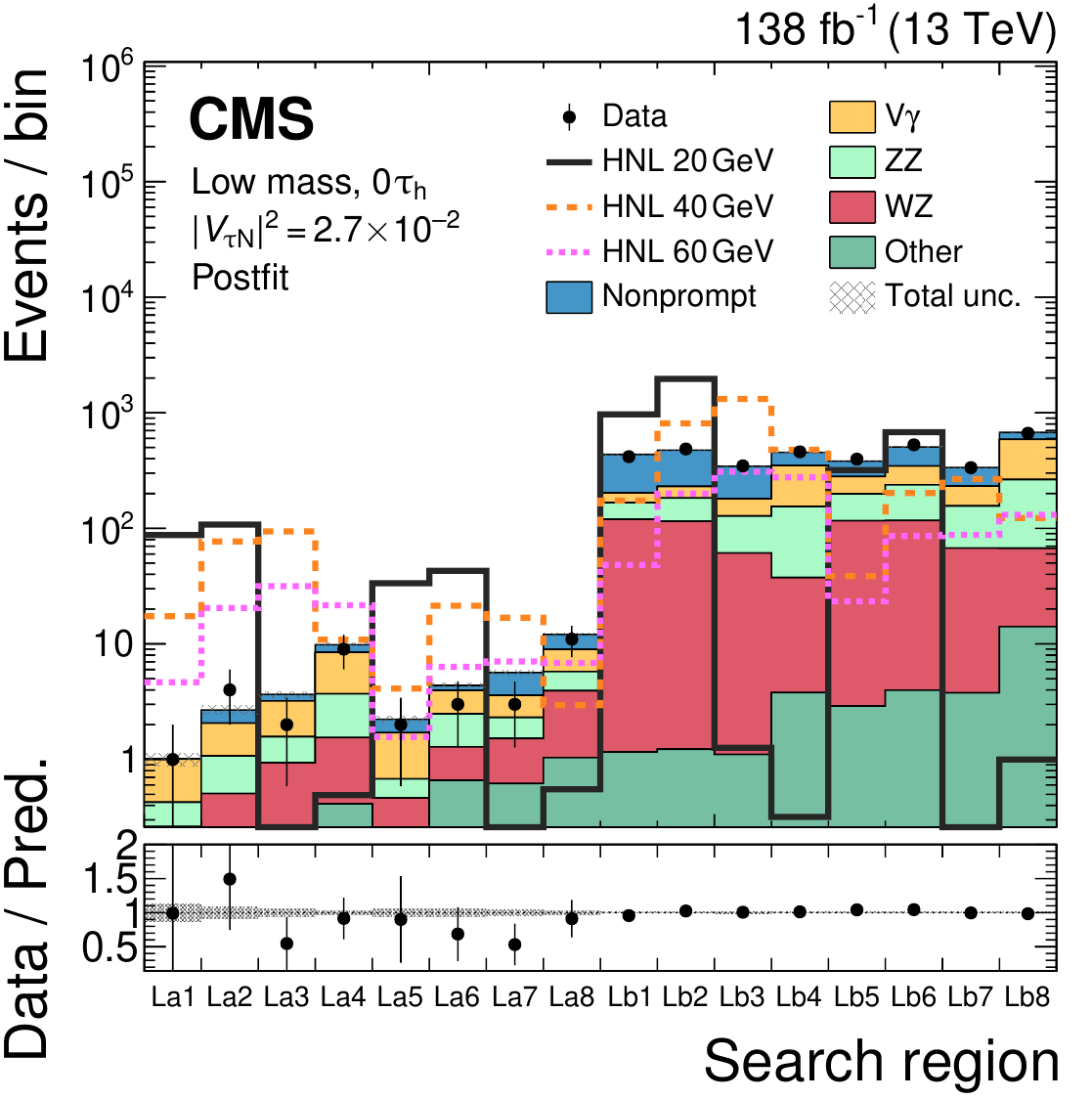}%
\hspace*{0.05\textwidth}%
\includegraphics[width=0.40\textwidth]{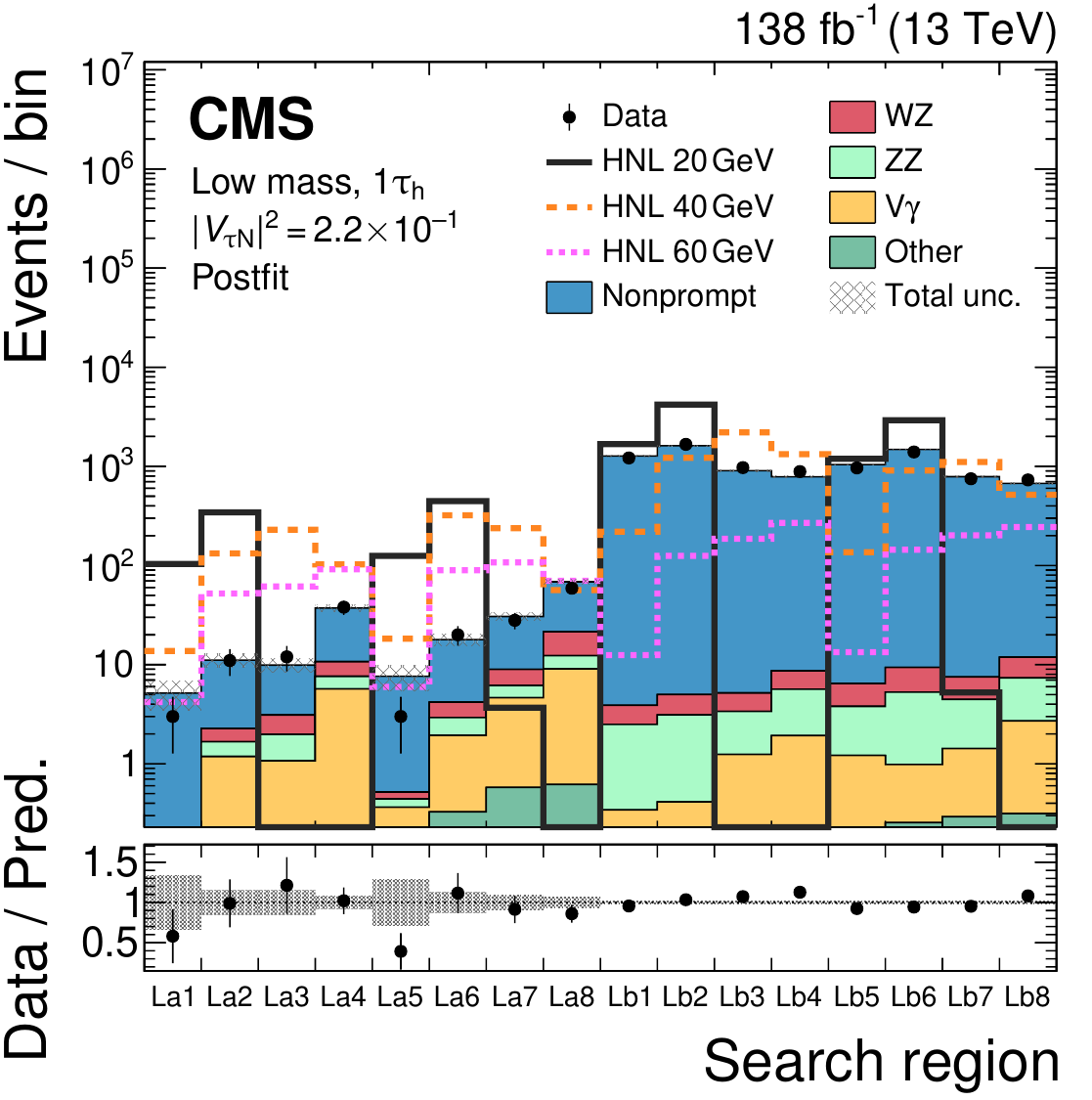} \\[1ex]
\includegraphics[width=0.40\textwidth]{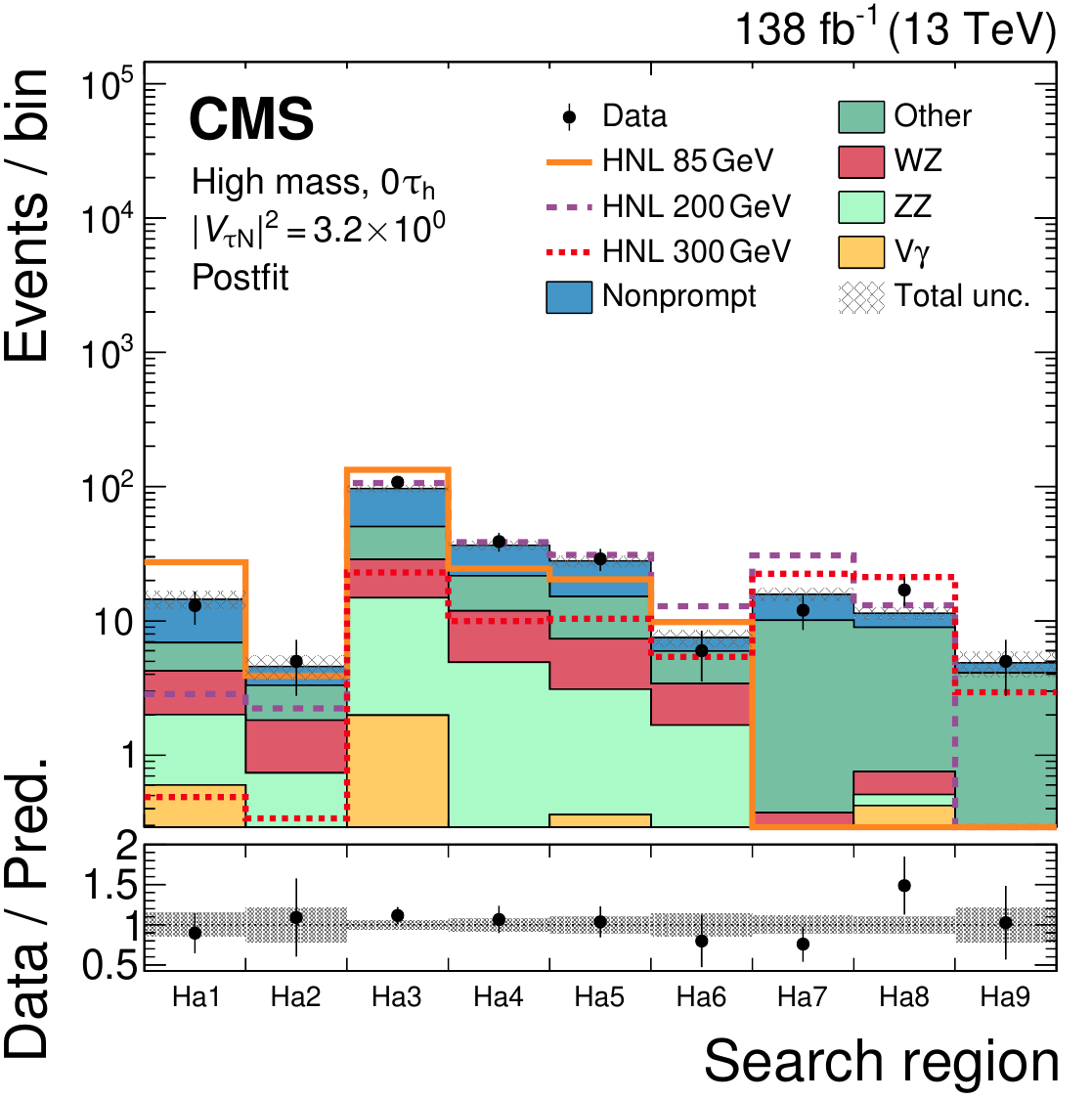}%
\hspace*{0.05\textwidth}%
\includegraphics[width=0.40\textwidth]{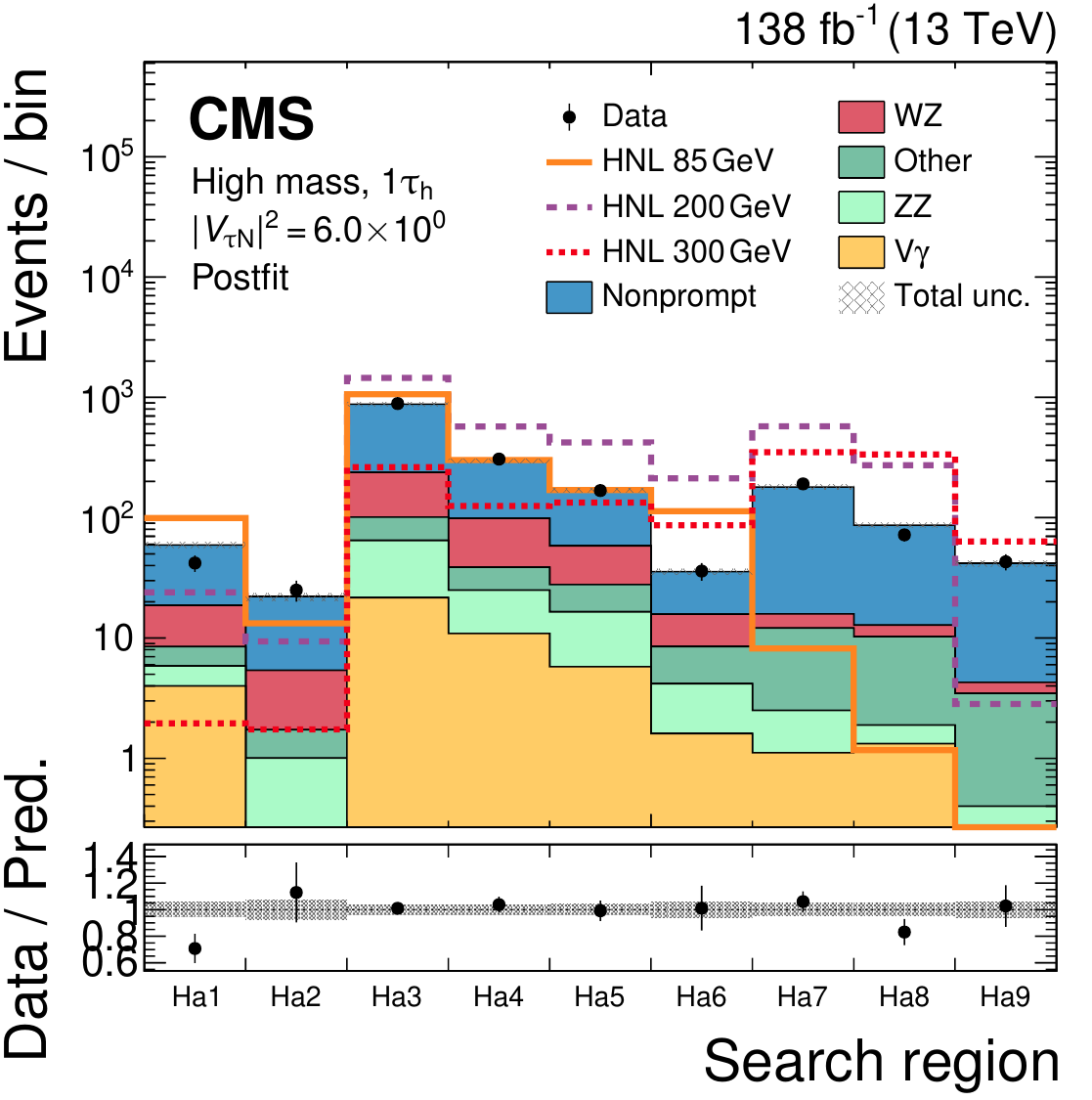} \\[1ex]
\includegraphics[width=0.40\textwidth]{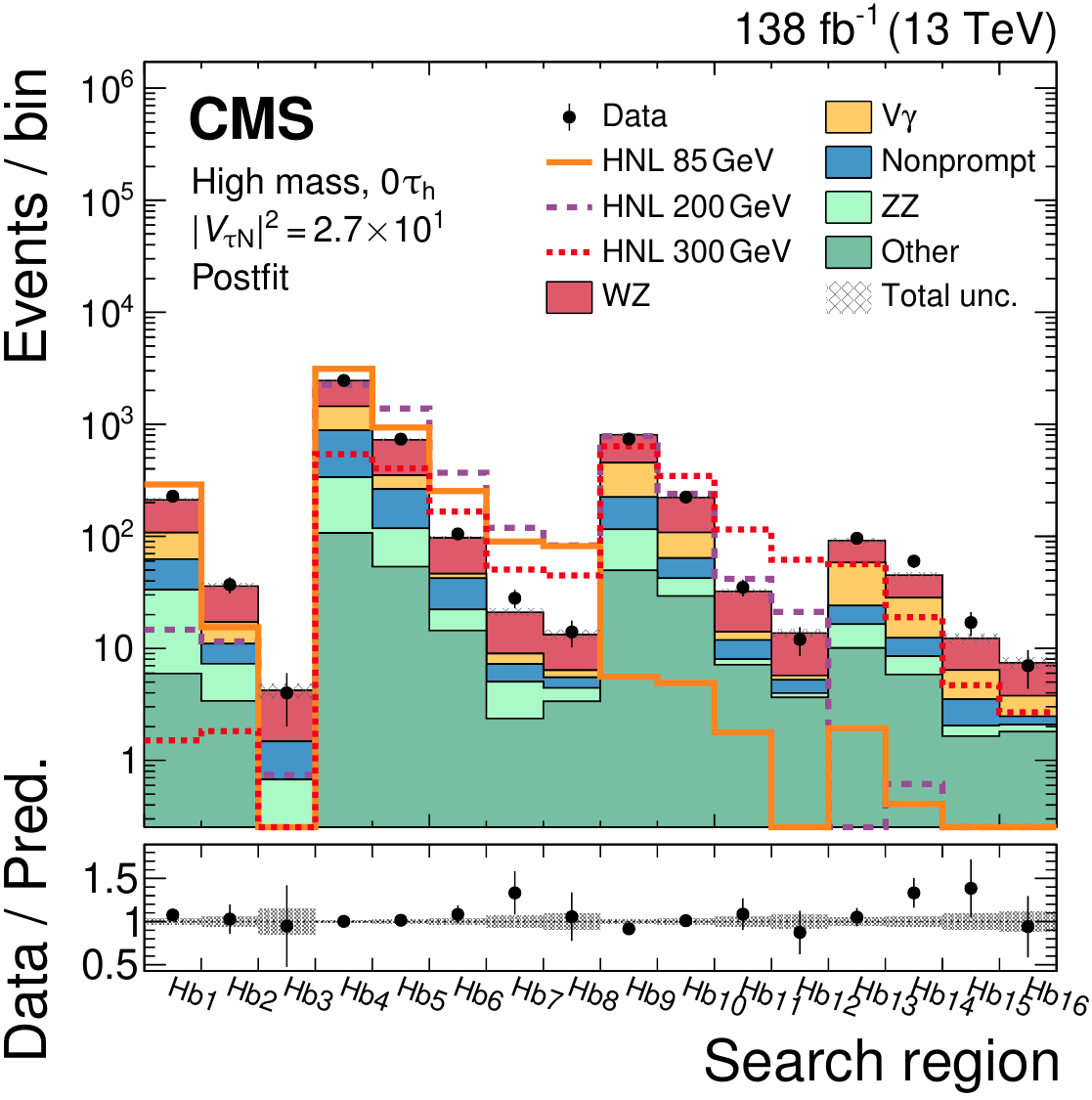}%
\hspace*{0.05\textwidth}%
\includegraphics[width=0.40\textwidth]{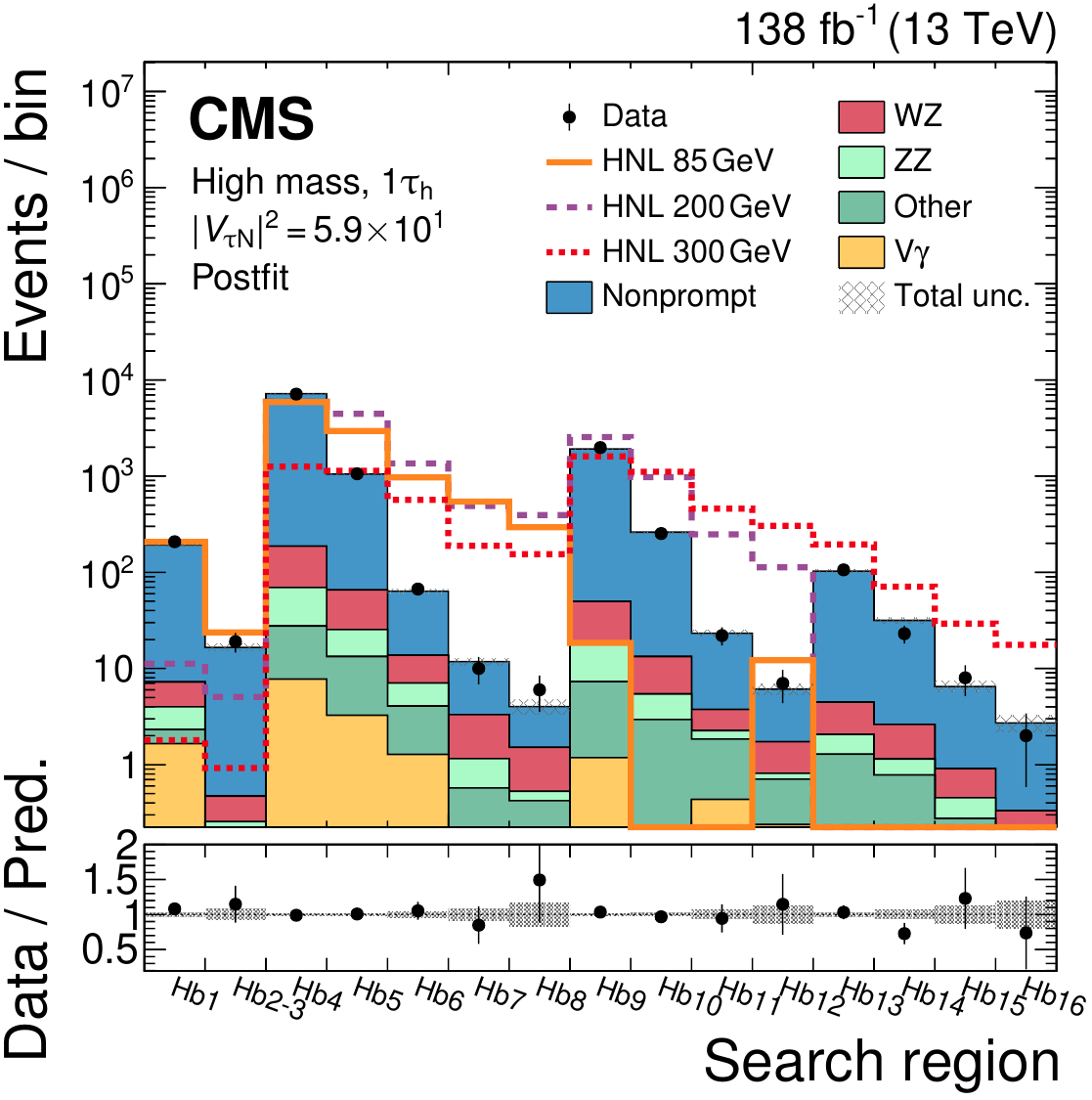}%
\caption{%
    Comparison of the number of observed (points) and predicted (coloured histograms) events in the SR bins, shown for the 0\PGth (left column) and 1\PGth (right column) categories combined.
    The La1--8 and Lb1--8 (upper row), Ha1--Ha9 (middle row), and Hb1--16 (lower row) are displayed.
    The predicted background yields are shown with the values of the normalizations and nuisance parameters obtained in background-only fits applied (``postfit'').
    The HNL predictions for three different \mhnl values with exclusive coupling to tau neutrinos are shown with coloured lines.
    The vertical bars on the points represent the statistical uncertainties in the data, and the hatched bands the total uncertainties in the background predictions as obtained from the fits.
    In the lower panels, the ratios of the event yield in data to the overall sum of the background predictions are shown.
}
\label{fig:sr}
\end{figure}

\begin{figure}[!htbp]
\centering
\includegraphics[width=0.42\textwidth]{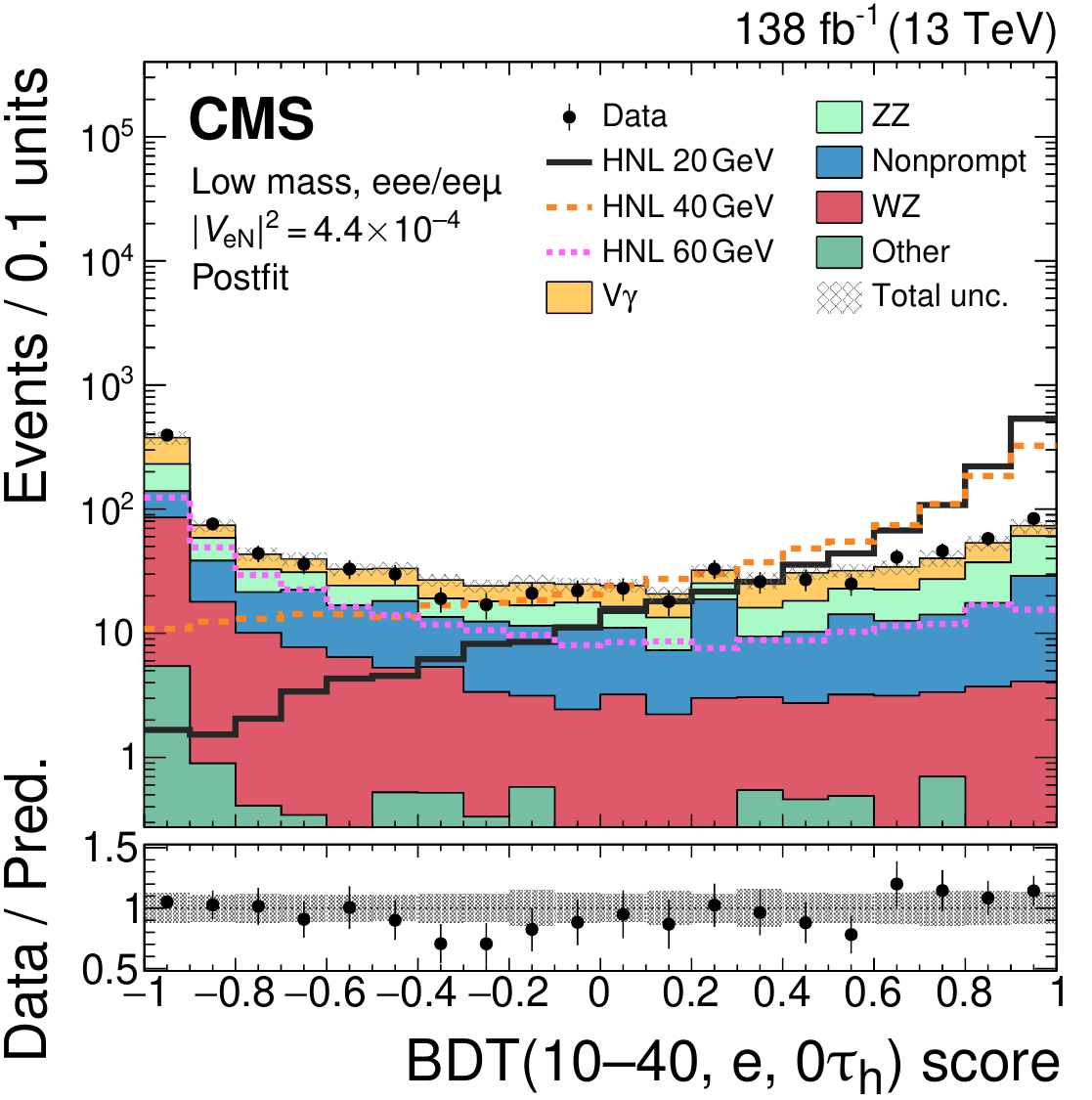}%
\hspace*{0.05\textwidth}%
\includegraphics[width=0.42\textwidth]{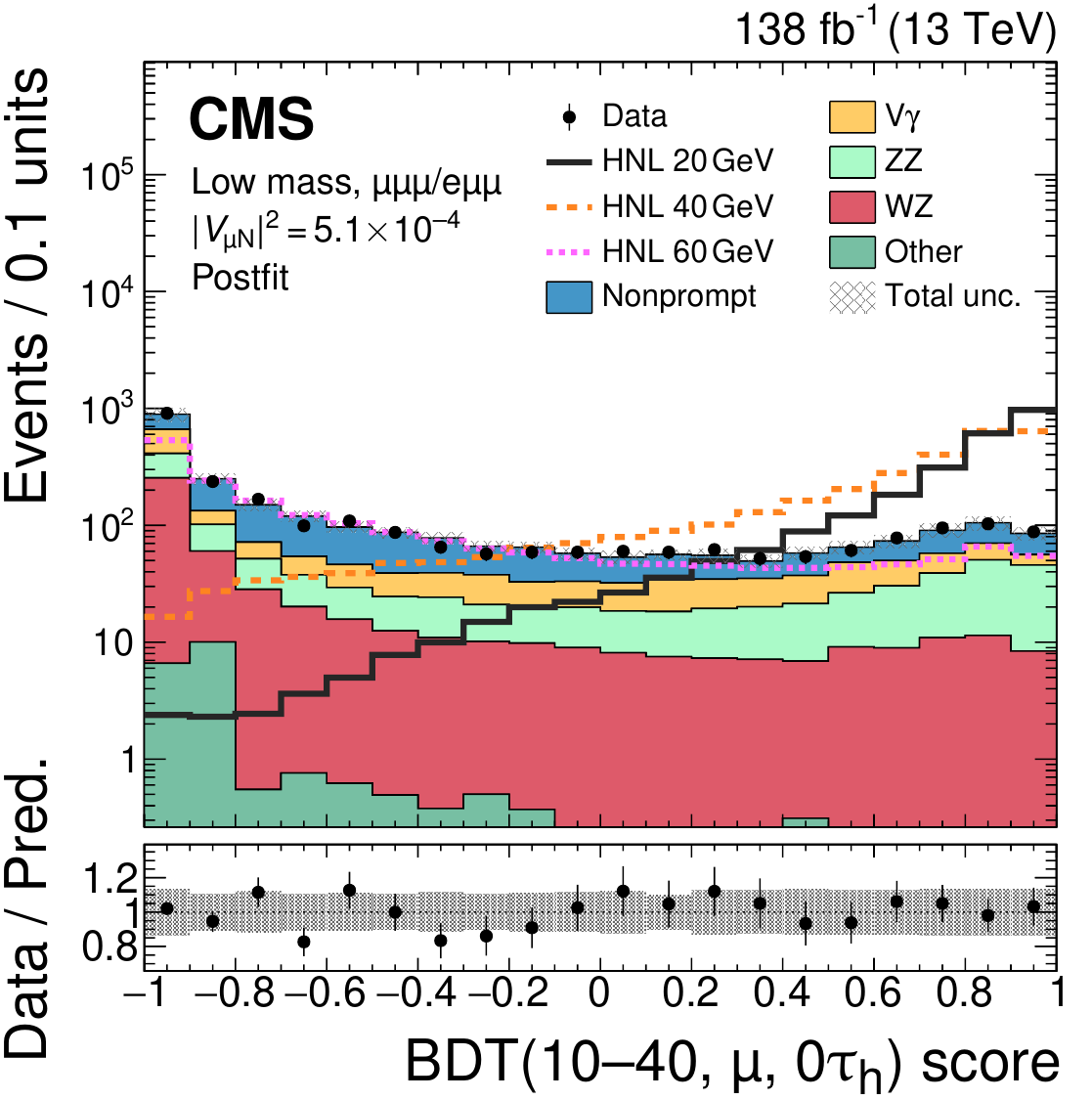} \\[1ex]
\includegraphics[width=0.42\textwidth]{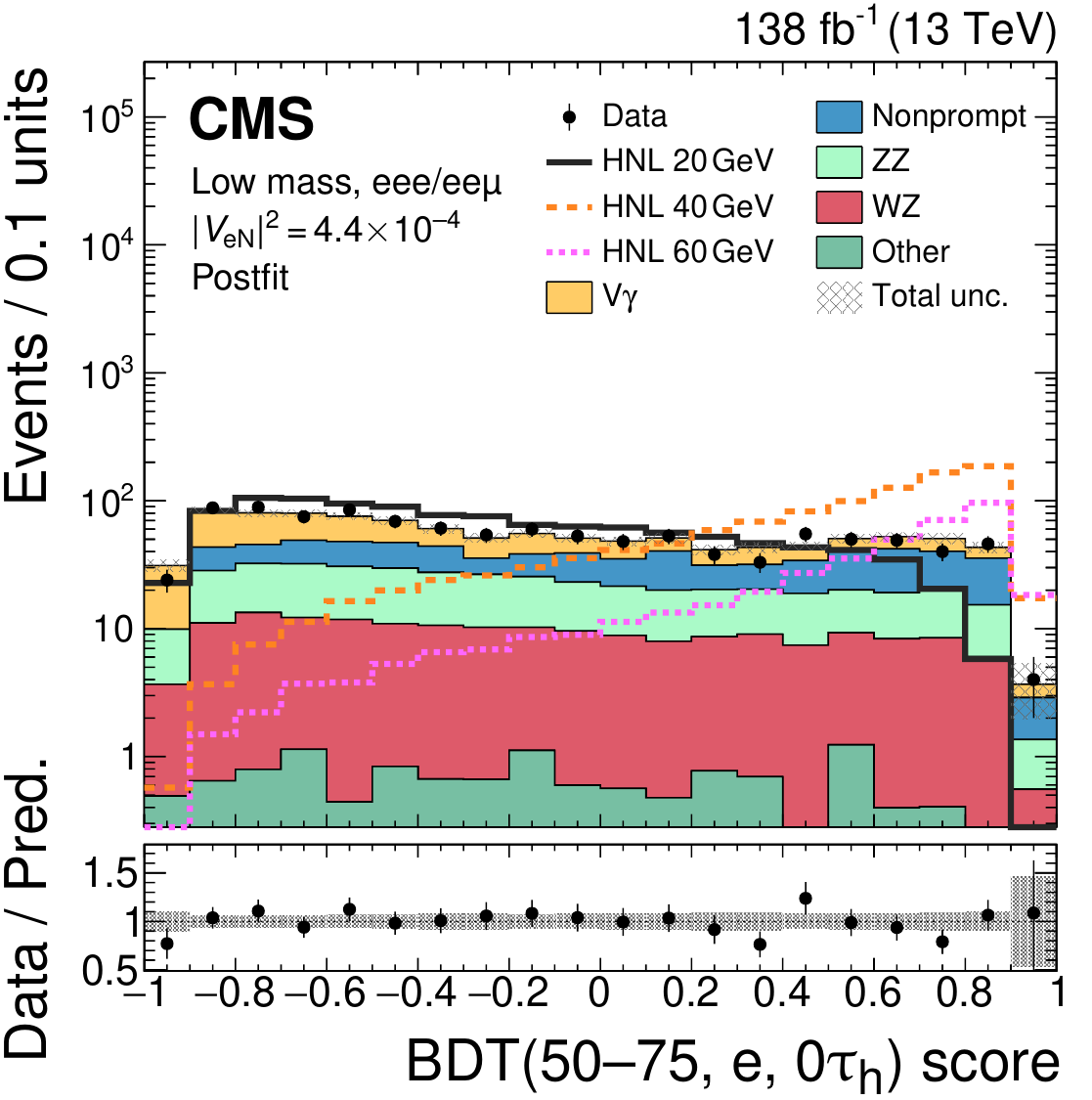}%
\hspace*{0.05\textwidth}%
\includegraphics[width=0.42\textwidth]{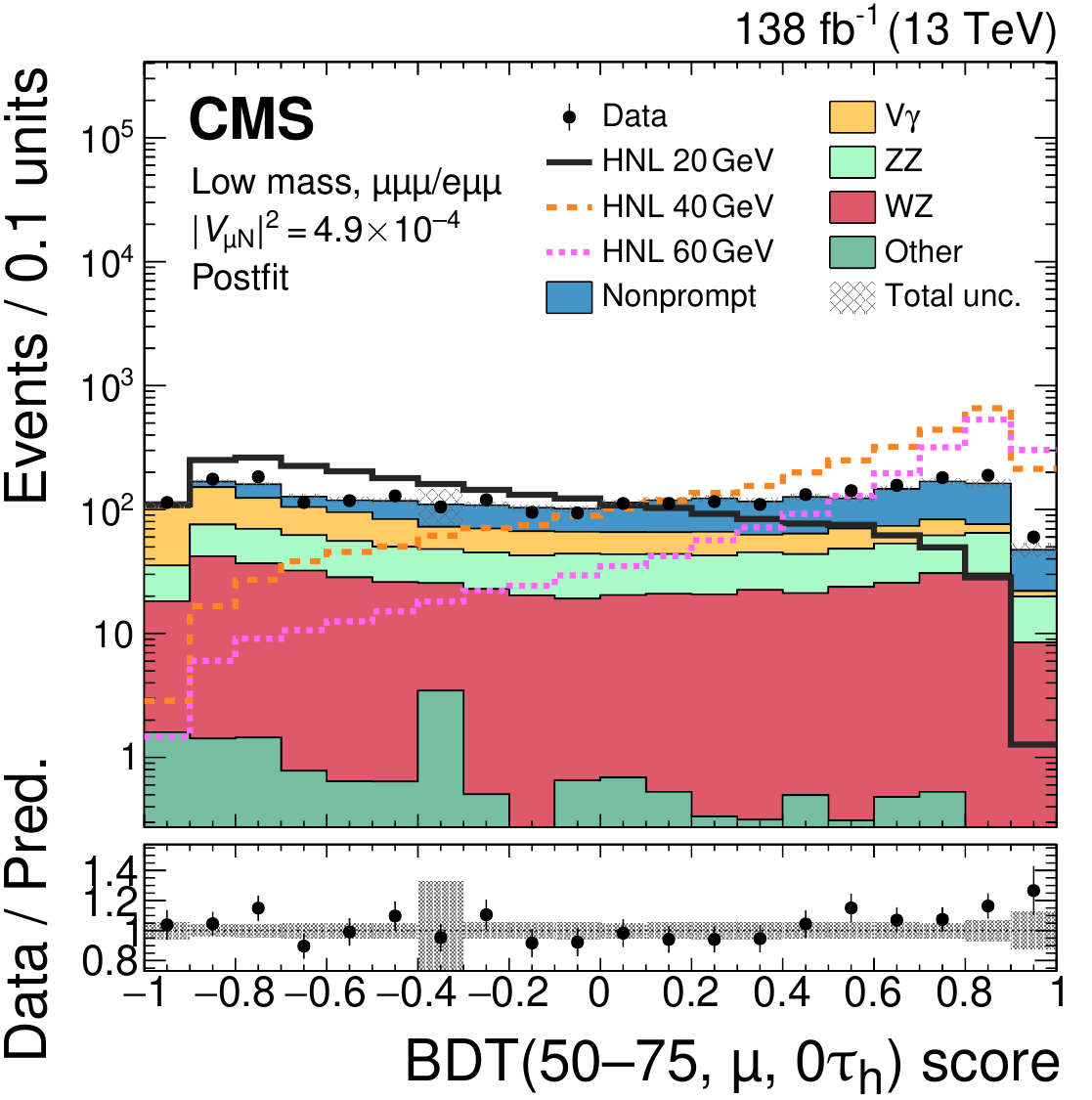}%
\caption{%
    Comparison of the observed (points) and predicted (coloured histograms) BDT output distributions of the low-mass selection, shown for the \EEE and \EEM channels combined (left column) and the \EMM and \MMM channels combined (right column).
    The output scores BDT(10--40, \Pe, 0\PGth) (upper left), BDT(10--40, \PGm, 0\PGth) (upper right), BDT(50--75, \Pe, 0\PGth) (lower left), and BDT(50--75, \PGm, 0\PGth) (lower right) are displayed.
    The predicted background yields are shown with the values of the normalizations and nuisance parameters obtained in background-only fits applied (``postfit'').
    The HNL predictions for three different \mhnl values with exclusive coupling to electron (left column) or muon (right column) neutrinos are shown with coloured lines.
    The vertical bars on the points represent the statistical uncertainties in the data, and the hatched bands the total uncertainties in the background predictions as obtained from the fits.
    In the lower panels, the ratios of the event yield in data to the overall sum of the background predictions are shown.
}
\label{fig:bdt_low_notau}
\end{figure}

\begin{figure}[!htbp]
\centering
\includegraphics[width=0.42\textwidth]{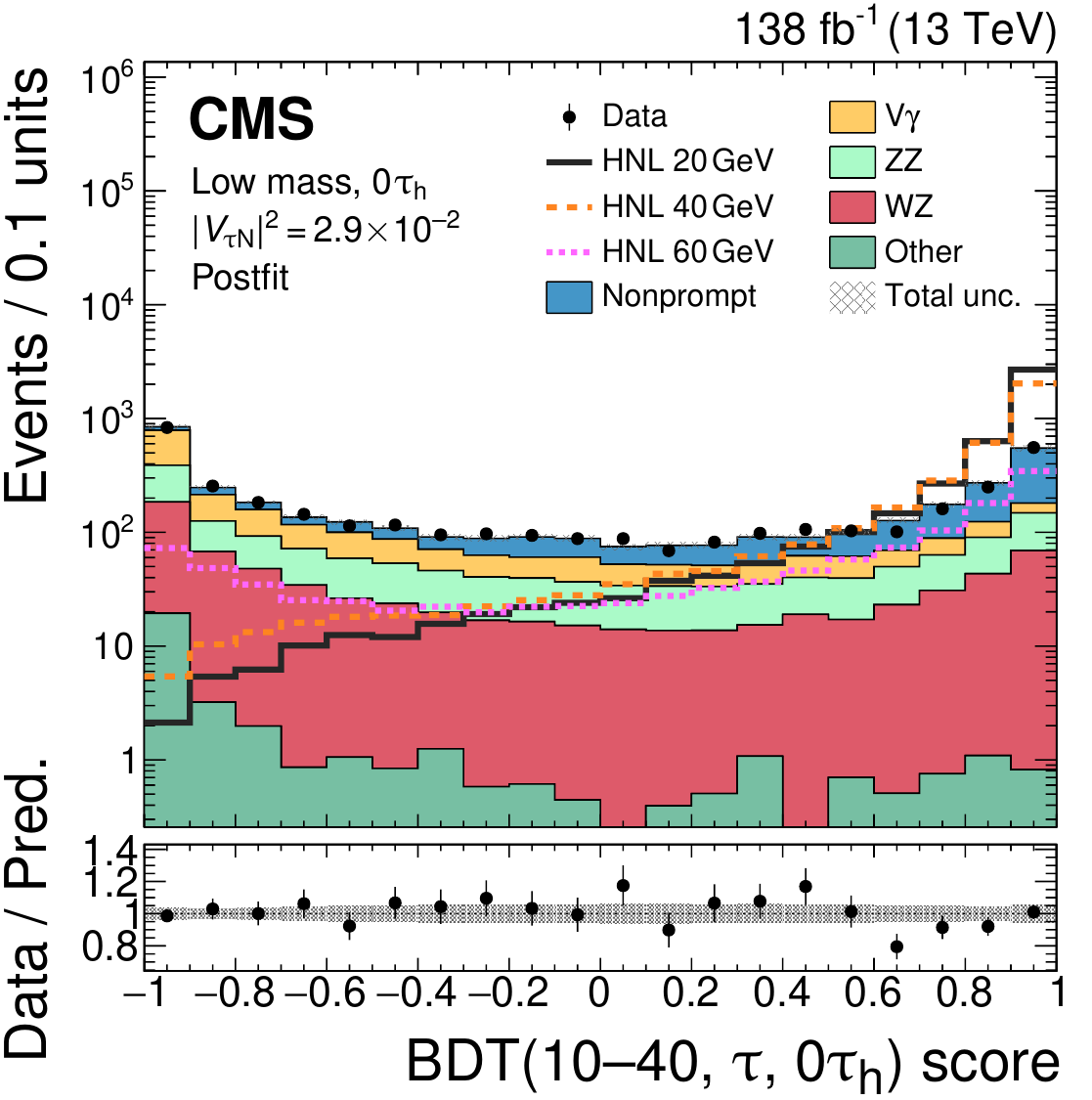}%
\hspace*{0.05\textwidth}%
\includegraphics[width=0.42\textwidth]{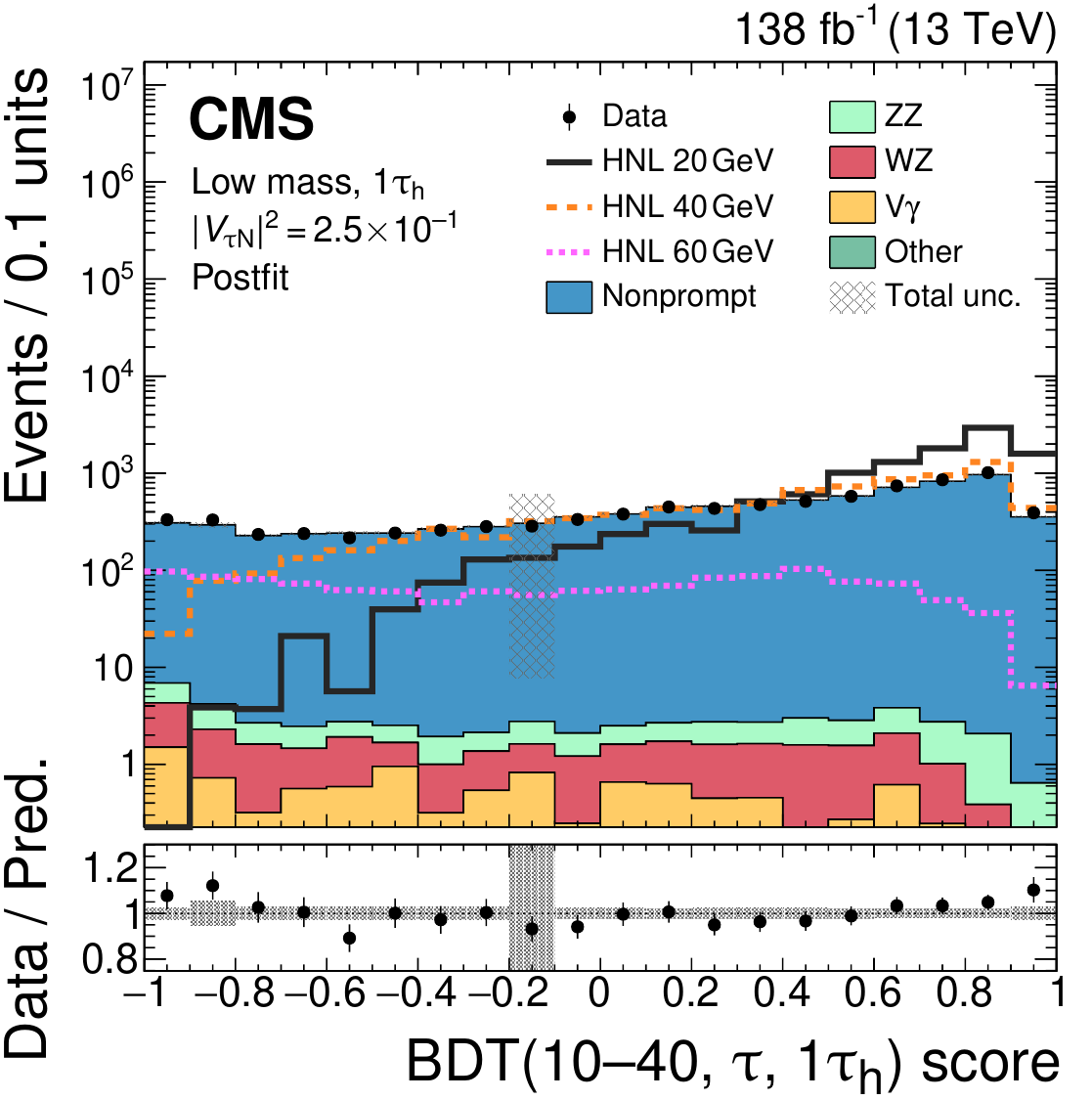} \\[1ex]
\includegraphics[width=0.42\textwidth]{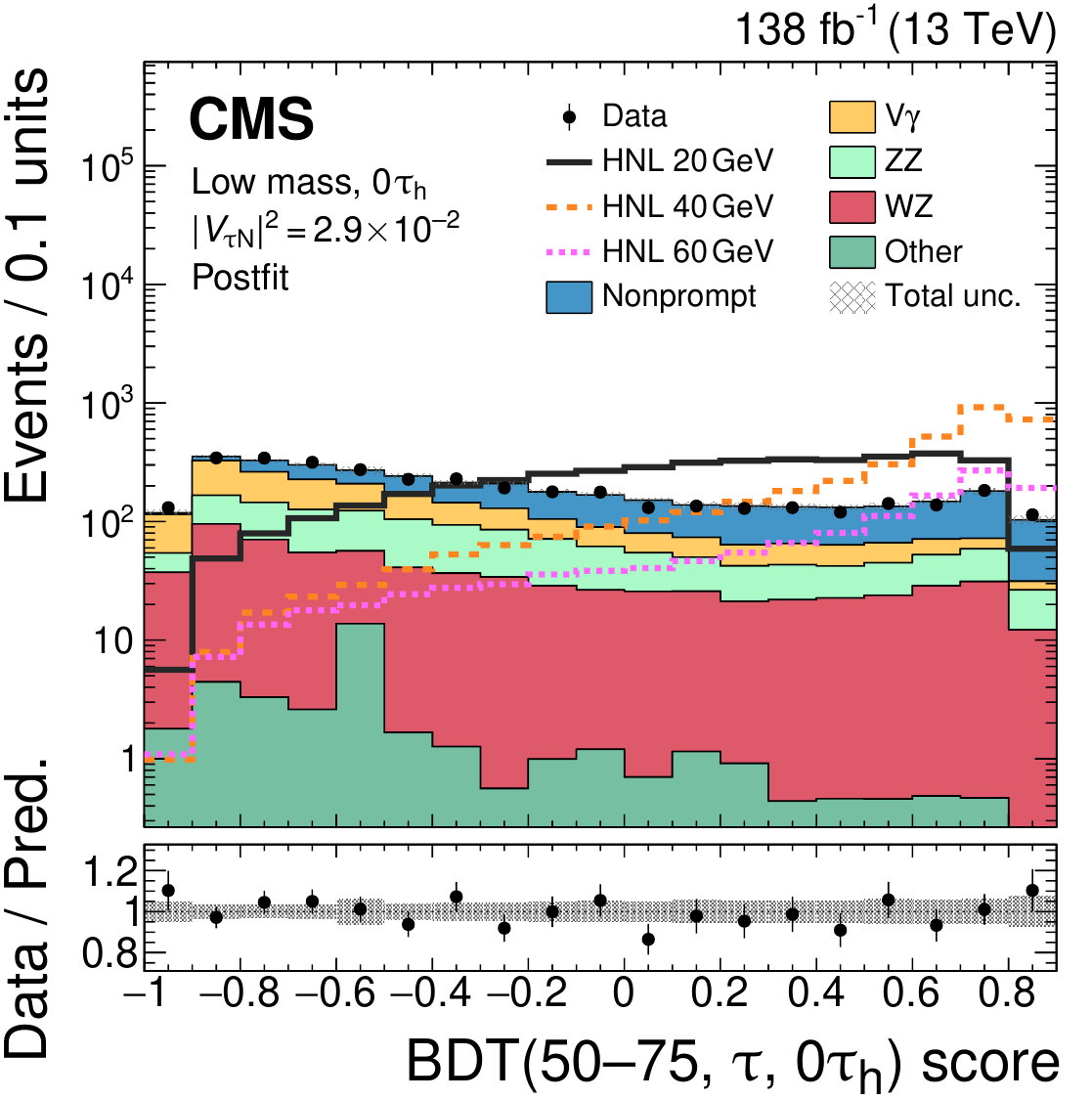}%
\hspace*{0.05\textwidth}%
\includegraphics[width=0.42\textwidth]{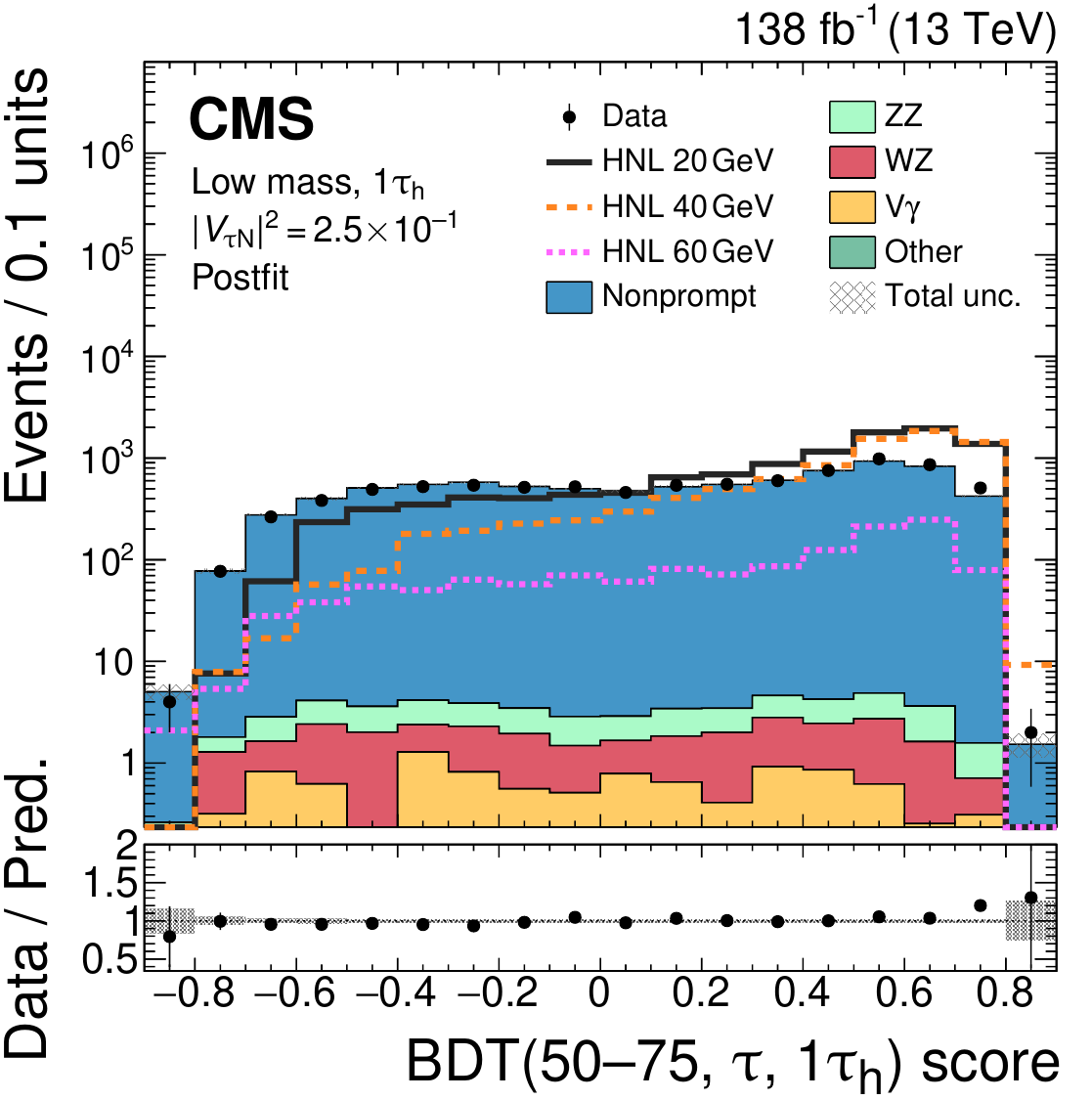}%
\caption{%
    Comparison of the observed (points) and predicted (coloured histograms) BDT output distributions of the low-mass selection, shown for the 0\PGth channels combined (left column) and the 1\PGth channels combined (right column).
    The output scores BDT(10--40, \PGt, 0\PGth) (upper left), BDT(10--40, \PGt, 1\PGth) (upper right), BDT(50--75, \PGt, 0\PGth) (lower left), and BDT(50--75, \PGt, 0\PGth) (lower right) are displayed.
    The predicted background yields are shown with the values of the normalizations and nuisance parameters obtained in background-only fits applied (``postfit'').
    The HNL predictions for three different \mhnl values with exclusive coupling to tau neutrinos are shown with coloured lines.
    The vertical bars on the points represent the statistical uncertainties in the data, and the hatched bands the total uncertainties in the background predictions as obtained from the fits.
    In the lower panels, the ratios of the event yield in data to the overall sum of the background predictions are shown.
}
\label{fig:bdt_low_tau}
\end{figure}

\begin{figure}[!p]
\centering
\includegraphics[width=0.42\textwidth]{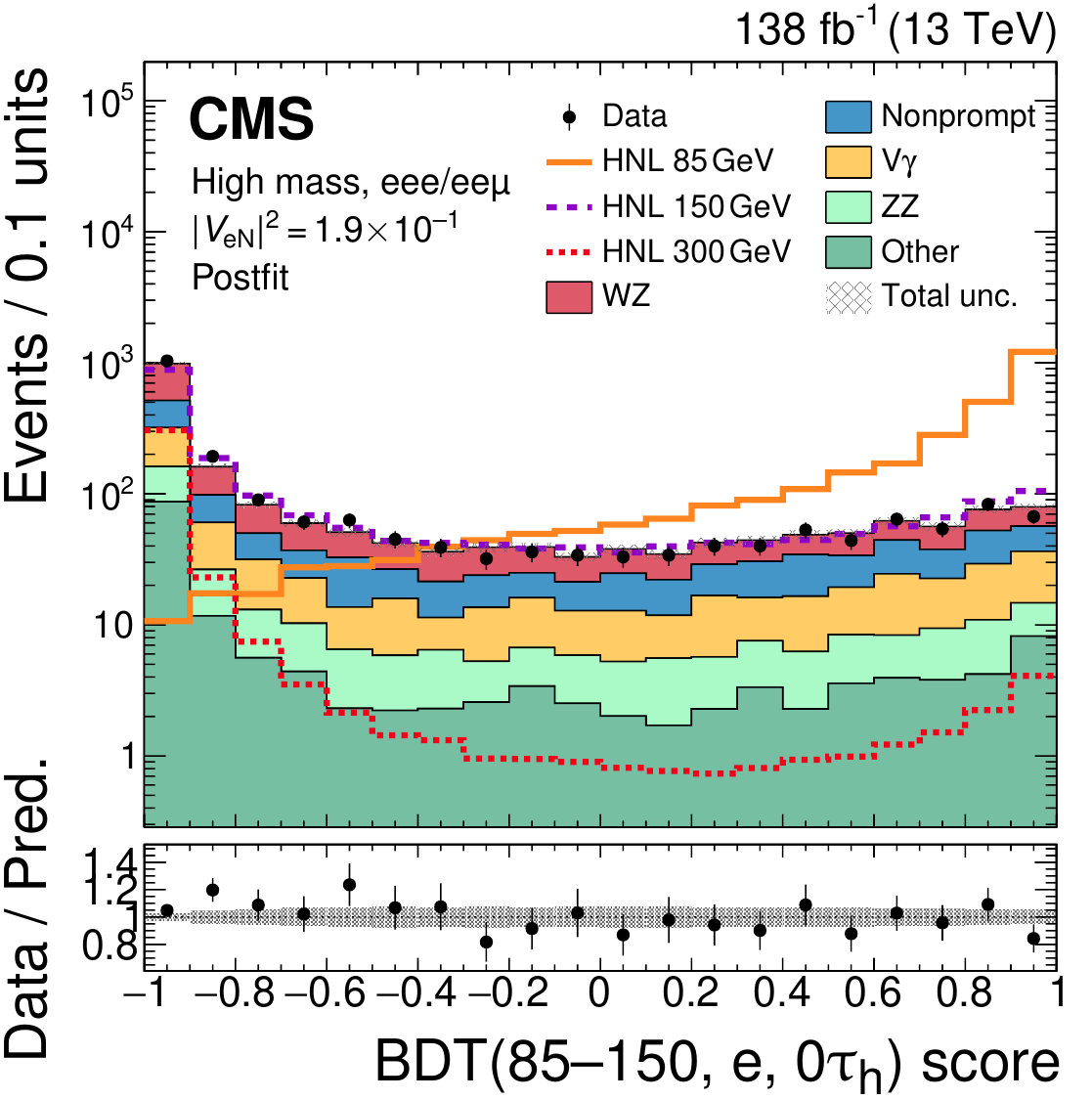}%
\hspace*{0.05\textwidth}%
\includegraphics[width=0.42\textwidth]{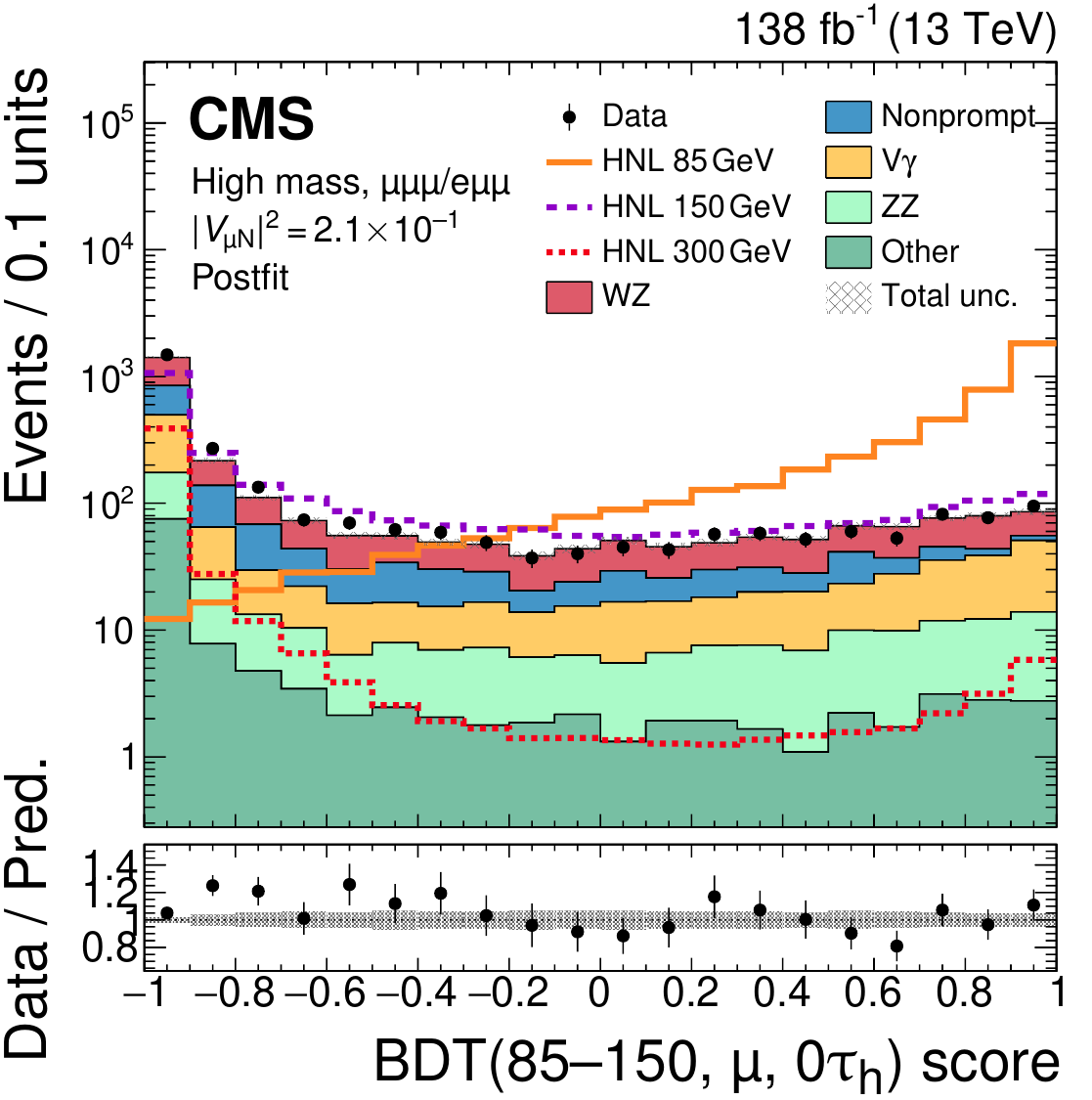} \\[1ex]
\includegraphics[width=0.42\textwidth]{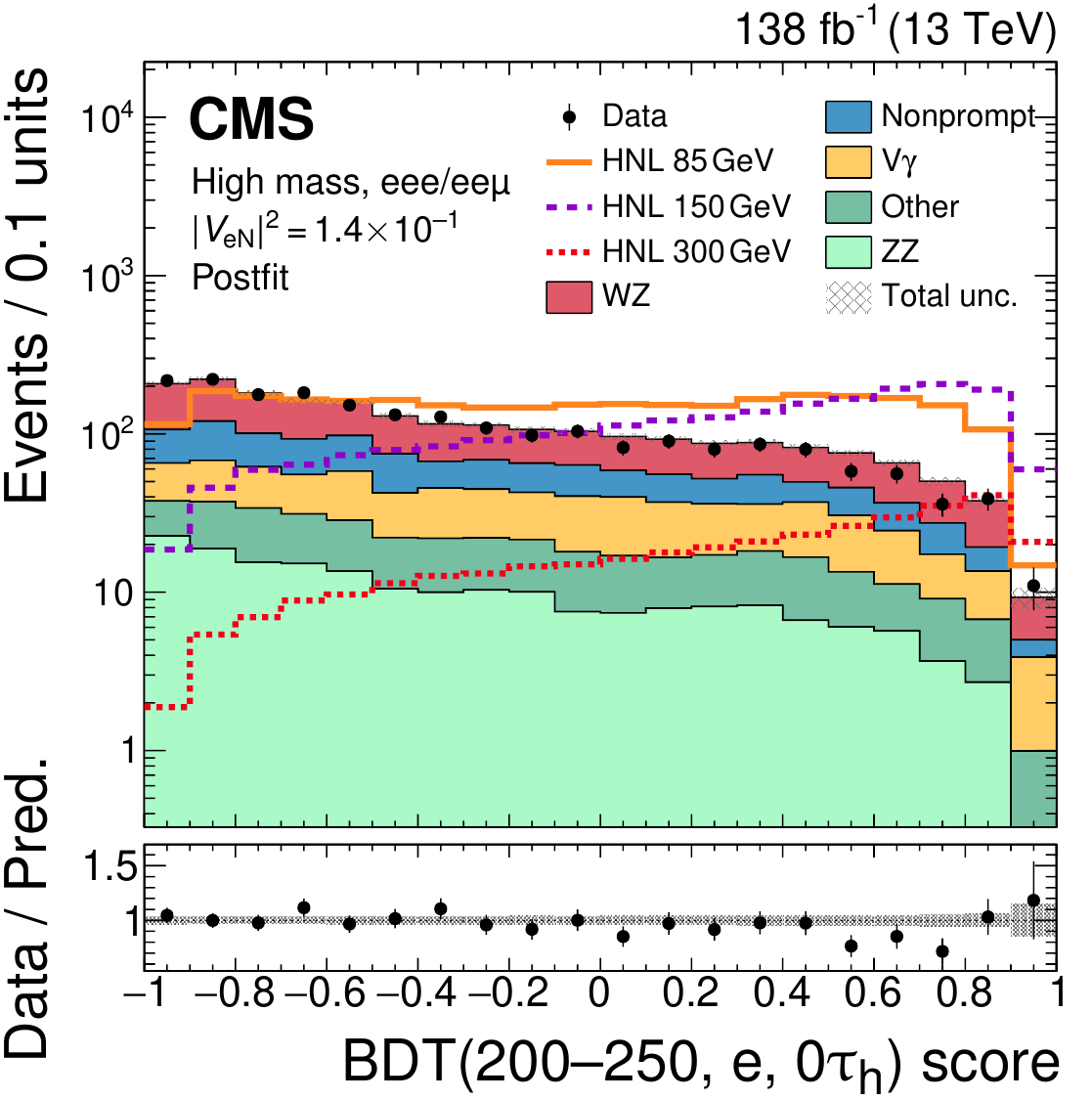}%
\hspace*{0.05\textwidth}%
\includegraphics[width=0.42\textwidth]{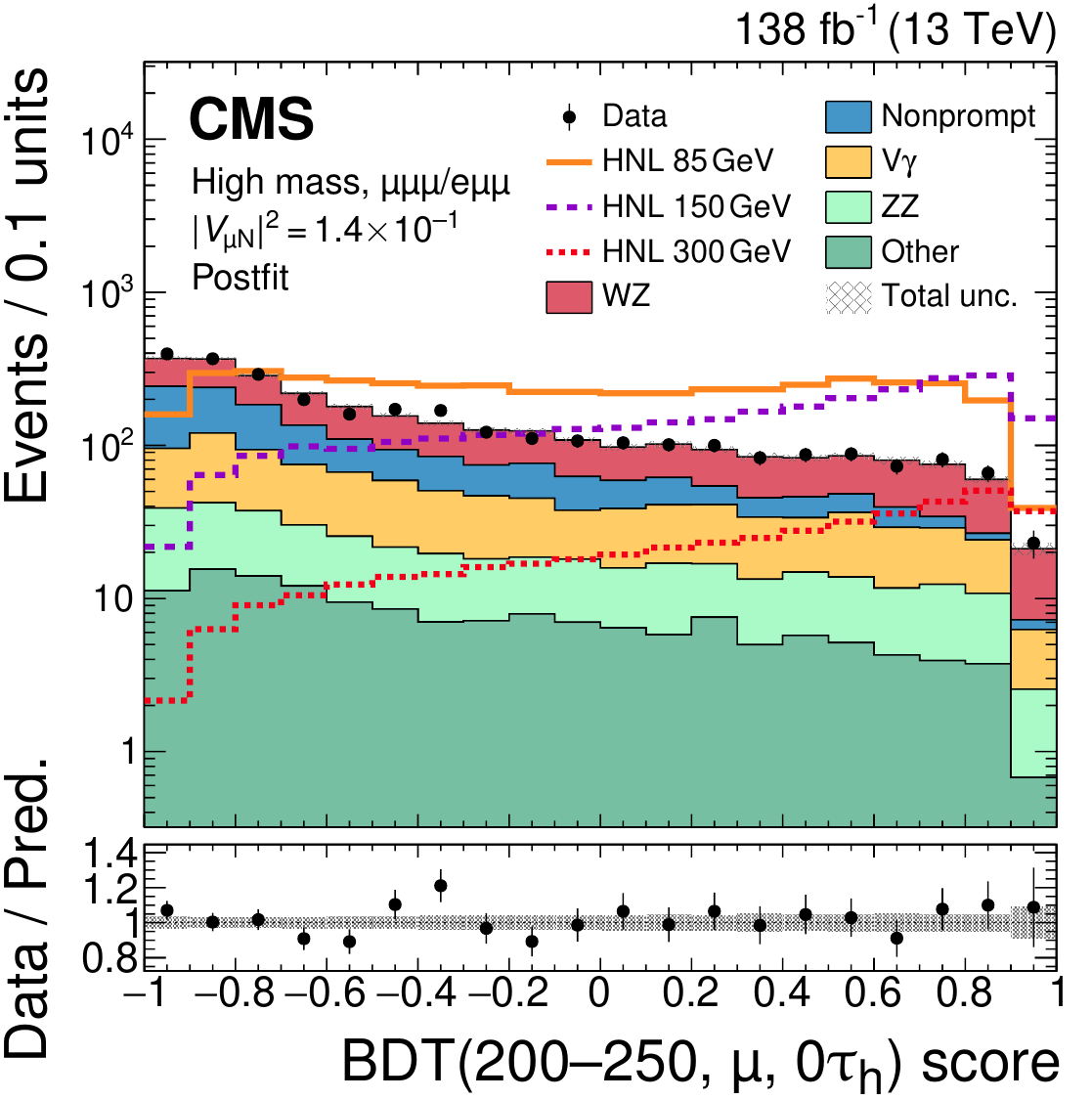} \\[1ex]
\includegraphics[width=0.42\textwidth]{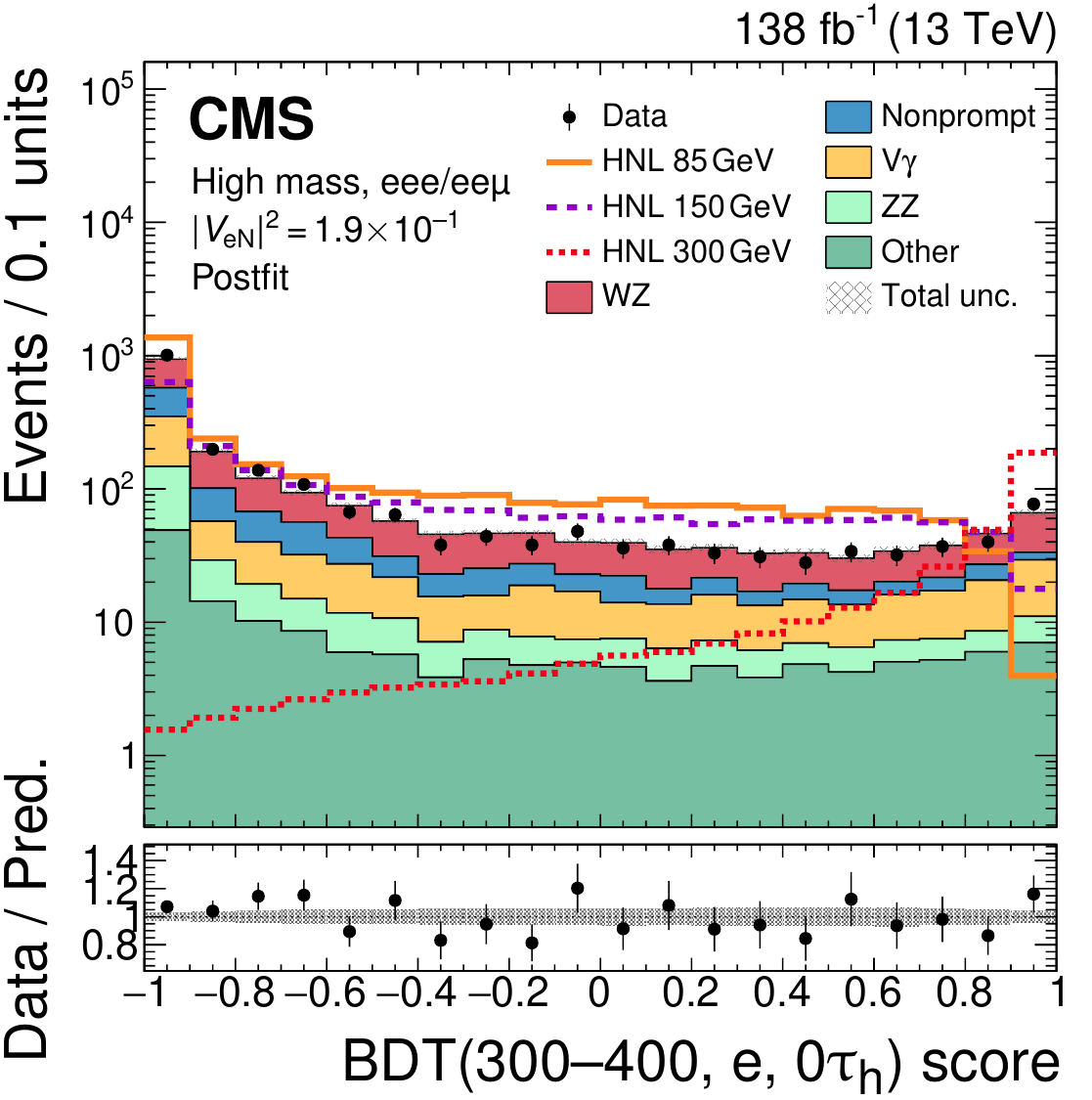}%
\hspace*{0.05\textwidth}%
\includegraphics[width=0.42\textwidth]{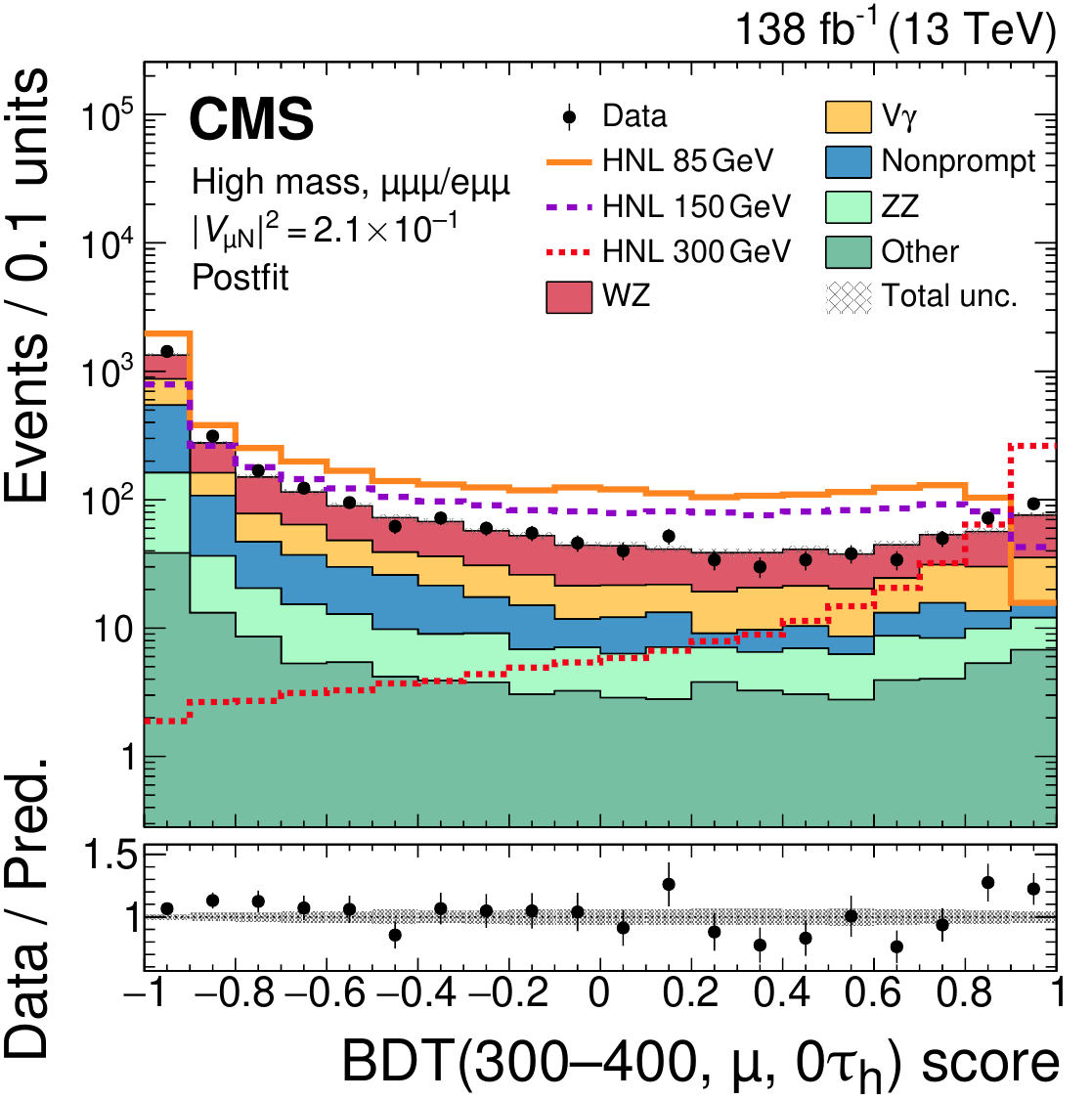}%
\caption{%
    Comparison of the observed (points) and predicted (coloured histograms) BDT output distributions of the high-mass selection, shown for the \EEE and \EEM channels combined (left column) and the \EMM and \MMM channels combined (right column).
    The output scores BDT(85--150, \Pe, 0\PGth) (upper left), BDT(85--150, \PGm, 0\PGth) (upper right), BDT(200--250, \Pe, 0\PGth) (middle left), BDT(200--250, \PGm, 0\PGth) (middle right), BDT(300--400, \Pe, 0\PGth) (lower left), and BDT(300--400, \PGm, 0\PGth) (lower right) are displayed.
    Notations as in Fig.~\ref{fig:bdt_low_notau}.
}
\label{fig:bdt_high}
\end{figure}

The statistical analysis is performed with the CMS tool \textsc{combine}~\cite{CMS:CAT-23-001}, which is based on the \textsc{RooFit}~\cite{Verkerke:2003ir} and \textsc{RooStats}~\cite{Moneta:2010pm} frameworks.
For each HNL signal scenario, a binned likelihood function $\likeli(\sigstr,\nuisan)$ is constructed from the product of Poisson probabilities to obtain the observed yields in the relevant distributions, given the HNL signal prediction scaled with a signal strength \sigstr and the SM background estimates.
Additional terms are included to account for the systematic uncertainty sources, where \nuisan denotes the full set of corresponding nuisance parameters~\cite{CMS:NOTE-2011-005}.
Statistical uncertainties in the predicted yields are implemented through a single nuisance parameter in each bin for all processes~\cite{Barlow:1993dm, Conway:2011in}.
We consider the cases of exclusive HNL couplings to electron, muon, and tau neutrinos separately, and use different distributions to construct \likeli depending on the coupling scenario and \mhnl, as listed in Table~\ref{tab:fitstrategy}.
To obtain background-only fits, the maximum likelihood estimator of \nuisan with a fixed $\sigstr=0$ is evaluated for specific fit setups.
In Fig.~\ref{fig:sr}, the number of observed events in data is compared with the background predictions in the SRs, separately combined for the 0\PGth and 1\PGth categories, after simultaneous background-only fits to the SRs of all flavour channels.
Furthermore, the distributions of the BDT output scores are compared between data and prediction in Figs.~\ref{fig:bdt_low_notau}--\ref{fig:bdt_high}, after the background-only fits corresponding to the fit setups in Table~\ref{tab:fitstrategy} where the corresponding BDT is used.
In all figures, the signal prediction for several HNL mass points is shown as well, with \Vhnlsq values chosen such that the predicted signal yield matches roughly the total background yield.

\begin{table}[!p]
\centering\renewcommand\arraystretch{1.1}
\topcaption{%
    Summary of the selections, categories, and distributions used in the maximum likelihood fits for the HNL signal points.
}
\begin{tabular}{>{(}r@{, }l<{)}cccl}
    \multicolumn{2}{r}{HNL model} & Selection & Categories & OSSF & Fitted distributions \\ \hline
    10--40\GeV & \Pe & low mass & 0\PGth & no & La1--8 \\
    \multicolumn{2}{c}{} & & 0\PGth & yes & BDT(10--40, \Pe, 0\PGth) \\[\cmsTabSkip]
    40--75\GeV & \Pe & low mass & 0\PGth & no & La1--8 \\
    \multicolumn{2}{c}{} & & 0\PGth & yes & BDT(50--75, \Pe, 0\PGth) \\[\cmsTabSkip]
    85--125\GeV & \Pe & high mass & 0\PGth & any & BDT(85--150, \Pe, 0\PGth) \\[\cmsTabSkip]
    125--250\GeV & \Pe & high mass & 0\PGth & any & BDT(200--250, \Pe, 0\PGth) \\[\cmsTabSkip]
    250--400\GeV & \Pe & high mass & 0\PGth & any & BDT(300--400, \Pe, 0\PGth) \\[\cmsTabSkip]
    $\geq$400\GeV & \Pe & high mass & 0\PGth & any & Ha1--9, Hb1--16  \\[2\cmsTabSkip]
    10--40\GeV & \PGm & low mass & 0\PGth & no & La1--8 \\
    \multicolumn{2}{c}{} & & 0\PGth & yes & BDT(10--40, \PGm, 0\PGth) \\[\cmsTabSkip]
    40--75\GeV & \PGm & low mass & 0\PGth & no & La1--8 \\
    \multicolumn{2}{c}{} & & 0\PGth & yes & BDT(50--75, \PGm, 0\PGth) \\[\cmsTabSkip]
    85--125\GeV & \PGm & high mass & 0\PGth & any & BDT(85--150, \PGm, 0\PGth) \\[\cmsTabSkip]
    125--200\GeV & \PGm & high mass & 0\PGth & any & BDT(200--250, \PGm, 0\PGth) \\[\cmsTabSkip]
    200--400\GeV & \PGm & high mass & 0\PGth & any & BDT(300--400, \PGm, 0\PGth) \\[\cmsTabSkip]
    $\geq$400\GeV & \PGm & high mass & 0\PGth & any & Ha1--9, Hb1--16 \\[2\cmsTabSkip]
    10--40\GeV & \PGt & low mass & 0\PGth & no & La1--8 \\
    \multicolumn{2}{c}{} & & 0\PGth & yes & BDT(10--40, \PGt, 0\PGth) \\
    \multicolumn{2}{c}{} & & 1\PGth & no & La1--8 \\
    \multicolumn{2}{c}{} & & 1\PGth & yes & BDT(10--40, \PGt, 1\PGth) \\[\cmsTabSkip]
    40--75\GeV & \PGt & low mass & 0\PGth & no & La1--8 \\
    \multicolumn{2}{c}{} & & 0\PGth & yes & BDT(50--75, \PGt, 0\PGth) \\
    \multicolumn{2}{c}{} & & 1\PGth & no & La1--8 \\
    \multicolumn{2}{c}{} & & 1\PGth & yes & BDT(50--75, \PGt, 1\PGth) \\[\cmsTabSkip]
    $\geq$85\GeV & \PGt & high mass & all & any & Ha1--9, Hb1--16 \\
\end{tabular}
\label{tab:fitstrategy}
\end{table}

The number of observed events in data is in good agreement with the SM background expectations within the statistical and systematic uncertainties.
No significant excess is found for any final state or in any SR.

\section{Interpretation}
\label{sec:interpretation}

To derive exclusion limits at 95\% confidence level (\CL) on HNL signal scenarios, we apply the modified frequentist \CLs approach~\cite{Junk:1999kv, Read:2002hq, CMS:NOTE-2011-005, Cowan:2010js}.
Distributions of the LHC test statistic~\cite{CMS:NOTE-2011-005}, based on the profile likelihood method, are evaluated in the asymptotic approximation~\cite{Cowan:2010js} and used to calculate the \CLs value~\cite{Junk:1999kv, Read:2002hq}.
We exclude a signal scenario if the signal strength of $\sigstr=1$ is excluded at 95\% \CL or greater.

\begin{figure}[!p]
\centering
\includegraphics[width=0.47\textwidth]{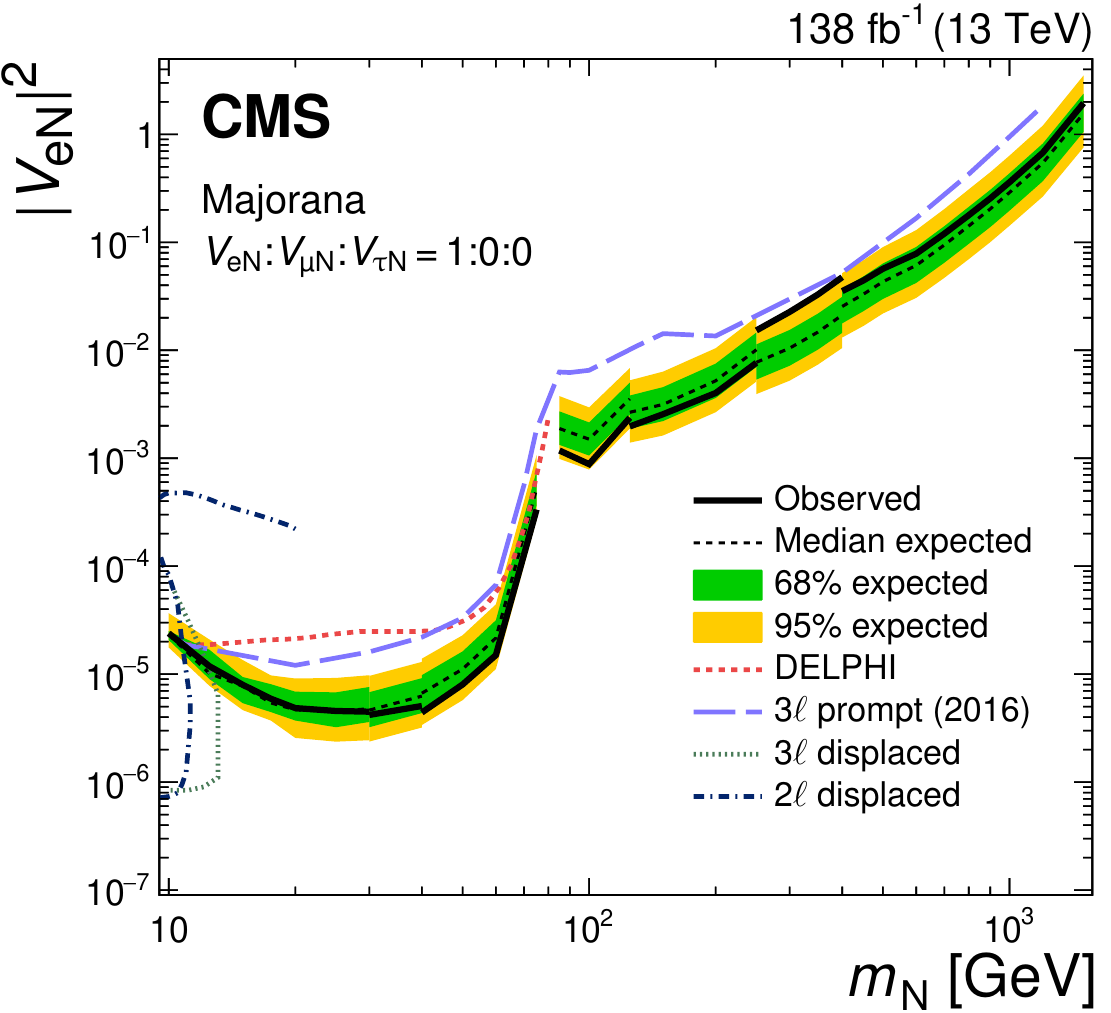}%
\hspace*{0.05\textwidth}%
\includegraphics[width=0.47\textwidth]{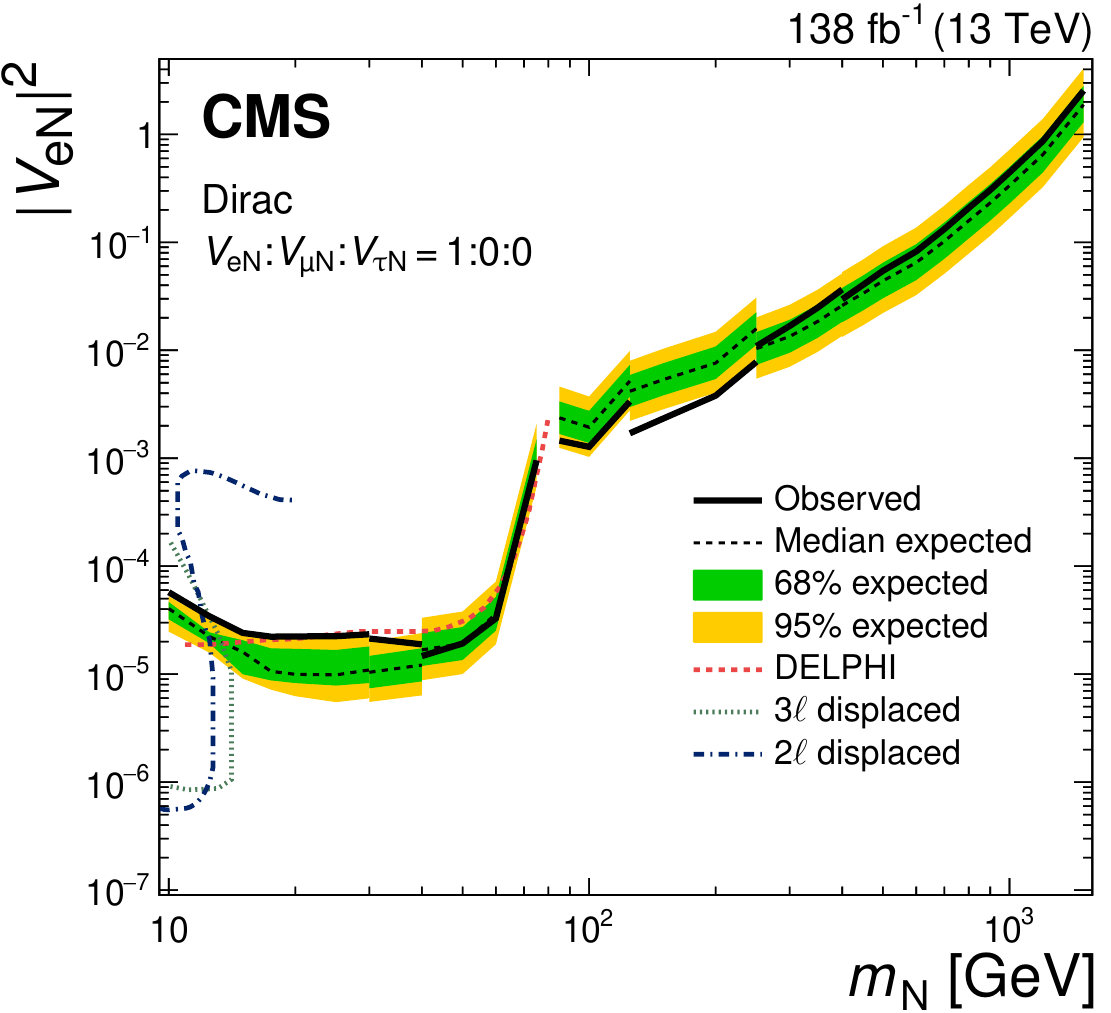} \\[1ex]
\includegraphics[width=0.47\textwidth]{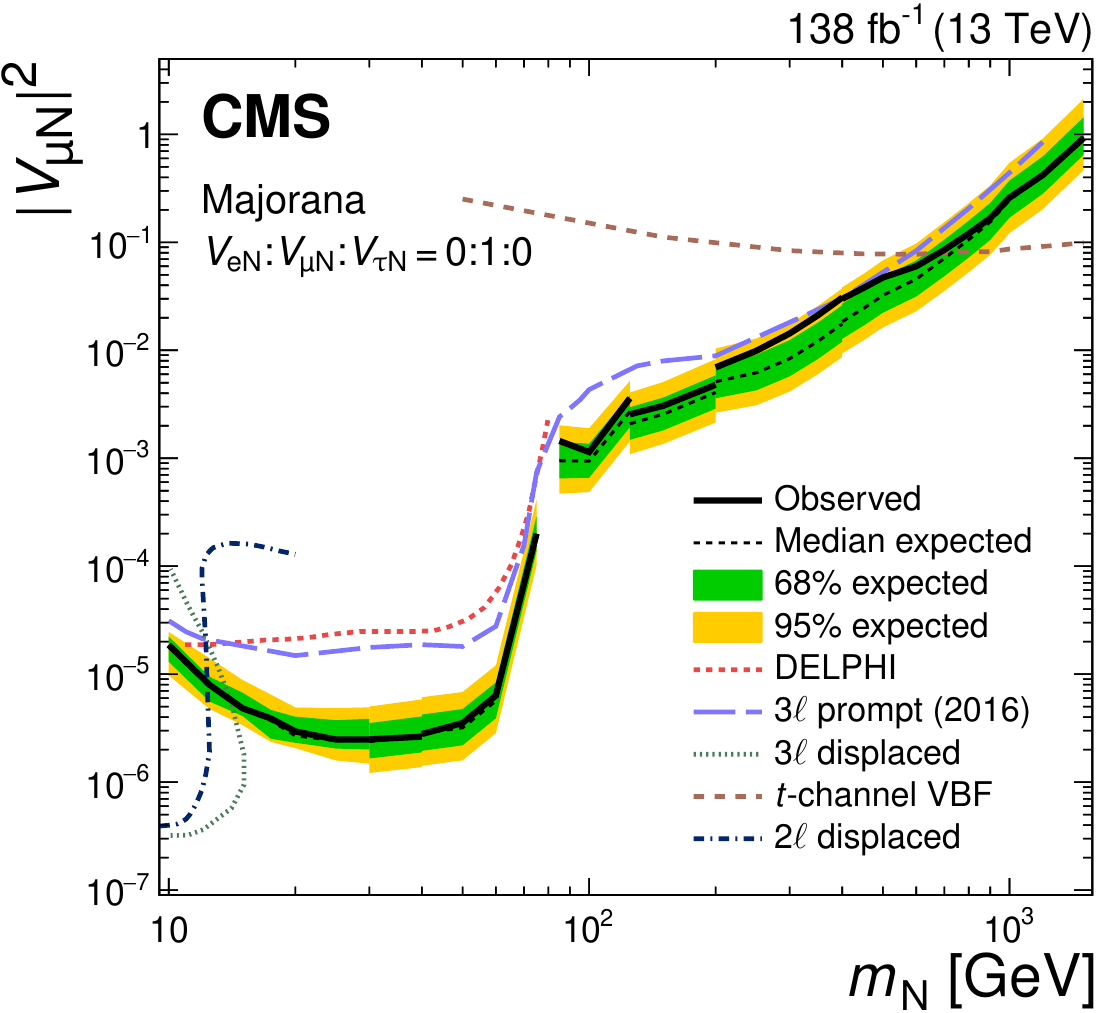}%
\hspace*{0.05\textwidth}%
\includegraphics[width=0.47\textwidth]{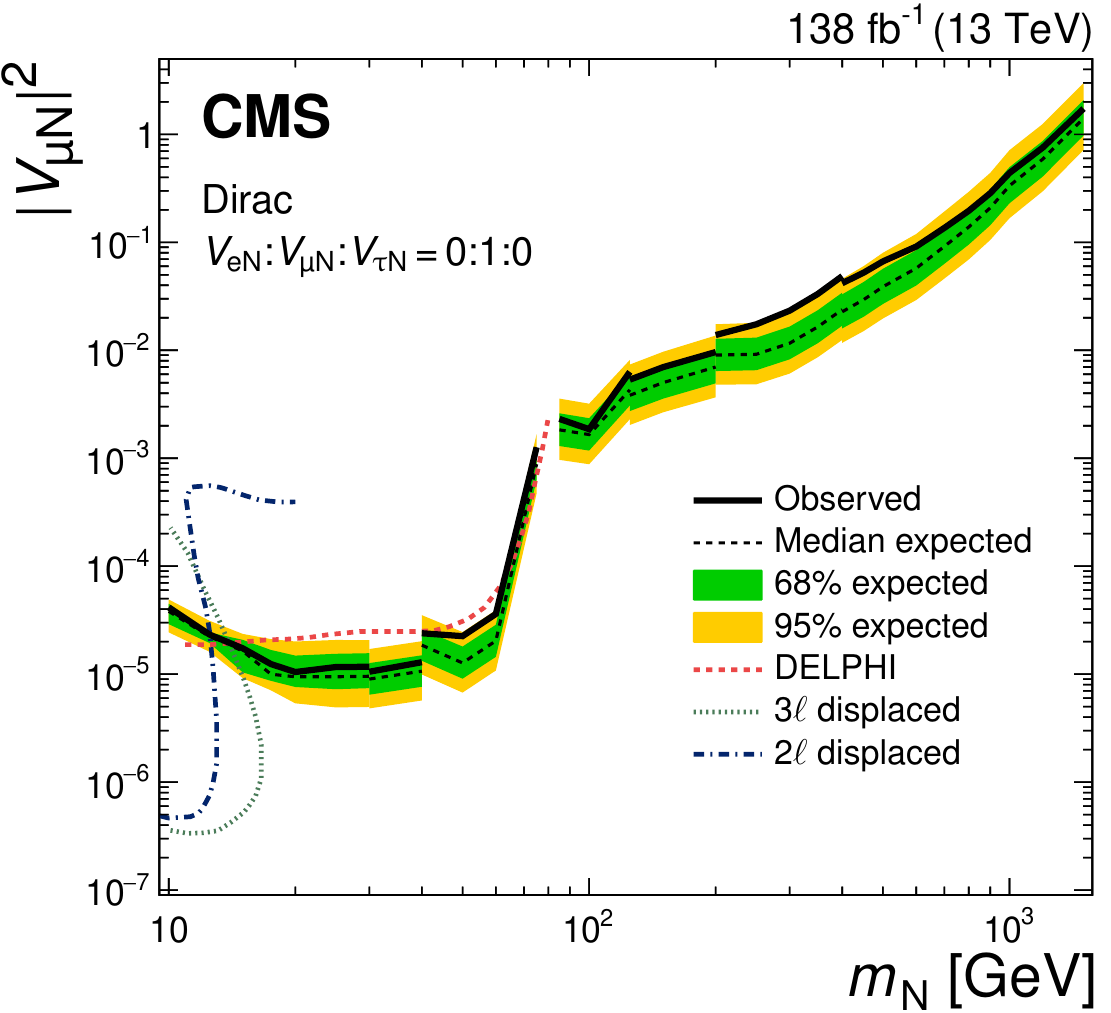} \\[1ex]
\includegraphics[width=0.47\textwidth]{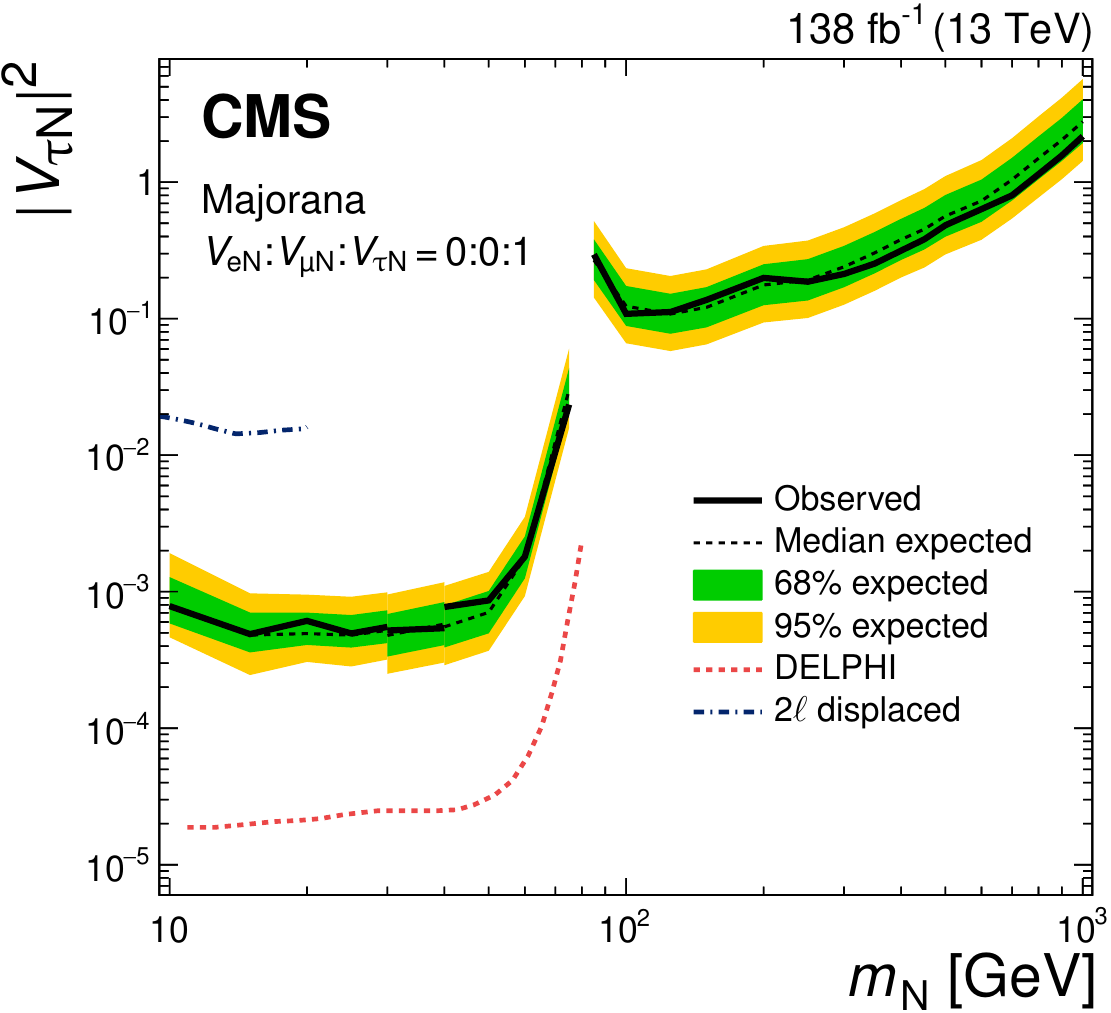}%
\hspace*{0.05\textwidth}%
\includegraphics[width=0.47\textwidth]{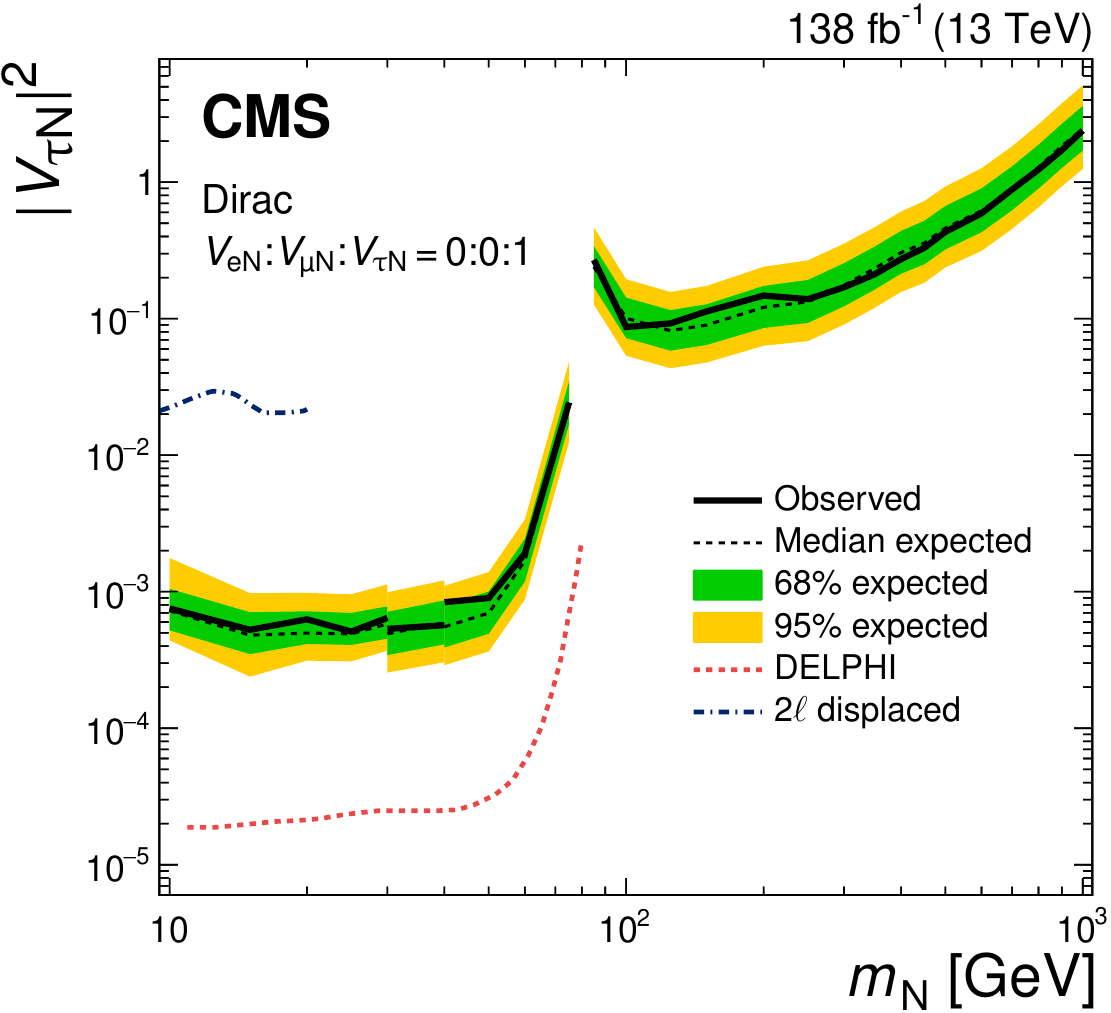}%
\caption{%
    The 95\% \CL limits on \Vhnlesq (upper row), \Vhnlmsq (middle row), and \Vhnltsq (lower row) as functions of \mhnl for a Majorana (left) and Dirac (right) HNL.
    The area above the solid (dashed) black curve indicates the observed (expected) exclusion region.
    Previous results from the DELPHI Collaboration~\cite{DELPHI:1996qcc} are shown for reference.
    The previous CMS result ``3\Pell prompt (2016)''~\cite{CMS:EXO-17-012} is shown to highlight the improvements achieved in our analysis, and the results ``3\Pell displaced''~\cite{CMS:EXO-20-009}, ``2\Pell displaced''~\cite{CMS:EXO-21-013}, and ``$t$-channel VBF''~\cite{CMS:EXO-21-003} are shown to highlight the complementarity to other search strategies.
}
\label{fig:limits}
\end{figure}

The limits are obtained under the assumption of a Majorana or a Dirac HNL, and are evaluated at a grid of points in the $\big(\mhnl,\Vhnlsq\big)$ parameter space.
For electron and muon neutrino couplings, masses of up to 1.5\TeV are considered, whereas tau neutrino couplings are evaluated only up to 1\TeV since the exclusion limit passes above $\Vhnltsq=1$ already at this mass point.
The results are shown in Fig.~\ref{fig:limits}.
The obtained limits are connected with straight lines between neighbouring mass points for which the same fit distributions are used.
For \mhnl values at which the fit distributions change, the limits are evaluated for both strategies and shown separately.
Since the BDTs are trained with nonoverlapping mass ranges, the sensitivity is generally different for two strategies evaluated at the same \mhnl value, and we thus obtain disjoint limit curves for several of these mass points.
The expected and observed exclusion limits generally agree within one standard deviation, with a few exceptions discussed in the following.

For exclusive couplings to electron neutrinos and $\mhnl<\mW$, we exclude \Vhnlesq values for Majorana (Dirac) HNL of 4.8\ten{-6} (2.23\ten{-5}) at a mass of 20\GeV, of 1.5\ten{-5} (3.3\ten{-5}) at 60\GeV, and of 3.3\ten{-4} (9.6\ten{-4}) at 75\GeV, which is the highest simulated mass point below \mW.
For masses below 30\GeV, the HNLs become long-lived and have a reduced selection efficiency, with, \eg, 16 times fewer events selected at $\mhnl=10\GeV$ and $\Vhnlesq=10^{-5}$ compared with a prompt HNL of the same mass, resulting in less stringent limits.
In the high-mass selection, we exclude \Vhnlesq values for Majorana (Dirac) HNL of 1.2\ten{-3} (1.5\ten{-3}) at 85\GeV, which is the lightest simulated mass point above \mW, of 8.8\ten{-4} (1.3\ten{-3}) at 100\GeV, and of 7.8\ten{-2} (8.2\ten{-2}) at 600\GeV.
The most notable discrepancies between observed and expected limits are for HNL masses between 250 and 400\GeV (125 and 250\GeV) in the case of a Majorana (Dirac) HNL, caused by a small excess (deficit) in the last bin of the corresponding BDT score distribution.
Similarly a discrepancy is observed at HNL masses below 40\GeV in case of a Dirac HNL.
This is due to a small excess in the relevant BDT distribution, which leads the sensitivity in the Dirac requirement of an OSSF electron pair in the final selection.
Compared with the results of the previous prompt HNL search presented in Ref.~\cite{CMS:EXO-17-012} for Majorana HNL, the limits improve by up to one order of magnitude.
For masses below 20\GeV, we exclude short-lived HNL scenarios not excluded by the displaced HNL searches presented in Refs.~\cite{CMS:EXO-20-009, CMS:EXO-21-013}.
The exclusion limits obtained by the DELPHI Collaboration~\cite{DELPHI:1996qcc} for $\mhnl<\mW$ are less stringent (similar) compared with our results for the case of Majorana (Dirac) HNL.

Using the low-mass selection and considering exclusive couplings to muon neutrinos, we find limits on \Vhnlmsq for Majorana (Dirac) HNL of 2.9\ten{-6} (1.0\ten{-5}) at $\mhnl=20\GeV$, of 6.3\ten{-6} (3.6\ten{-5}) at 60\GeV, and of 2.0\ten{-4} (1.3\ten{-3}) at 75\GeV.
Below 30\GeV, the limits are less stringent because of the impact of the long HNL lifetime, with 34 times fewer events selected at $\mhnl=10\GeV$ and $\Vhnlmsq=10^{-5}$ compared with a prompt HNL.
Above \mW, we exclude \Vhnlmsq values for Majorana (Dirac) HNL of 1.4\ten{-3} (2.3\ten{-3}) at a mass of 85\GeV, of 1.1\ten{-3} (1.9\ten{-3}) at 100\GeV, and of 6.0\ten{-2} (9.1\ten{-2}) at 600\GeV.
We improve the exclusion limits from the previous prompt HNL search~\cite{CMS:EXO-17-012} by up to one order of magnitude, and complement the limits from previous displaced HNL searches~\cite{CMS:EXO-20-009, CMS:EXO-21-013} for short-lived HNLs below 20\GeV.
For $\mhnl<\mW$, our exclusion limits are more stringent (similar) in the case of Majorana (Dirac) HNLs compared with the results of the DELPHI Collaboration~\cite{DELPHI:1996qcc}.
Compared with a CMS result that searches for high-mass Majorana HNLs with muon neutrino couplings in $t$-channel VBF production~\cite{CMS:EXO-21-003}, our results provide stricter exclusion limits up to $\mhnl\approx700\GeV$.

The case of HNLs at the \GeVns scale with exclusive tau neutrino couplings was probed before only by the DELPHI Collaboration for $\mhnl<\mW$~\cite{DELPHI:1996qcc}, by the BaBar Collaboration for $\mhnl<1.3\GeV$~\cite{BaBar:2022cqj}, and recently by the CMS Collaboration in displaced HNL searches for $\mhnl<20\GeV$~\cite{CMS:EXO-21-013}.
Using the low-mass selection, we exclude Majorana (Dirac) HNLs with \Vhnltsq values of 6.1\ten{-4} (6.3\ten{-4}) at a mass of 20\GeV, of 1.8\ten{-3} (1.9\ten{-3}) at 60\GeV, and of 2.3\ten{-2} (2.4\ten{-2}) at 75\GeV.
The DELPHI limits for $\mhnl<\mW$ are up to two orders of magnitude more stringent than our results.
Above \mW, the tau neutrino couplings are probed for the first time, and we find limits for Majorana (Dirac) HNLs of 3.0\ten{-1} (2.7\ten{-1}) at 85\GeV, 1.1\ten{-1} (8.6\ten{-2}) at 100\GeV, and 5.9\ten{-1} (5.0\ten{-2}) at 600\GeV.

\section{Summary}

A search for heavy neutral leptons (HNLs) produced in proton-proton collisions at \sqrts has been presented.
The data were collected with the CMS experiment at the LHC and correspond to an integrated luminosity of 138\fbinv.
Events with three charged leptons (electrons, muons, and hadronically decaying tau leptons) are selected, and dedicated identification criteria based on machine learning techniques are applied to reduce the contribution from nonprompt leptons not originating from the hard scattering process.
Remaining standard model (SM) background contributions with nonprompt leptons are estimated from control samples in data, whereas other SM contributions that mostly stem from diboson production are estimated from Monte Carlo event simulations.
A combination of categorization by kinematic properties and machine learning discriminants achieves optimal separation of the predicted signal and SM background contributions.

No significant deviations from the SM predictions are observed.
Exclusion limits at 95\% confidence level are evaluated, assuming exclusive HNL couplings to a single generation of SM neutrinos in the mass range 10\GeV--1.5\TeV, for both Majorana and Dirac HNLs.
These results exceed previous experimental constraints over large parts of the mass range.
Constraints on tau neutrino couplings for HNL masses above the \PW boson mass are presented for the first time.

\begin{acknowledgments}
We congratulate our colleagues in the CERN accelerator departments for the excellent performance of the LHC and thank the technical and administrative staffs at CERN and at other CMS institutes for their contributions to the success of the CMS effort. In addition, we gratefully acknowledge the computing centres and personnel of the Worldwide LHC Computing Grid and other centres for delivering so effectively the computing infrastructure essential to our analyses. Finally, we acknowledge the enduring support for the construction and operation of the LHC, the CMS detector, and the supporting computing infrastructure provided by the following funding agencies: SC (Armenia), BMBWF and FWF (Austria); FNRS and FWO (Belgium); CNPq, CAPES, FAPERJ, FAPERGS, and FAPESP (Brazil); MES and BNSF (Bulgaria); CERN; CAS, MoST, and NSFC (China); MINCIENCIAS (Colombia); MSES and CSF (Croatia); RIF (Cyprus); SENESCYT (Ecuador); ERC PRG, RVTT3 and MoER TK202 (Estonia); Academy of Finland, MEC, and HIP (Finland); CEA and CNRS/IN2P3 (France); SRNSF (Georgia); BMBF, DFG, and HGF (Germany); GSRI (Greece); NKFIH (Hungary); DAE and DST (India); IPM (Iran); SFI (Ireland); INFN (Italy); MSIP and NRF (Republic of Korea); MES (Latvia); LMTLT (Lithuania); MOE and UM (Malaysia); BUAP, CINVESTAV, CONACYT, LNS, SEP, and UASLP-FAI (Mexico); MOS (Montenegro); MBIE (New Zealand); PAEC (Pakistan); MES and NSC (Poland); FCT (Portugal); MESTD (Serbia); MCIN/AEI and PCTI (Spain); MOSTR (Sri Lanka); Swiss Funding Agencies (Switzerland); MST (Taipei); MHESI and NSTDA (Thailand); TUBITAK and TENMAK (Turkey); NASU (Ukraine); STFC (United Kingdom); DOE and NSF (USA).

\hyphenation{Rachada-pisek} Individuals have received support from the Marie-Curie programme and the European Research Council and Horizon 2020 Grant, contract Nos.\ 675440, 724704, 752730, 758316, 765710, 824093, 101115353, and COST Action CA16108 (European Union); the Leventis Foundation; the Alfred P.\ Sloan Foundation; the Alexander von Humboldt Foundation; the Science Committee, project no. 22rl-037 (Armenia); the Belgian Federal Science Policy Office; the Fonds pour la Formation \`a la Recherche dans l'Industrie et dans l'Agriculture (FRIA-Belgium); the Agentschap voor Innovatie door Wetenschap en Technologie (IWT-Belgium); the F.R.S.-FNRS and FWO (Belgium) under the ``Excellence of Science -- EOS" -- be.h project n.\ 30820817; the Beijing Municipal Science \& Technology Commission, No. Z191100007219010 and Fundamental Research Funds for the Central Universities (China); the Ministry of Education, Youth and Sports (MEYS) of the Czech Republic; the Shota Rustaveli National Science Foundation, grant FR-22-985 (Georgia); the Deutsche Forschungsgemeinschaft (DFG), under Germany's Excellence Strategy -- EXC 2121 ``Quantum Universe" -- 390833306, and under project number 400140256 - GRK2497; the Hellenic Foundation for Research and Innovation (HFRI), Project Number 2288 (Greece); the Hungarian Academy of Sciences, the New National Excellence Program - \'UNKP, the NKFIH research grants K 124845, K 124850, K 128713, K 128786, K 129058, K 131991, K 133046, K 138136, K 143460, K 143477, 2020-2.2.1-ED-2021-00181, and TKP2021-NKTA-64 (Hungary); the Council of Science and Industrial Research, India; ICSC -- National Research Centre for High Performance Computing, Big Data and Quantum Computing, funded by the NextGenerationEU program (Italy); the Latvian Council of Science; the Ministry of Education and Science, project no. 2022/WK/14, and the National Science Center, contracts Opus 2021/41/B/ST2/01369 and 2021/43/B/ST2/01552 (Poland); the Funda\c{c}\~ao para a Ci\^encia e a Tecnologia, grant CEECIND/01334/2018 (Portugal); the National Priorities Research Program by Qatar National Research Fund; MCIN/AEI/10.13039/501100011033, ERDF ``a way of making Europe", and the Programa Estatal de Fomento de la Investigaci{\'o}n Cient{\'i}fica y T{\'e}cnica de Excelencia Mar\'{\i}a de Maeztu, grant MDM-2017-0765 and Programa Severo Ochoa del Principado de Asturias (Spain); the Chulalongkorn Academic into Its 2nd Century Project Advancement Project, and the National Science, Research and Innovation Fund via the Program Management Unit for Human Resources \& Institutional Development, Research and Innovation, grant B37G660013 (Thailand); the Kavli Foundation; the Nvidia Corporation; the SuperMicro Corporation; the Welch Foundation, contract C-1845; and the Weston Havens Foundation (USA).
\end{acknowledgments}

\bibliography{auto_generated}

\providecommand{\href}[2]{#2}\begingroup\raggedright\begin{thebibliography}{100}%
\makeatletter
\providecommand{\hrefCMSnoop }[0]{\@secondoftwo}%
\makeatother
\providecommand{\doi}{\texttt{doi:}\begingroup \urlstyle{tt}\Url}

\bibitem{Super-Kamiokande:1998kpq}
\hrefCMSnoop {}{{Super-Kamiokande} Collaboration, ``Evidence for oscillation of
  atmospheric neutrinos'',} \textit{ Phys. Rev. Lett.} \textbf{ 81} (1998)
  1562,
  \href{http://dx.doi.org/10.1103/PhysRevLett.81.1562}{\doi{10.1103/PhysRevLett.81.1562}},
  \href{http://www.arXiv.org/abs/hep-ex/9807003}{\texttt{arXiv:hep-ex/9807003}}.

\bibitem{SNO:2002tuh}
\hrefCMSnoop {}{{SNO} Collaboration, ``Direct evidence for neutrino flavor
  transformation from neutral-current interactions in the {Sudbury Neutrino
  Observatory}'',} \textit{ Phys. Rev. Lett.} \textbf{ 89} (2002) 011301,
  \href{http://dx.doi.org/10.1103/PhysRevLett.89.011301}{\doi{10.1103/PhysRevLett.89.011301}},
  \href{http://www.arXiv.org/abs/nucl-ex/0204008}{\texttt{arXiv:nucl-ex/0204008}}.

\bibitem{KamLAND:2002uet}
\hrefCMSnoop {}{{KamLAND} Collaboration, ``First results from {KamLAND}:
  Evidence for reactor antineutrino disappearance'',} \textit{ Phys. Rev.
  Lett.} \textbf{ 90} (2003) 021802,
  \href{http://dx.doi.org/10.1103/PhysRevLett.90.021802}{\doi{10.1103/PhysRevLett.90.021802}},
  \href{http://www.arXiv.org/abs/hep-ex/0212021}{\texttt{arXiv:hep-ex/0212021}}.

\bibitem{Bilenky:2016pep}
\hrefCMSnoop {}{S.~Bilenky, ``Neutrino oscillations: From a historical
  perspective to the present status'',} \textit{ Nucl. Phys. B} \textbf{ 908}
  (2016) 2,
  \href{http://dx.doi.org/10.1016/j.nuclphysb.2016.01.025}{\doi{10.1016/j.nuclphysb.2016.01.025}},
  \href{http://www.arXiv.org/abs/1602.00170}{\texttt{arXiv:1602.00170}}.

\bibitem{Formaggio:2021nfz}
\hrefCMSnoop {}{J.~Formaggio, A.~de~Gouv{\^e}a, and R.~Robertson, ``Direct
  measurements of neutrino mass'',} \textit{ Phys. Rept.} \textbf{ 914} (2021)
  1,
  \href{http://dx.doi.org/10.1016/j.physrep.2021.02.002}{\doi{10.1016/j.physrep.2021.02.002}},
  \href{http://www.arXiv.org/abs/2102.00594}{\texttt{arXiv:2102.00594}}.

\bibitem{KATRIN:2021uub}
\hrefCMSnoop {}{{KATRIN} Collaboration, ``Direct neutrino-mass measurement with
  sub-electronvolt sensitivity'',} \textit{ Nature Phys.} \textbf{ 18} (2022)
  160,
  \href{http://dx.doi.org/10.1038/s41567-021-01463-1}{\doi{10.1038/s41567-021-01463-1}},
  \href{http://www.arXiv.org/abs/2105.08533}{\texttt{arXiv:2105.08533}}.

\bibitem{Planck:2018vyg}
\hrefCMSnoop {}{{Planck} Collaboration, ``Planck 2018 results. {VI}.
  cosmological parameters'',} \textit{ Astron. Astrophys.} \textbf{ 641} (2020)
  A6,
  \href{http://dx.doi.org/10.1051/0004-6361/201833910}{\doi{10.1051/0004-6361/201833910}},
  \href{http://www.arXiv.org/abs/1807.06209}{\texttt{arXiv:1807.06209}}.
  [Erratum: \DOI{10.1051/0004-6361/201833910e}].

\bibitem{eBOSS:2020yzd}
\hrefCMSnoop {}{{eBOSS} Collaboration, ``Completed {SDSS-IV} extended baryon
  oscillation spectroscopic survey: Cosmological implications from two decades
  of spectroscopic surveys at the {Apache Point Observatory}'',} \textit{ Phys.
  Rev. D} \textbf{ 103} (2021) 083533,
  \href{http://dx.doi.org/10.1103/PhysRevD.103.083533}{\doi{10.1103/PhysRevD.103.083533}},
  \href{http://www.arXiv.org/abs/2007.08991}{\texttt{arXiv:2007.08991}}.

\bibitem{Sakr:2022ans}
\hrefCMSnoop {}{Z.~Sakr, ``A short review on the latest neutrinos mass and
  number constraints from cosmological observables'',} \textit{ Universe}
  \textbf{ 8} (2022) 284,
  \href{http://dx.doi.org/10.3390/universe8050284}{\doi{10.3390/universe8050284}}.

\bibitem{Minkowski:1977sc}
\hrefCMSnoop {}{P.~Minkowski, ``${\PGm\to\Pe\PGg}$ at a rate of one out of
  $10^9$ muon decays?'',} \textit{ Phys. Lett. B} \textbf{ 67} (1977) 421,
  \href{http://dx.doi.org/10.1016/0370-2693(77)90435-X}{\doi{10.1016/0370-2693(77)90435-X}}.

\bibitem{Yanagida:1979as}
\hrefCMSnoop {}{T.~Yanagida, ``Horizontal gauge symmetry and masses of
  neutrinos'',} in \textit{ {Proc. Workshop on the Unified Theories and the
  Baryon Number in the Universe: Tsukuba, Japan, February 13--14, 1979}}.
\newblock 1979.
\newblock [Conf. Proc. C 7902131 (1979) 95].

\bibitem{Gell-Mann:1979vob}
\hrefCMSnoop {}{M.~Gell-Mann, P.~Ramond, and R.~Slansky, ``Complex spinors and
  unified theories'',} in \textit{ {Proc. Supergravity Workshop: Stony Brook
  NY, USA, September 27--28, 1979}}.
\newblock 1979.
\newblock \href{http://www.arXiv.org/abs/1306.4669}{\texttt{arXiv:1306.4669}}.
\newblock [Conf. Proc. C 790927 (1979) 315].

\bibitem{Glashow:1979nm}
\hrefCMSnoop {}{S.~Glashow, ``The future of elementary particle physics'',}
  \textit{ NATO Sci. Ser. B} \textbf{ 61} (1980) 687,
  \href{http://dx.doi.org/10.1007/978-1-4684-7197-7_15}{\doi{10.1007/978-1-4684-7197-7_15}}.

\bibitem{Mohapatra:1979ia}
\hrefCMSnoop {}{R.~Mohapatra and G.~Senjanovi{\'c}, ``Neutrino mass and
  spontaneous parity nonconservation'',} \textit{ Phys. Rev. Lett.} \textbf{
  44} (1980) 912,
  \href{http://dx.doi.org/10.1103/PhysRevLett.44.912}{\doi{10.1103/PhysRevLett.44.912}}.

\bibitem{Schechter:1980gr}
\hrefCMSnoop {}{J.~Schechter and J.~Valle, ``Neutrino masses in
  $\mathrm{SU}(2)\otimes\mathrm{U}(1)$ theories'',} \textit{ Phys. Rev. D}
  \textbf{ 22} (1980) 2227,
  \href{http://dx.doi.org/10.1103/PhysRevD.22.2227}{\doi{10.1103/PhysRevD.22.2227}}.

\bibitem{Shrock:1980ct}
\hrefCMSnoop {}{R.~Shrock, ``General theory of weak leptonic and semileptonic
  decays. {I}. leptonic pseudoscalar meson decays, with associated tests for,
  and bounds on, neutrino masses and lepton mixing'',} \textit{ Phys. Rev. D}
  \textbf{ 24} (1981) 1232,
  \href{http://dx.doi.org/10.1103/PhysRevD.24.1232}{\doi{10.1103/PhysRevD.24.1232}}.

\bibitem{Cai:2017mow}
\hrefCMSnoop {}{Y.~Cai, T.~Han, T.~Li, and R.~Ruiz, ``Lepton number violation:
  Seesaw models and their collider tests'',} \textit{ Front. Phys.} \textbf{ 6}
  (2018) 40,
  \href{http://dx.doi.org/10.3389/fphy.2018.00040}{\doi{10.3389/fphy.2018.00040}},
  \href{http://www.arXiv.org/abs/1711.02180}{\texttt{arXiv:1711.02180}}.

\bibitem{Dodelson:1993je}
\hrefCMSnoop {}{S.~Dodelson and L.~Widrow, ``Sterile neutrinos as dark
  matter'',} \textit{ Phys. Rev. Lett.} \textbf{ 72} (1994) 17,
  \href{http://dx.doi.org/10.1103/PhysRevLett.72.17}{\doi{10.1103/PhysRevLett.72.17}},
  \href{http://www.arXiv.org/abs/hep-ph/9303287}{\texttt{arXiv:hep-ph/9303287}}.

\bibitem{Boyarsky:2018tvu}
A.~Boyarsky\hrefCMSnoop {}{ { et~al.}, ``Sterile neutrino dark matter'',}
  \textit{ Prog. Part. Nucl. Phys.} \textbf{ 104} (2019) 1,
  \href{http://dx.doi.org/10.1016/j.ppnp.2018.07.004}{\doi{10.1016/j.ppnp.2018.07.004}},
  \href{http://www.arXiv.org/abs/1807.07938}{\texttt{arXiv:1807.07938}}.

\bibitem{Fukugita:1986hr}
\hrefCMSnoop {}{M.~Fukugita and T.~Yanagida, ``Baryogenesis without grand
  unification'',} \textit{ Phys. Lett. B} \textbf{ 174} (1986) 45,
  \href{http://dx.doi.org/10.1016/0370-2693(86)91126-3}{\doi{10.1016/0370-2693(86)91126-3}}.

\bibitem{Chun:2017spz}
E.~Chun\hrefCMSnoop {}{ { et~al.}, ``Probing leptogenesis'',} \textit{ Int. J.
  Mod. Phys. A} \textbf{ 33} (2018) 1842005,
  \href{http://dx.doi.org/10.1142/S0217751X18420058}{\doi{10.1142/S0217751X18420058}},
  \href{http://www.arXiv.org/abs/1711.02865}{\texttt{arXiv:1711.02865}}.

\bibitem{Drewes:2021nqr}
\hrefCMSnoop {}{M.~Drewes, Y.~Georis, and J.~Klari{\'c}, ``Mapping the viable
  parameter space for testable leptogenesis'',} \textit{ Phys. Rev. Lett.}
  \textbf{ 128} (2022) 051801,
  \href{http://dx.doi.org/10.1103/PhysRevLett.128.051801}{\doi{10.1103/PhysRevLett.128.051801}},
  \href{http://www.arXiv.org/abs/2106.16226}{\texttt{arXiv:2106.16226}}.

\bibitem{Beacham:2019nyx}
\hrefCMSnoop {}{J.~Beacham { et~al.}, ``Physics beyond colliders at {CERN}:
  Beyond the standard model working group report'',} \textit{ J. Phys. G}
  \textbf{ 47} (2020) 010501,
  \href{http://dx.doi.org/10.1088/1361-6471/ab4cd2}{\doi{10.1088/1361-6471/ab4cd2}},
  \href{http://www.arXiv.org/abs/1901.09966}{\texttt{arXiv:1901.09966}}.

\bibitem{Drewes:2022akb}
\hrefCMSnoop {}{M.~Drewes, J.~Klari{\'c}, and J.~L{\'o}pez-Pav{\'o}n, ``New
  benchmark models for heavy neutral lepton searches'',} \textit{ Eur. Phys. J.
  C} \textbf{ 82} (2022) 1176,
  \href{http://dx.doi.org/10.1140/epjc/s10052-022-11100-7}{\doi{10.1140/epjc/s10052-022-11100-7}},
  \href{http://www.arXiv.org/abs/2207.02742}{\texttt{arXiv:2207.02742}}.

\bibitem{delAguila:2008cj}
\hrefCMSnoop {}{F.~del Aguila and J.~Aguilar-Saavedra, ``Distinguishing seesaw
  models at {LHC} with multi-lepton signals'',} \textit{ Nucl. Phys. B}
  \textbf{ 813} (2009) 22,
  \href{http://dx.doi.org/10.1016/j.nuclphysb.2008.12.029}{\doi{10.1016/j.nuclphysb.2008.12.029}},
  \href{http://www.arXiv.org/abs/0808.2468}{\texttt{arXiv:0808.2468}}.

\bibitem{Atre:2009rg}
\hrefCMSnoop {}{A.~Atre, T.~Han, S.~Pascoli, and B.~Zhang, ``The search for
  heavy {Majorana} neutrinos'',} \textit{ JHEP} \textbf{ 05} (2009) 030,
  \href{http://dx.doi.org/10.1088/1126-6708/2009/05/030}{\doi{10.1088/1126-6708/2009/05/030}},
  \href{http://www.arXiv.org/abs/0901.3589}{\texttt{arXiv:0901.3589}}.

\bibitem{Tello:2010am}
V.~Tello\hrefCMSnoop {}{ { et~al.}, ``Left-right symmetry: from {LHC} to
  neutrinoless double beta decay'',} \textit{ Phys. Rev. Lett.} \textbf{ 106}
  (2011) 151801,
  \href{http://dx.doi.org/10.1103/PhysRevLett.106.151801}{\doi{10.1103/PhysRevLett.106.151801}},
  \href{http://www.arXiv.org/abs/1011.3522}{\texttt{arXiv:1011.3522}}.

\bibitem{Das:2012ze}
\hrefCMSnoop {}{A.~Das and N.~Okada, ``Inverse seesaw neutrino signatures at
  the {LHC} and {ILC}'',} \textit{ Phys. Rev. D} \textbf{ 88} (2013) 113001,
  \href{http://dx.doi.org/10.1103/PhysRevD.88.113001}{\doi{10.1103/PhysRevD.88.113001}},
  \href{http://www.arXiv.org/abs/1207.3734}{\texttt{arXiv:1207.3734}}.

\bibitem{Deppisch:2015qwa}
\hrefCMSnoop {}{F.~Deppisch, P.~Bhupal~Dev, and A.~Pilaftsis, ``Neutrinos and
  collider physics'',} \textit{ New J. Phys.} \textbf{ 17} (2015) 075019,
  \href{http://dx.doi.org/10.1088/1367-2630/17/7/075019}{\doi{10.1088/1367-2630/17/7/075019}},
  \href{http://www.arXiv.org/abs/1502.06541}{\texttt{arXiv:1502.06541}}.

\bibitem{Das:2017nvm}
\hrefCMSnoop {}{A.~Das and N.~Okada, ``Bounds on heavy {Majorana} neutrinos in
  type-{I} seesaw and implications for collider searches'',} \textit{ Phys.
  Lett. B} \textbf{ 774} (2017) 32,
  \href{http://dx.doi.org/10.1016/j.physletb.2017.09.042}{\doi{10.1016/j.physletb.2017.09.042}},
  \href{http://www.arXiv.org/abs/1702.04668}{\texttt{arXiv:1702.04668}}.

\bibitem{Das:2017gke}
\hrefCMSnoop {}{A.~Das, P.~Konar, and A.~Thalapillil, ``Jet substructure
  shedding light on heavy {Majorana} neutrinos at the {LHC}'',} \textit{ JHEP}
  \textbf{ 02} (2018) 083,
  \href{http://dx.doi.org/10.1007/JHEP02(2018)083}{\doi{10.1007/JHEP02(2018)083}},
  \href{http://www.arXiv.org/abs/1709.09712}{\texttt{arXiv:1709.09712}}.

\bibitem{Bhardwaj:2018lma}
\hrefCMSnoop {}{A.~Bhardwaj, A.~Das, P.~Konar, and A.~Thalapillil, ``Looking
  for minimal inverse seesaw scenarios at the {LHC} with jet substructure
  techniques'',} \textit{ J. Phys. G} \textbf{ 47} (2020) 075002,
  \href{http://dx.doi.org/10.1088/1361-6471/ab7769}{\doi{10.1088/1361-6471/ab7769}},
  \href{http://www.arXiv.org/abs/1801.00797}{\texttt{arXiv:1801.00797}}.

\bibitem{Pascoli:2018heg}
\hrefCMSnoop {}{S.~Pascoli, R.~Ruiz, and C.~Weiland, ``Heavy neutrinos with
  dynamic jet vetoes: multilepton searches at $\sqrt{s}=14$, 27, and
  {100\TeV}'',} \textit{ JHEP} \textbf{ 06} (2019) 049,
  \href{http://dx.doi.org/10.1007/JHEP06(2019)049}{\doi{10.1007/JHEP06(2019)049}},
  \href{http://www.arXiv.org/abs/1812.08750}{\texttt{arXiv:1812.08750}}.

\bibitem{Abdullahi:2022jlv}
\hrefCMSnoop {}{A.~Abdullahi { et~al.}, ``The present and future status of
  heavy neutral leptons'',} \textit{ J. Phys. G} \textbf{ 50} (2023) 020501,
  \href{http://dx.doi.org/10.1088/1361-6471/ac98f9}{\doi{10.1088/1361-6471/ac98f9}},
  \href{http://www.arXiv.org/abs/2203.08039}{\texttt{arXiv:2203.08039}}.

\bibitem{Antel:2023hkf}
\hrefCMSnoop {}{C.~Antel { et~al.}, ``Feebly interacting particles: {FIPs} 2022
  workshop report'',} \textit{ Eur. Phys. J. C} \textbf{ 83} (2023) 1122,
  \href{http://dx.doi.org/10.1140/epjc/s10052-023-12168-5}{\doi{10.1140/epjc/s10052-023-12168-5}},
  \href{http://www.arXiv.org/abs/2305.01715}{\texttt{arXiv:2305.01715}}.

\bibitem{Keung:1983uu}
\hrefCMSnoop {}{W.-Y. Keung and G.~Senjanovic, ``Majorana neutrinos and the
  production of the right-handed charged gauge boson'',} \textit{ Phys. Rev.
  Lett.} \textbf{ 50} (1983) 1427,
  \href{http://dx.doi.org/10.1103/PhysRevLett.50.1427}{\doi{10.1103/PhysRevLett.50.1427}}.

\bibitem{Petcov:1984nf}
\hrefCMSnoop {}{S.~Petcov, ``Possible signature for production of {Majorana}
  particles in {\EE} and ${\Pp\PAp}$ collisions'',} \textit{ Phys. Lett. B}
  \textbf{ 139} (1984) 421,
  \href{http://dx.doi.org/10.1016/0370-2693(84)91844-6}{\doi{10.1016/0370-2693(84)91844-6}}.

\bibitem{Datta:1993nm}
\hrefCMSnoop {}{A.~Datta, M.~Guchait, and A.~Pilaftsis, ``Probing lepton number
  violation via {Majorana} neutrinos at hadron supercolliders'',} \textit{
  Phys. Rev. D} \textbf{ 50} (1994) 3195,
  \href{http://dx.doi.org/10.1103/PhysRevD.50.3195}{\doi{10.1103/PhysRevD.50.3195}},
  \href{http://www.arXiv.org/abs/hep-ph/9311257}{\texttt{arXiv:hep-ph/9311257}}.

\bibitem{Dev:2013wba}
\hrefCMSnoop {}{P.~Bhupal~Dev, A.~Pilaftsis, and U.-k. Yang, ``New production
  mechanism for heavy neutrinos at the {LHC}'',} \textit{ Phys. Rev. Lett.}
  \textbf{ 112} (2014) 081801,
  \href{http://dx.doi.org/10.1103/PhysRevLett.112.081801}{\doi{10.1103/PhysRevLett.112.081801}},
  \href{http://www.arXiv.org/abs/1308.2209}{\texttt{arXiv:1308.2209}}.

\bibitem{Alva:2014gxa}
\hrefCMSnoop {}{D.~Alva, T.~Han, and R.~Ruiz, ``Heavy {Majorana} neutrinos from
  ${\PW\PGg}$ fusion at hadron colliders'',} \textit{ JHEP} \textbf{ 02} (2015)
  072,
  \href{http://dx.doi.org/10.1007/JHEP02(2015)072}{\doi{10.1007/JHEP02(2015)072}},
  \href{http://www.arXiv.org/abs/1411.7305}{\texttt{arXiv:1411.7305}}.

\bibitem{Degrande:2016aje}
\hrefCMSnoop {}{C.~Degrande, O.~Mattelaer, R.~Ruiz, and J.~Turner,
  ``Fully-automated precision predictions for heavy neutrino production
  mechanisms at hadron colliders'',} \textit{ Phys. Rev. D} \textbf{ 94} (2016)
  053002,
  \href{http://dx.doi.org/10.1103/PhysRevD.94.053002}{\doi{10.1103/PhysRevD.94.053002}},
  \href{http://www.arXiv.org/abs/1602.06957}{\texttt{arXiv:1602.06957}}.

\bibitem{CMS:EXO-11-076}
\hrefCMSnoop {}{{CMS Collaboration}, ``Search for heavy {Majorana} neutrinos in
  ${\PGmpm\PGmpm}+{}$jets and ${\Pepm\Pepm}+{}$jets events in ${\Pp\Pp}$
  collisions at $\sqrt{s}={7\TeV}$'',} \textit{ Phys. Lett. B} \textbf{ 717}
  (2012) 109,
  \href{http://dx.doi.org/10.1016/j.physletb.2012.09.012}{\doi{10.1016/j.physletb.2012.09.012}},
  \href{http://www.arXiv.org/abs/1207.6079}{\texttt{arXiv:1207.6079}}.

\bibitem{CMS:EXO-12-057}
\hrefCMSnoop {}{{CMS Collaboration}, ``Search for heavy {Majorana} neutrinos in
  ${\PGmpm\PGmpm}+{}$jets events in proton-proton collisions at
  $\sqrt{s}={8\TeV}$'',} \textit{ Phys. Lett. B} \textbf{ 748} (2015) 144,
  \href{http://dx.doi.org/10.1016/j.physletb.2015.06.070}{\doi{10.1016/j.physletb.2015.06.070}},
  \href{http://www.arXiv.org/abs/1501.05566}{\texttt{arXiv:1501.05566}}.

\bibitem{ATLAS:2015gtp}
\hrefCMSnoop {}{{ATLAS Collaboration}, ``Search for heavy {Majorana} neutrinos
  with the {ATLAS} detector in ${\Pp\Pp}$ collisions at $\sqrt{s}={8\TeV}$'',}
  \textit{ JHEP} \textbf{ 07} (2015) 162,
  \href{http://dx.doi.org/10.1007/JHEP07(2015)162}{\doi{10.1007/JHEP07(2015)162}},
  \href{http://www.arXiv.org/abs/1506.06020}{\texttt{arXiv:1506.06020}}.

\bibitem{CMS:EXO-14-014}
\hrefCMSnoop {}{{CMS Collaboration}, ``Search for heavy {Majorana} neutrinos in
  ${\Pepm\Pepm}+{}$jets and ${\Pepm\PGmpm}+{}$jets events in proton-proton
  collisions at $\sqrt{s}={8\TeV}$'',} \textit{ JHEP} \textbf{ 04} (2016) 169,
  \href{http://dx.doi.org/10.1007/JHEP04(2016)169}{\doi{10.1007/JHEP04(2016)169}},
  \href{http://www.arXiv.org/abs/1603.02248}{\texttt{arXiv:1603.02248}}.

\bibitem{CMS:EXO-17-012}
\hrefCMSnoop {}{{CMS Collaboration}, ``Search for heavy neutral leptons in
  events with three charged leptons in proton-proton collisions at
  $\sqrt{s}={13\TeV}$'',} \textit{ Phys. Rev. Lett.} \textbf{ 120} (2018)
  221801,
  \href{http://dx.doi.org/10.1103/PhysRevLett.120.221801}{\doi{10.1103/PhysRevLett.120.221801}},
  \href{http://www.arXiv.org/abs/1802.02965}{\texttt{arXiv:1802.02965}}.

\bibitem{CMS:EXO-17-028}
\hrefCMSnoop {}{{CMS Collaboration}, ``Search for heavy {Majorana} neutrinos in
  same-sign dilepton channels in proton-proton collisions at
  $\sqrt{s}={13\TeV}$'',} \textit{ JHEP} \textbf{ 01} (2019) 122,
  \href{http://dx.doi.org/10.1007/JHEP01(2019)122}{\doi{10.1007/JHEP01(2019)122}},
  \href{http://www.arXiv.org/abs/1806.10905}{\texttt{arXiv:1806.10905}}.

\bibitem{ATLAS:2019kpx}
\hrefCMSnoop {}{{ATLAS Collaboration}, ``Search for heavy neutral leptons in
  decays of {\PW} bosons produced in {13\TeV} ${\Pp\Pp}$ collisions using
  prompt and displaced signatures with the {ATLAS} detector'',} \textit{ JHEP}
  \textbf{ 10} (2019) 265,
  \href{http://dx.doi.org/10.1007/JHEP10(2019)265}{\doi{10.1007/JHEP10(2019)265}},
  \href{http://www.arXiv.org/abs/1905.09787}{\texttt{arXiv:1905.09787}}.

\bibitem{LHCb:2020wxx}
\hrefCMSnoop {}{{LHCb Collaboration}, ``Search for heavy neutral leptons in
  ${\PWp\to\PGmp\PGmpm\,\text{jet}}$ decays'',} \textit{ Eur. Phys. J. C}
  \textbf{ 81} (2021) 248,
  \href{http://dx.doi.org/10.1140/epjc/s10052-021-08973-5}{\doi{10.1140/epjc/s10052-021-08973-5}},
  \href{http://www.arXiv.org/abs/2011.05263}{\texttt{arXiv:2011.05263}}.

\bibitem{CMS:EXO-20-009}
\hrefCMSnoop {}{{CMS Collaboration}, ``Search for long-lived heavy neutral
  leptons with displaced vertices in proton-proton collisions at
  $\sqrt{s}={13\TeV}$'',} \textit{ JHEP} \textbf{ 07} (2022) 081,
  \href{http://dx.doi.org/10.1007/JHEP07(2022)081}{\doi{10.1007/JHEP07(2022)081}},
  \href{http://www.arXiv.org/abs/2201.05578}{\texttt{arXiv:2201.05578}}.

\bibitem{ATLAS:2022atq}
\hrefCMSnoop {}{{ATLAS Collaboration}, ``Search for heavy neutral leptons in
  decays of {\PW} bosons using a dilepton displaced vertex in
  $\sqrt{s}={13\TeV}$ ${\Pp\Pp}$ collisions with the {ATLAS} detector'',}
  \textit{ Phys. Rev. Lett.} \textbf{ 131} (2023) 061803,
  \href{http://dx.doi.org/10.1103/PhysRevLett.131.061803}{\doi{10.1103/PhysRevLett.131.061803}},
  \href{http://www.arXiv.org/abs/2204.11988}{\texttt{arXiv:2204.11988}}.

\bibitem{CMS:EXO-21-013}
\hrefCMSnoop {}{{CMS Collaboration}, ``Search for long-lived heavy neutral
  leptons with lepton flavour conserving or violating decays to a jet and a
  charged lepton'',} \textit{ JHEP} \textbf{ 03} (2024) 105,
  \href{http://dx.doi.org/10.1007/JHEP03(2024)105}{\doi{10.1007/JHEP03(2024)105}},
  \href{http://www.arXiv.org/abs/2312.07484}{\texttt{arXiv:2312.07484}}.

\bibitem{CMS:EXO-23-006}
\hrefCMSnoop {}{{CMS Collaboration}, ``Review of searches for vector-like
  quarks, vector-like leptons, and heavy neutral leptons in proton-proton
  collisions at $\sqrt{s}={13\TeV}$ at the {CMS} experiment'',} 2024.
  \href{http://www.arXiv.org/abs/2405.17605}{\texttt{arXiv:2405.17605}}.
  Submitted to \textit{Phys. Rept.}

\bibitem{Abada:2018sfh}
\hrefCMSnoop {}{A.~Abada, N.~Bernal, M.~Losada, and X.~Marcano, ``Inclusive
  displaced vertex searches for heavy neutral leptons at the {LHC}'',} \textit{
  JHEP} \textbf{ 01} (2019) 093,
  \href{http://dx.doi.org/10.1007/JHEP01(2019)093}{\doi{10.1007/JHEP01(2019)093}},
  \href{http://www.arXiv.org/abs/1807.10024}{\texttt{arXiv:1807.10024}}.

\bibitem{Tastet:2021vwp}
\hrefCMSnoop {}{J.-L. Tastet, O.~Ruchayskiy, and I.~Timiryasov,
  ``Reinterpreting the {ATLAS} bounds on heavy neutral leptons in a realistic
  neutrino oscillation model'',} \textit{ JHEP} \textbf{ 12} (2021) 182,
  \href{http://dx.doi.org/10.1007/JHEP12(2021)182}{\doi{10.1007/JHEP12(2021)182}},
  \href{http://www.arXiv.org/abs/2107.12980}{\texttt{arXiv:2107.12980}}.

\bibitem{hepdata}
\hrefCMSnoop {}{}{HEPData} record for this analysis, 2024.
\newblock
  \href{http://dx.doi.org/10.17182/hepdata.146676}{\doi{10.17182/hepdata.146676}}.

\bibitem{Asaka:2005an}
\hrefCMSnoop {}{T.~Asaka, S.~Blanchet, and M.~Shaposhnikov, ``The {\PGn}{MSM},
  dark matter and neutrino masses'',} \textit{ Phys. Lett. B} \textbf{ 631}
  (2005) 151,
  \href{http://dx.doi.org/10.1016/j.physletb.2005.09.070}{\doi{10.1016/j.physletb.2005.09.070}},
  \href{http://www.arXiv.org/abs/hep-ph/0503065}{\texttt{arXiv:hep-ph/0503065}}.

\bibitem{CMS:Detector-2008}
\hrefCMSnoop {}{{CMS Collaboration}, ``The {CMS} experiment at the {CERN}
  {LHC}'',} \textit{ JINST} \textbf{ 3} (2008) S08004,
  \href{http://dx.doi.org/10.1088/1748-0221/3/08/S08004}{\doi{10.1088/1748-0221/3/08/S08004}}.

\bibitem{CMS:PRF-21-001}
\hrefCMSnoop {}{{CMS Collaboration}, ``Development of the {CMS} detector for
  the {CERN} {LHC} {Run} 3'',} \textit{ JINST} \textbf{ 19} (2024) P05064,
  \href{http://dx.doi.org/10.1088/1748-0221/19/05/P05064}{\doi{10.1088/1748-0221/19/05/P05064}},
  \href{http://www.arXiv.org/abs/2309.05466}{\texttt{arXiv:2309.05466}}.

\bibitem{CMS:TRG-17-001}
\hrefCMSnoop {}{{CMS Collaboration}, ``Performance of the {CMS} {\Lone} trigger
  in proton-proton collisions at $\sqrt{s}={13\TeV}$'',} \textit{ JINST}
  \textbf{ 15} (2020) P10017,
  \href{http://dx.doi.org/10.1088/1748-0221/15/10/P10017}{\doi{10.1088/1748-0221/15/10/P10017}},
  \href{http://www.arXiv.org/abs/2006.10165}{\texttt{arXiv:2006.10165}}.

\bibitem{CMS:TRG-12-001}
\hrefCMSnoop {}{{CMS Collaboration}, ``The {CMS} trigger system'',} \textit{
  JINST} \textbf{ 12} (2017) P01020,
  \href{http://dx.doi.org/10.1088/1748-0221/12/01/P01020}{\doi{10.1088/1748-0221/12/01/P01020}},
  \href{http://www.arXiv.org/abs/1609.02366}{\texttt{arXiv:1609.02366}}.

\bibitem{CMS:PRF-14-001}
\hrefCMSnoop {}{{CMS Collaboration}, ``Particle-flow reconstruction and global
  event description with the {CMS} detector'',} \textit{ JINST} \textbf{ 12}
  (2017) P10003,
  \href{http://dx.doi.org/10.1088/1748-0221/12/10/P10003}{\doi{10.1088/1748-0221/12/10/P10003}},
  \href{http://www.arXiv.org/abs/1706.04965}{\texttt{arXiv:1706.04965}}.

\bibitem{CMS:TDR-15-02}
\hrefCMSnoop {}{{CMS Collaboration}, ``Technical proposal for the {Phase-II}
  upgrade of the {Compact Muon Solenoid}'',} CMS Technical Proposal
  CERN-LHCC-2015-010, CMS-TDR-15-02, 2015.
\newblock
  \href{http://dx.doi.org/10.17181/CERN.VU8I.D59J}{\doi{10.17181/CERN.VU8I.D59J}}.

\bibitem{Cacciari:2008gp}
\hrefCMSnoop {}{M.~Cacciari, G.~P. Salam, and G.~Soyez, ``The anti-\kt jet
  clustering algorithm'',} \textit{ JHEP} \textbf{ 04} (2008) 063,
  \href{http://dx.doi.org/10.1088/1126-6708/2008/04/063}{\doi{10.1088/1126-6708/2008/04/063}},
  \href{http://www.arXiv.org/abs/0802.1189}{\texttt{arXiv:0802.1189}}.

\bibitem{Cacciari:2011ma}
\hrefCMSnoop {}{M.~Cacciari, G.~P. Salam, and G.~Soyez, ``{\FASTJET} user
  manual'',} \textit{ Eur. Phys. J. C} \textbf{ 72} (2012) 1896,
  \href{http://dx.doi.org/10.1140/epjc/s10052-012-1896-2}{\doi{10.1140/epjc/s10052-012-1896-2}},
  \href{http://www.arXiv.org/abs/1111.6097}{\texttt{arXiv:1111.6097}}.

\bibitem{CMS:JME-13-004}
\hrefCMSnoop {}{{CMS Collaboration}, ``Jet energy scale and resolution in the
  {CMS} experiment in ${\Pp\Pp}$ collisions at {8\TeV}'',} \textit{ JINST}
  \textbf{ 12} (2017) P02014,
  \href{http://dx.doi.org/10.1088/1748-0221/12/02/P02014}{\doi{10.1088/1748-0221/12/02/P02014}},
  \href{http://www.arXiv.org/abs/1607.03663}{\texttt{arXiv:1607.03663}}.

\bibitem{CMS:JME-16-003}
\href {https://cds.cern.ch/record/2256875}{{CMS Collaboration}, ``Jet
  algorithms performance in {13\TeV} data'',} CMS Physics Analysis Summary
  CMS-PAS-JME-16-003, 2017.

\bibitem{CMS:JME-17-001}
\hrefCMSnoop {}{{CMS Collaboration}, ``Performance of missing transverse
  momentum reconstruction in proton-proton collisions at $\sqrt{s}={13\TeV}$
  using the {CMS} detector'',} \textit{ JINST} \textbf{ 14} (2019) P07004,
  \href{http://dx.doi.org/10.1088/1748-0221/14/07/P07004}{\doi{10.1088/1748-0221/14/07/P07004}},
  \href{http://www.arXiv.org/abs/1903.06078}{\texttt{arXiv:1903.06078}}.

\bibitem{CMS:BTV-16-002}
\hrefCMSnoop {}{{CMS Collaboration}, ``Identification of heavy-flavour jets
  with the {CMS} detector in ${\Pp\Pp}$ collisions at {13\TeV}'',} \textit{
  JINST} \textbf{ 13} (2018) P05011,
  \href{http://dx.doi.org/10.1088/1748-0221/13/05/P05011}{\doi{10.1088/1748-0221/13/05/P05011}},
  \href{http://www.arXiv.org/abs/1712.07158}{\texttt{arXiv:1712.07158}}.

\bibitem{Bols:2020bkb}
E.~Bols\hrefCMSnoop {}{ { et~al.}, ``Jet flavour classification using
  {DeepJet}'',} \textit{ JINST} \textbf{ 15} (2020) P12012,
  \href{http://dx.doi.org/10.1088/1748-0221/15/12/P12012}{\doi{10.1088/1748-0221/15/12/P12012}},
  \href{http://www.arXiv.org/abs/2008.10519}{\texttt{arXiv:2008.10519}}.

\bibitem{CMS:DP-2023-005}
\href {https://cds.cern.ch/record/2854609}{{CMS Collaboration}, ``Performance
  summary of {AK4} jet {\PQb} tagging with data from proton-proton collisions
  at {13\TeV} with the {CMS} detector'',} CMS Detector Performance Note
  CMS-DP-2023-005, 2023.

\bibitem{GEANT4:2002zbu}
\hrefCMSnoop {}{{GEANT4} Collaboration, ``{\GEANTfour}---a simulation
  toolkit'',} \textit{ Nucl. Instrum. Meth. A} \textbf{ 506} (2003) 250,
  \href{http://dx.doi.org/10.1016/S0168-9002(03)01368-8}{\doi{10.1016/S0168-9002(03)01368-8}}.

\bibitem{CMS:JME-18-001}
\hrefCMSnoop {}{{CMS Collaboration}, ``Pileup mitigation at {CMS} in {13\TeV}
  data'',} \textit{ JINST} \textbf{ 15} (2020) P09018,
  \href{http://dx.doi.org/10.1088/1748-0221/15/09/P09018}{\doi{10.1088/1748-0221/15/09/P09018}},
  \href{http://www.arXiv.org/abs/2003.00503}{\texttt{arXiv:2003.00503}}.

\bibitem{Alwall:2014hca}
J.~Alwall\hrefCMSnoop {}{ { et~al.}, ``The automated computation of tree-level
  and next-to-leading order differential cross sections, and their matching to
  parton shower simulations'',} \textit{ JHEP} \textbf{ 07} (2014) 079,
  \href{http://dx.doi.org/10.1007/JHEP07(2014)079}{\doi{10.1007/JHEP07(2014)079}},
  \href{http://www.arXiv.org/abs/1405.0301}{\texttt{arXiv:1405.0301}}.

\bibitem{Artoisenet:2012st}
\hrefCMSnoop {}{P.~Artoisenet, R.~Frederix, O.~Mattelaer, and R.~Rietkerk,
  ``Automatic spin-entangled decays of heavy resonances in {Monte Carlo}
  simulations'',} \textit{ JHEP} \textbf{ 03} (2013) 015,
  \href{http://dx.doi.org/10.1007/JHEP03(2013)015}{\doi{10.1007/JHEP03(2013)015}},
  \href{http://www.arXiv.org/abs/1212.3460}{\texttt{arXiv:1212.3460}}.

\bibitem{NNPDF:2017mvq}
\hrefCMSnoop {}{{NNPDF} Collaboration, ``Parton distributions from
  high-precision collider data'',} \textit{ Eur. Phys. J. C} \textbf{ 77}
  (2017) 663,
  \href{http://dx.doi.org/10.1140/epjc/s10052-017-5199-5}{\doi{10.1140/epjc/s10052-017-5199-5}},
  \href{http://www.arXiv.org/abs/1706.00428}{\texttt{arXiv:1706.00428}}.

\bibitem{Manohar:2016nzj}
\hrefCMSnoop {}{A.~Manohar, P.~Nason, G.~P. Salam, and G.~Zanderighi, ``How
  bright is the proton? {A} precise determination of the photon parton
  distribution function'',} \textit{ Phys. Rev. Lett.} \textbf{ 117} (2016)
  242002,
  \href{http://dx.doi.org/10.1103/PhysRevLett.117.242002}{\doi{10.1103/PhysRevLett.117.242002}},
  \href{http://www.arXiv.org/abs/1607.04266}{\texttt{arXiv:1607.04266}}.

\bibitem{Manohar:2017eqh}
\hrefCMSnoop {}{A.~V. Manohar, P.~Nason, G.~P. Salam, and G.~Zanderighi, ``The
  photon content of the proton'',} \textit{ JHEP} \textbf{ 12} (2017) 046,
  \href{http://dx.doi.org/10.1007/JHEP12(2017)046}{\doi{10.1007/JHEP12(2017)046}},
  \href{http://www.arXiv.org/abs/1708.01256}{\texttt{arXiv:1708.01256}}.

\bibitem{Bertone:2017bme}
\hrefCMSnoop {}{{NNPDF} Collaboration, ``Illuminating the photon content of the
  proton within a global {PDF} analysis'',} \textit{ SciPost Phys.} \textbf{ 5}
  (2018) 008,
  \href{http://dx.doi.org/10.21468/SciPostPhys.5.1.008}{\doi{10.21468/SciPostPhys.5.1.008}},
  \href{http://www.arXiv.org/abs/1712.07053}{\texttt{arXiv:1712.07053}}.

\bibitem{Bondarenko:2018ptm}
\hrefCMSnoop {}{K.~Bondarenko, A.~Boyarsky, D.~Gorbunov, and O.~Ruchayskiy,
  ``Phenomenology of {\GeVns}-scale heavy neutral leptons'',} \textit{ JHEP}
  \textbf{ 11} (2018) 032,
  \href{http://dx.doi.org/10.1007/JHEP11(2018)032}{\doi{10.1007/JHEP11(2018)032}},
  \href{http://www.arXiv.org/abs/1805.08567}{\texttt{arXiv:1805.08567}}.

\bibitem{Nason:2004rx}
\hrefCMSnoop {}{P.~Nason, ``A new method for combining {NLO} {QCD} with shower
  {Monte Carlo} algorithms'',} \textit{ JHEP} \textbf{ 11} (2004) 040,
  \href{http://dx.doi.org/10.1088/1126-6708/2004/11/040}{\doi{10.1088/1126-6708/2004/11/040}},
  \href{http://www.arXiv.org/abs/hep-ph/0409146}{\texttt{arXiv:hep-ph/0409146}}.

\bibitem{Frixione:2007nw}
\hrefCMSnoop {}{S.~Frixione, G.~Ridolfi, and P.~Nason, ``A positive-weight
  next-to-leading-order {Monte Carlo} for heavy flavour hadroproduction'',}
  \textit{ JHEP} \textbf{ 09} (2007) 126,
  \href{http://dx.doi.org/10.1088/1126-6708/2007/09/126}{\doi{10.1088/1126-6708/2007/09/126}},
  \href{http://www.arXiv.org/abs/0707.3088}{\texttt{arXiv:0707.3088}}.

\bibitem{Frixione:2007vw}
\hrefCMSnoop {}{S.~Frixione, P.~Nason, and C.~Oleari, ``Matching {NLO} {QCD}
  computations with parton shower simulations: the {\POWHEG} method'',}
  \textit{ JHEP} \textbf{ 11} (2007) 070,
  \href{http://dx.doi.org/10.1088/1126-6708/2007/11/070}{\doi{10.1088/1126-6708/2007/11/070}},
  \href{http://www.arXiv.org/abs/0709.2092}{\texttt{arXiv:0709.2092}}.

\bibitem{Alioli:2009je}
\hrefCMSnoop {}{S.~Alioli, P.~Nason, C.~Oleari, and E.~Re, ``{NLO} single-top
  production matched with shower in {\POWHEG}: $s$- and $t$-channel
  contributions'',} \textit{ JHEP} \textbf{ 09} (2009) 111,
  \href{http://dx.doi.org/10.1088/1126-6708/2009/09/111}{\doi{10.1088/1126-6708/2009/09/111}},
  \href{http://www.arXiv.org/abs/0907.4076}{\texttt{arXiv:0907.4076}}.
  [Erratum: \DOI{10.1007/JHEP02(2010)011}].

\bibitem{Nason:2009ai}
\hrefCMSnoop {}{P.~Nason and C.~Oleari, ``{NLO} {Higgs} boson production via
  vector-boson fusion matched with shower in {\POWHEG}'',} \textit{ JHEP}
  \textbf{ 02} (2010) 037,
  \href{http://dx.doi.org/10.1007/JHEP02(2010)037}{\doi{10.1007/JHEP02(2010)037}},
  \href{http://www.arXiv.org/abs/0911.5299}{\texttt{arXiv:0911.5299}}.

\bibitem{Alioli:2010xd}
\hrefCMSnoop {}{S.~Alioli, P.~Nason, C.~Oleari, and E.~Re, ``A general
  framework for implementing {NLO} calculations in shower {Monte Carlo}
  programs: the {\POWHEG} \textsc{box}'',} \textit{ JHEP} \textbf{ 06} (2010)
  043,
  \href{http://dx.doi.org/10.1007/JHEP06(2010)043}{\doi{10.1007/JHEP06(2010)043}},
  \href{http://www.arXiv.org/abs/1002.2581}{\texttt{arXiv:1002.2581}}.

\bibitem{Re:2010bp}
\hrefCMSnoop {}{E.~Re, ``Single-top ${\PW\PQt}$-channel production matched with
  parton showers using the {\POWHEG} method'',} \textit{ Eur. Phys. J. C}
  \textbf{ 71} (2011) 1547,
  \href{http://dx.doi.org/10.1140/epjc/s10052-011-1547-z}{\doi{10.1140/epjc/s10052-011-1547-z}},
  \href{http://www.arXiv.org/abs/1009.2450}{\texttt{arXiv:1009.2450}}.

\bibitem{Bagnaschi:2011tu}
\hrefCMSnoop {}{E.~Bagnaschi, G.~Degrassi, P.~Slavich, and A.~Vicini, ``Higgs
  production via gluon fusion in the {\POWHEG} approach in the {SM} and in the
  {MSSM}'',} \textit{ JHEP} \textbf{ 02} (2012) 088,
  \href{http://dx.doi.org/10.1007/JHEP02(2012)088}{\doi{10.1007/JHEP02(2012)088}},
  \href{http://www.arXiv.org/abs/1111.2854}{\texttt{arXiv:1111.2854}}.

\bibitem{Nason:2013ydw}
\hrefCMSnoop {}{P.~Nason and G.~Zanderighi, ``${\PWp\PWm}$, ${\PW\PZ}$ and
  ${\PZ\PZ}$ production in the {\POWHEG}-\textsc{box-v2}'',} \textit{ Eur.
  Phys. J. C} \textbf{ 74} (2014) 2702,
  \href{http://dx.doi.org/10.1140/epjc/s10052-013-2702-5}{\doi{10.1140/epjc/s10052-013-2702-5}},
  \href{http://www.arXiv.org/abs/1311.1365}{\texttt{arXiv:1311.1365}}.

\bibitem{Campbell:1999ah}
\hrefCMSnoop {}{J.~M. Campbell and R.~K. Ellis, ``An update on vector boson
  pair production at hadron colliders'',} \textit{ Phys. Rev. D} \textbf{ 60}
  (1999) 113006,
  \href{http://dx.doi.org/10.1103/PhysRevD.60.113006}{\doi{10.1103/PhysRevD.60.113006}},
  \href{http://www.arXiv.org/abs/hep-ph/9905386}{\texttt{arXiv:hep-ph/9905386}}.

\bibitem{Campbell:2011bn}
\hrefCMSnoop {}{J.~M. Campbell, R.~K. Ellis, and C.~Williams, ``Vector boson
  pair production at the {LHC}'',} \textit{ JHEP} \textbf{ 07} (2011) 018,
  \href{http://dx.doi.org/10.1007/JHEP07(2011)018}{\doi{10.1007/JHEP07(2011)018}},
  \href{http://www.arXiv.org/abs/1105.0020}{\texttt{arXiv:1105.0020}}.

\bibitem{Campbell:2015qma}
\hrefCMSnoop {}{J.~M. Campbell, R.~K. Ellis, and W.~T. Giele, ``A
  multi-threaded version of {\MCFM}'',} \textit{ Eur. Phys. J. C} \textbf{ 75}
  (2015) 246,
  \href{http://dx.doi.org/10.1140/epjc/s10052-015-3461-2}{\doi{10.1140/epjc/s10052-015-3461-2}},
  \href{http://www.arXiv.org/abs/1503.06182}{\texttt{arXiv:1503.06182}}.

\bibitem{Sjostrand:2014zea}
T.~Sj{\"o}strand\hrefCMSnoop {}{ { et~al.}, ``An introduction to
  {\PYTHIA8.2}'',} \textit{ Comput. Phys. Commun.} \textbf{ 191} (2015) 159,
  \href{http://dx.doi.org/10.1016/j.cpc.2015.01.024}{\doi{10.1016/j.cpc.2015.01.024}},
  \href{http://www.arXiv.org/abs/1410.3012}{\texttt{arXiv:1410.3012}}.

\bibitem{CMS:GEN-17-001}
\hrefCMSnoop {}{{CMS Collaboration}, ``Extraction and validation of a new set
  of {CMS} {\PYTHIA8} tunes from underlying-event measurements'',} \textit{
  Eur. Phys. J. C} \textbf{ 80} (2020) 4,
  \href{http://dx.doi.org/10.1140/epjc/s10052-019-7499-4}{\doi{10.1140/epjc/s10052-019-7499-4}},
  \href{http://www.arXiv.org/abs/1903.12179}{\texttt{arXiv:1903.12179}}.

\bibitem{Alwall:2007fs}
J.~Alwall\hrefCMSnoop {}{ { et~al.}, ``Comparative study of various algorithms
  for the merging of parton showers and matrix elements in hadronic
  collisions'',} \textit{ Eur. Phys. J. C} \textbf{ 53} (2008) 473,
  \href{http://dx.doi.org/10.1140/epjc/s10052-007-0490-5}{\doi{10.1140/epjc/s10052-007-0490-5}},
  \href{http://www.arXiv.org/abs/0706.2569}{\texttt{arXiv:0706.2569}}.

\bibitem{Frederix:2012ps}
\hrefCMSnoop {}{R.~Frederix and S.~Frixione, ``Merging meets matching in
  {\MCATNLO}'',} \textit{ JHEP} \textbf{ 12} (2012) 061,
  \href{http://dx.doi.org/10.1007/JHEP12(2012)061}{\doi{10.1007/JHEP12(2012)061}},
  \href{http://www.arXiv.org/abs/1209.6215}{\texttt{arXiv:1209.6215}}.

\bibitem{Bolognesi:2012mm}
S.~Bolognesi\hrefCMSnoop {}{ { et~al.}, ``On the spin and parity of a
  single-produced resonance at the {LHC}'',} \textit{ Phys. Rev. D} \textbf{
  86} (2012) 095031,
  \href{http://dx.doi.org/10.1103/PhysRevD.86.095031}{\doi{10.1103/PhysRevD.86.095031}},
  \href{http://www.arXiv.org/abs/1208.4018}{\texttt{arXiv:1208.4018}}.

\bibitem{CMS:EGM-17-001}
\hrefCMSnoop {}{{CMS Collaboration}, ``Electron and photon reconstruction and
  identification with the {CMS} experiment at the {CERN} {LHC}'',} \textit{
  JINST} \textbf{ 16} (2021) P05014,
  \href{http://dx.doi.org/10.1088/1748-0221/16/05/P05014}{\doi{10.1088/1748-0221/16/05/P05014}},
  \href{http://www.arXiv.org/abs/2012.06888}{\texttt{arXiv:2012.06888}}.

\bibitem{CMS:DP-2020-021}
\href {https://cds.cern.ch/record/2717925}{{CMS Collaboration}, ``{ECAL} 2016
  refined calibration and \mbox{Run 2} summary plots'',} CMS Detector
  Performance Note CMS-DP-2020-021, 2020.

\bibitem{CMS:MUO-16-001}
\hrefCMSnoop {}{{CMS Collaboration}, ``Performance of the {CMS} muon detector
  and muon reconstruction with proton-proton collisions at
  $\sqrt{s}={13\TeV}$'',} \textit{ JINST} \textbf{ 13} (2018) P06015,
  \href{http://dx.doi.org/10.1088/1748-0221/13/06/P06015}{\doi{10.1088/1748-0221/13/06/P06015}},
  \href{http://www.arXiv.org/abs/1804.04528}{\texttt{arXiv:1804.04528}}.

\bibitem{Rehermann:2010vq}
\hrefCMSnoop {}{K.~Rehermann and B.~Tweedie, ``Efficient identification of
  boosted semileptonic top quarks at the {LHC}'',} \textit{ JHEP} \textbf{ 03}
  (2011) 059,
  \href{http://dx.doi.org/10.1007/JHEP03(2011)059}{\doi{10.1007/JHEP03(2011)059}},
  \href{http://www.arXiv.org/abs/1007.2221}{\texttt{arXiv:1007.2221}}.

\bibitem{CMS:TOP-22-013}
\hrefCMSnoop {}{{CMS Collaboration}, ``Observation of four top quark production
  in proton-proton collisions at $\sqrt{s}={13\TeV}$'',} \textit{ Phys. Lett.
  B} \textbf{ 847} (2023) 138290,
  \href{http://dx.doi.org/10.1016/j.physletb.2023.138290}{\doi{10.1016/j.physletb.2023.138290}},
  \href{http://www.arXiv.org/abs/2305.13439}{\texttt{arXiv:2305.13439}}.

\bibitem{CMS:HIG-17-018}
\hrefCMSnoop {}{{CMS Collaboration}, ``Evidence for associated production of a
  {Higgs} boson with a top quark pair in final states with electrons, muons,
  and hadronically decaying {\PGt} leptons at $\sqrt{s}={13\TeV}$'',} \textit{
  JHEP} \textbf{ 08} (2018) 066,
  \href{http://dx.doi.org/10.1007/JHEP08(2018)066}{\doi{10.1007/JHEP08(2018)066}},
  \href{http://www.arXiv.org/abs/1803.05485}{\texttt{arXiv:1803.05485}}.

\bibitem{CMS:TOP-18-008}
\hrefCMSnoop {}{{CMS Collaboration}, ``Observation of single top quark
  production in association with a {\PZ} boson in proton-proton collisions at
  $\sqrt{s}={13\TeV}$'',} \textit{ Phys. Rev. Lett.} \textbf{ 122} (2019)
  132003,
  \href{http://dx.doi.org/10.1103/PhysRevLett.122.132003}{\doi{10.1103/PhysRevLett.122.132003}},
  \href{http://www.arXiv.org/abs/1812.05900}{\texttt{arXiv:1812.05900}}.

\bibitem{CMS:HIG-19-008}
\hrefCMSnoop {}{{CMS Collaboration}, ``Measurement of the {Higgs} boson
  production rate in association with top quarks in final states with
  electrons, muons, and hadronically decaying tau leptons at
  $\sqrt{s}={13\TeV}$'',} \textit{ Eur. Phys. J. C} \textbf{ 81} (2021) 378,
  \href{http://dx.doi.org/10.1140/epjc/s10052-021-09014-x}{\doi{10.1140/epjc/s10052-021-09014-x}},
  \href{http://www.arXiv.org/abs/2011.03652}{\texttt{arXiv:2011.03652}}.

\bibitem{CMS:SUS-19-012}
\hrefCMSnoop {}{{CMS Collaboration}, ``Search for electroweak production of
  charginos and neutralinos in proton-proton collisions at
  $\sqrt{s}={13\TeV}$'',} \textit{ JHEP} \textbf{ 04} (2022) 147,
  \href{http://dx.doi.org/10.1007/JHEP04(2022)147}{\doi{10.1007/JHEP04(2022)147}},
  \href{http://www.arXiv.org/abs/2106.14246}{\texttt{arXiv:2106.14246}}.

\bibitem{CMS:SMP-20-012}
\hrefCMSnoop {}{{CMS Collaboration}, ``Measurements of the electroweak diboson
  production cross sections in proton-proton collisions at
  $\sqrt{s}={5.02\TeV}$ using leptonic decays'',} \textit{ Phys. Rev. Lett.}
  \textbf{ 127} (2021) 191801,
  \href{http://dx.doi.org/10.1103/PhysRevLett.127.191801}{\doi{10.1103/PhysRevLett.127.191801}},
  \href{http://www.arXiv.org/abs/2107.01137}{\texttt{arXiv:2107.01137}}.

\bibitem{CMS:TOP-20-010}
\hrefCMSnoop {}{{CMS Collaboration}, ``Inclusive and differential cross section
  measurements of single top quark production in association with a {\PZ} boson
  in proton-proton collisions at $\sqrt{s}={13\TeV}$'',} \textit{ JHEP}
  \textbf{ 02} (2022) 107,
  \href{http://dx.doi.org/10.1007/JHEP02(2022)107}{\doi{10.1007/JHEP02(2022)107}},
  \href{http://www.arXiv.org/abs/2111.02860}{\texttt{arXiv:2111.02860}}.

\bibitem{CMS:MUO-22-001}
\hrefCMSnoop {}{{CMS Collaboration}, ``Muon identification using multivariate
  techniques in the {CMS} experiment in proton-proton collisions at
  $\sqrt{s}={13\TeV}$'',} \textit{ JINST} \textbf{ 19} (2024) P02031,
  \href{http://dx.doi.org/10.1088/1748-0221/19/02/P02031}{\doi{10.1088/1748-0221/19/02/P02031}},
  \href{http://www.arXiv.org/abs/2310.03844}{\texttt{arXiv:2310.03844}}.

\bibitem{CMS:TAU-16-003}
\hrefCMSnoop {}{{CMS Collaboration}, ``Performance of reconstruction and
  identification of {\PGt} leptons decaying to hadrons and {\PGnGt} in
  ${\Pp\Pp}$ collisions at $\sqrt{s}={13\TeV}$'',} \textit{ JINST} \textbf{ 13}
  (2018) P10005,
  \href{http://dx.doi.org/10.1088/1748-0221/13/10/P10005}{\doi{10.1088/1748-0221/13/10/P10005}},
  \href{http://www.arXiv.org/abs/1809.02816}{\texttt{arXiv:1809.02816}}.

\bibitem{CMS:TAU-20-001}
\hrefCMSnoop {}{{CMS Collaboration}, ``Identification of hadronic tau lepton
  decays using a deep neural network'',} \textit{ JINST} \textbf{ 17} (2022)
  P07023,
  \href{http://dx.doi.org/10.1088/1748-0221/17/07/P07023}{\doi{10.1088/1748-0221/17/07/P07023}},
  \href{http://www.arXiv.org/abs/2201.08458}{\texttt{arXiv:2201.08458}}.

\bibitem{ParticleDataGroup:2022pth}
\hrefCMSnoop {}{{Particle Data Group}, R.~L. Workman { et~al.}, ``Review of
  particle physics'',} \textit{ Prog. Theor. Exp. Phys.} \textbf{ 2022} (2022)
  083C01,
  \href{http://dx.doi.org/10.1093/ptep/ptac097}{\doi{10.1093/ptep/ptac097}}.

\bibitem{CMS:EGM-13-001}
\hrefCMSnoop {}{{CMS Collaboration}, ``Performance of electron reconstruction
  and selection with the {CMS} detector in proton-proton collisions at
  $\sqrt{s}={8\TeV}$'',} \textit{ JINST} \textbf{ 10} (2015) P06005,
  \href{http://dx.doi.org/10.1088/1748-0221/10/06/P06005}{\doi{10.1088/1748-0221/10/06/P06005}},
  \href{http://www.arXiv.org/abs/1502.02701}{\texttt{arXiv:1502.02701}}.

\bibitem{CMS:CFT-09-014}
\hrefCMSnoop {}{{CMS Collaboration}, ``Performance of {CMS} muon reconstruction
  in cosmic-ray events'',} \textit{ JINST} \textbf{ 5} (2010) T03022,
  \href{http://dx.doi.org/10.1088/1748-0221/5/03/T03022}{\doi{10.1088/1748-0221/5/03/T03022}},
  \href{http://www.arXiv.org/abs/0911.4994}{\texttt{arXiv:0911.4994}}.

\bibitem{CMS:MUO-17-001}
\hrefCMSnoop {}{{CMS Collaboration}, ``Performance of the reconstruction and
  identification of high-momentum muons in proton-proton collisions at
  $\sqrt{s}={13\TeV}$'',} \textit{ JINST} \textbf{ 15} (2020) P02027,
  \href{http://dx.doi.org/10.1088/1748-0221/15/02/P02027}{\doi{10.1088/1748-0221/15/02/P02027}},
  \href{http://www.arXiv.org/abs/1912.03516}{\texttt{arXiv:1912.03516}}.

\bibitem{Voss:2007jxm}
\hrefCMSnoop {}{H.~Voss, A.~H{\"o}cker, J.~Stelzer, and F.~Tegenfeldt,
  ``{TMVA}, the toolkit for multivariate data analysis with {ROOT}'',} in
  \textit{ {Proc. 11th International Workshop on Advanced Computing and
  Analysis Techniques in Physics Research (ACAT 2017): Amsterdam, The
  Netherlands, April 23--27, 2007}}.
\newblock 2007.
\newblock
  \href{http://www.arXiv.org/abs/physics/0703039}{\texttt{arXiv:physics/0703039}}.
\newblock [PoS (ACAT2007) 040].
  \href{http://dx.doi.org/10.22323/1.050.0040}{\doi{10.22323/1.050.0040}}.

\bibitem{CMS:SUS-15-008}
\hrefCMSnoop {}{{CMS Collaboration}, ``Search for new physics in same-sign
  dilepton events in proton-proton collisions at $\sqrt{s}={13\TeV}$'',}
  \textit{ Eur. Phys. J. C} \textbf{ 76} (2016) 439,
  \href{http://dx.doi.org/10.1140/epjc/s10052-016-4261-z}{\doi{10.1140/epjc/s10052-016-4261-z}},
  \href{http://www.arXiv.org/abs/1605.03171}{\texttt{arXiv:1605.03171}}.

\bibitem{CMS:TOP-21-011}
\hrefCMSnoop {}{{CMS Collaboration}, ``Measurement of the cross section of top
  quark-antiquark pair production in association with a {\PW} boson in
  proton-proton collisions at $\sqrt{s}={13\TeV}$'',} \textit{ JHEP} \textbf{
  07} (2023) 219,
  \href{http://dx.doi.org/10.1007/JHEP07(2023)219}{\doi{10.1007/JHEP07(2023)219}},
  \href{http://www.arXiv.org/abs/2208.06485}{\texttt{arXiv:2208.06485}}.

\bibitem{CMS:LUM-17-003}
\hrefCMSnoop {}{{CMS Collaboration}, ``Precision luminosity measurement in
  proton-proton collisions at $\sqrt{s}={13\TeV}$ in 2015 and 2016 at {CMS}'',}
  \textit{ Eur. Phys. J. C} \textbf{ 81} (2021) 800,
  \href{http://dx.doi.org/10.1140/epjc/s10052-021-09538-2}{\doi{10.1140/epjc/s10052-021-09538-2}},
  \href{http://www.arXiv.org/abs/2104.01927}{\texttt{arXiv:2104.01927}}.

\bibitem{CMS:LUM-17-004}
\href {https://cds.cern.ch/record/2621960}{{CMS Collaboration}, ``{CMS}
  luminosity measurement for the 2017 data-taking period at
  $\sqrt{s}={13\TeV}$'',} CMS Physics Analysis Summary CMS-PAS-LUM-17-004,
  2018.

\bibitem{CMS:LUM-18-002}
\href {https://cds.cern.ch/record/2676164}{{CMS Collaboration}, ``{CMS}
  luminosity measurement for the 2018 data-taking period at
  $\sqrt{s}={13\TeV}$'',} CMS Physics Analysis Summary CMS-PAS-LUM-18-002,
  2019.

\bibitem{CMS:EWK-10-002}
\hrefCMSnoop {}{{CMS Collaboration}, ``Measurements of inclusive {\PW} and
  {\PZ} cross sections in ${\Pp\Pp}$ collisions at $\sqrt{s}={7\TeV}$'',}
  \textit{ JHEP} \textbf{ 01} (2011) 080,
  \href{http://dx.doi.org/10.1007/JHEP01(2011)080}{\doi{10.1007/JHEP01(2011)080}},
  \href{http://www.arXiv.org/abs/1012.2466}{\texttt{arXiv:1012.2466}}.

\bibitem{Lazopoulos:2007ix}
\hrefCMSnoop {}{A.~Lazopoulos, K.~Melnikov, and F.~Petriello, ``{QCD}
  corrections to tri-boson production'',} \textit{ Phys. Rev. D} \textbf{ 76}
  (2007) 014001,
  \href{http://dx.doi.org/10.1103/PhysRevD.76.014001}{\doi{10.1103/PhysRevD.76.014001}},
  \href{http://www.arXiv.org/abs/hep-ph/0703273}{\texttt{arXiv:hep-ph/0703273}}.

\bibitem{Binoth:2008kt}
\hrefCMSnoop {}{T.~Binoth, G.~Ossola, C.~G. Papadopoulos, and R.~Pittau,
  ``{NLO} {QCD} corrections to tri-boson production'',} \textit{ JHEP} \textbf{
  06} (2008) 082,
  \href{http://dx.doi.org/10.1088/1126-6708/2008/06/082}{\doi{10.1088/1126-6708/2008/06/082}},
  \href{http://www.arXiv.org/abs/0804.0350}{\texttt{arXiv:0804.0350}}.

\bibitem{Hankele:2007sb}
\hrefCMSnoop {}{V.~Hankele and D.~Zeppenfeld, ``{QCD} corrections to hadronic
  ${\PW\PW\PZ}$ production with leptonic decays'',} \textit{ Phys. Lett. B}
  \textbf{ 661} (2008) 103,
  \href{http://dx.doi.org/10.1016/j.physletb.2008.02.014}{\doi{10.1016/j.physletb.2008.02.014}},
  \href{http://www.arXiv.org/abs/0712.3544}{\texttt{arXiv:0712.3544}}.

\bibitem{Campanario:2008yg}
F.~Campanario\hrefCMSnoop {}{ { et~al.}, ``{QCD} corrections to charged triple
  vector boson production with leptonic decay'',} \textit{ Phys. Rev. D}
  \textbf{ 78} (2008) 094012,
  \href{http://dx.doi.org/10.1103/PhysRevD.78.094012}{\doi{10.1103/PhysRevD.78.094012}},
  \href{http://www.arXiv.org/abs/0809.0790}{\texttt{arXiv:0809.0790}}.

\bibitem{Dittmaier:2017bnh}
\hrefCMSnoop {}{S.~Dittmaier, A.~Huss, and G.~Knippen, ``Next-to-leading-order
  {QCD} and electroweak corrections to ${\PW\PW\PW}$ production at
  proton-proton colliders'',} \textit{ JHEP} \textbf{ 09} (2017) 034,
  \href{http://dx.doi.org/10.1007/JHEP09(2017)034}{\doi{10.1007/JHEP09(2017)034}},
  \href{http://www.arXiv.org/abs/1705.03722}{\texttt{arXiv:1705.03722}}.

\bibitem{CMS:HIG-22-001}
\hrefCMSnoop {}{{CMS Collaboration}, ``A portrait of the {Higgs} boson by the
  {CMS} experiment ten years after the discovery'',} \textit{ Nature} \textbf{
  607} (2022) 60,
  \href{http://dx.doi.org/10.1038/s41586-022-04892-x}{\doi{10.1038/s41586-022-04892-x}},
  \href{http://www.arXiv.org/abs/2207.00043}{\texttt{arXiv:2207.00043}}.
  [Author correction: \DOI{10.1038/s41586-023-06164-8}].

\bibitem{Melnikov:2006kv}
\hrefCMSnoop {}{K.~Melnikov and F.~Petriello, ``Electroweak gauge boson
  production at hadron colliders through $\mathcal{O}({\alpS^2})$'',} \textit{
  Phys. Rev. D} \textbf{ 74} (2006) 114017,
  \href{http://dx.doi.org/10.1103/PhysRevD.74.114017}{\doi{10.1103/PhysRevD.74.114017}},
  \href{http://www.arXiv.org/abs/hep-ph/0609070}{\texttt{arXiv:hep-ph/0609070}}.

\bibitem{Gavin:2010az}
\hrefCMSnoop {}{R.~Gavin, Y.~Li, F.~Petriello, and S.~Quackenbush,
  ``{\FEWZ2.0}: A code for hadronic {\PZ} production at next-to-next-to-leading
  order'',} \textit{ Comput. Phys. Commun.} \textbf{ 182} (2011) 2388,
  \href{http://dx.doi.org/10.1016/j.cpc.2011.06.008}{\doi{10.1016/j.cpc.2011.06.008}},
  \href{http://www.arXiv.org/abs/1011.3540}{\texttt{arXiv:1011.3540}}.

\bibitem{Gavin:2012sy}
\hrefCMSnoop {}{R.~Gavin, Y.~Li, F.~Petriello, and S.~Quackenbush, ``{\PW}
  physics at the {LHC} with {\FEWZ2.1}'',} \textit{ Comput. Phys. Commun.}
  \textbf{ 184} (2013) 208,
  \href{http://dx.doi.org/10.1016/j.cpc.2012.09.005}{\doi{10.1016/j.cpc.2012.09.005}},
  \href{http://www.arXiv.org/abs/1201.5896}{\texttt{arXiv:1201.5896}}.

\bibitem{Li:2012wna}
\hrefCMSnoop {}{Y.~Li and F.~Petriello, ``Combining {QCD} and electroweak
  corrections to dilepton production in {\FEWZ}'',} \textit{ Phys. Rev. D}
  \textbf{ 86} (2012) 094034,
  \href{http://dx.doi.org/10.1103/PhysRevD.86.094034}{\doi{10.1103/PhysRevD.86.094034}},
  \href{http://www.arXiv.org/abs/1208.5967}{\texttt{arXiv:1208.5967}}.

\bibitem{CMS:CAT-23-001}
\hrefCMSnoop {}{{CMS Collaboration}, ``The {CMS} statistical analysis and
  combination tool: \textsc{combine}'',} 2024.
  \href{http://www.arXiv.org/abs/2404.06614}{\texttt{arXiv:2404.06614}}.
  Submitted to \textit{Comput. Softw. Big Sci.}

\bibitem{Verkerke:2003ir}
\href
  {https://www.slac.stanford.edu/econf/C0303241/proc/papers/MOLT007.PDF}{W.~Verkerke
  and D.~Kirkby, ``The \textsc{RooFit} toolkit for data modeling'',} in
  \textit{ {Proc. 13th International Conference on Computing in High Energy and
  Nuclear Physics (CHEP 2003): La Jolla CA, United States, March 24--28,
  2003}}.
\newblock 2003.
\newblock
  \href{http://www.arXiv.org/abs/physics/0306116}{\texttt{arXiv:physics/0306116}}.
\newblock [eConf C0303241 (2003) MOLT007].

\bibitem{Moneta:2010pm}
L.~Moneta\hrefCMSnoop {}{ { et~al.}, ``The \textsc{RooStats} project'',} in
  \textit{ {Proc. 13th International Workshop on Advanced Computing and
  Analysis Techniques in Physics Research (ACAT 2010): Jaipur, India, February
  22--27, 2010}}.
\newblock 2010.
\newblock \href{http://www.arXiv.org/abs/1009.1003}{\texttt{arXiv:1009.1003}}.
\newblock [PoS (ACAT2010) 057].
  \href{http://dx.doi.org/10.22323/1.093.0057}{\doi{10.22323/1.093.0057}}.

\bibitem{CMS:NOTE-2011-005}
\href {https://cds.cern.ch/record/1379837}{{ATLAS and CMS Collaborations, and
  LHC Higgs Combination Group}, ``Procedure for the {LHC} {Higgs} boson search
  combination in {Summer} 2011'',} Technical Report CMS-NOTE-2011-005,
  ATL-PHYS-PUB-2011-11, 2011.

\bibitem{Barlow:1993dm}
\hrefCMSnoop {}{R.~Barlow and C.~Beeston, ``Fitting using finite {Monte Carlo}
  samples'',} \textit{ Comput. Phys. Commun.} \textbf{ 77} (1993) 219,
  \href{http://dx.doi.org/10.1016/0010-4655(93)90005-W}{\doi{10.1016/0010-4655(93)90005-W}}.

\bibitem{Conway:2011in}
\hrefCMSnoop {}{J.~S. Conway, ``Incorporating nuisance parameters in
  likelihoods for multisource spectra'',} in \textit{ {Proc. 2011 Workshop on
  Statistical Issues Related to Discovery Claims in Search Experiments and
  Unfolding (PHYSTAT 2011): Geneva, Switzerland, January 17--20, 2011}}.
\newblock 2011.
\newblock \href{http://www.arXiv.org/abs/1103.0354}{\texttt{arXiv:1103.0354}}.
\newblock
  \href{http://dx.doi.org/10.5170/CERN-2011-006.115}{\doi{10.5170/CERN-2011-006.115}}.

\bibitem{Junk:1999kv}
\hrefCMSnoop {}{T.~Junk, ``Confidence level computation for combining searches
  with small statistics'',} \textit{ Nucl. Instrum. Meth. A} \textbf{ 434}
  (1999) 435,
  \href{http://dx.doi.org/10.1016/S0168-9002(99)00498-2}{\doi{10.1016/S0168-9002(99)00498-2}},
  \href{http://www.arXiv.org/abs/hep-ex/9902006}{\texttt{arXiv:hep-ex/9902006}}.

\bibitem{Read:2002hq}
\hrefCMSnoop {}{A.~L. Read, ``Presentation of search results: The {\CLs}
  technique'',} \textit{ J. Phys. G} \textbf{ 28} (2002) 2693,
  \href{http://dx.doi.org/10.1088/0954-3899/28/10/313}{\doi{10.1088/0954-3899/28/10/313}}.

\bibitem{Cowan:2010js}
\hrefCMSnoop {}{G.~Cowan, K.~Cranmer, E.~Gross, and O.~Vitells, ``Asymptotic
  formulae for likelihood-based tests of new physics'',} \textit{ Eur. Phys. J.
  C} \textbf{ 71} (2011) 1554,
  \href{http://dx.doi.org/10.1140/epjc/s10052-011-1554-0}{\doi{10.1140/epjc/s10052-011-1554-0}},
  \href{http://www.arXiv.org/abs/1007.1727}{\texttt{arXiv:1007.1727}}.
  [Erratum: \DOI{10.1140/epjc/s10052-013-2501-z}].

\bibitem{DELPHI:1996qcc}
\hrefCMSnoop {}{{DELPHI} Collaboration, ``Search for neutral heavy leptons
  produced in {\PZ} decays'',} \textit{ Z. Phys. C} \textbf{ 74} (1997) 57,
  \href{http://dx.doi.org/10.1007/s002880050370}{\doi{10.1007/s002880050370}}.
  [Erratum: \DOI{10.1007/BF03546181}].

\bibitem{CMS:EXO-21-003}
\hrefCMSnoop {}{{CMS Collaboration}, ``Probing heavy {Majorana} neutrinos and
  the {Weinberg} operator through vector boson fusion processes in
  proton-proton collisions at $\sqrt{s}={13\TeV}$'',} \textit{ Phys. Rev.
  Lett.} \textbf{ 131} (2023) 011803,
  \href{http://dx.doi.org/10.1103/PhysRevLett.131.011803}{\doi{10.1103/PhysRevLett.131.011803}},
  \href{http://www.arXiv.org/abs/2206.08956}{\texttt{arXiv:2206.08956}}.

\bibitem{BaBar:2022cqj}
\hrefCMSnoop {}{{BaBar} Collaboration, ``Search for heavy neutral leptons using
  tau lepton decays at {BaBaR}'',} \textit{ Phys. Rev. D} \textbf{ 107} (2023)
  052009,
  \href{http://dx.doi.org/10.1103/PhysRevD.107.052009}{\doi{10.1103/PhysRevD.107.052009}},
  \href{http://www.arXiv.org/abs/2207.09575}{\texttt{arXiv:2207.09575}}.

\end{thebibliography}\endgroup
\cleardoublepage \appendix\section{The CMS Collaboration \label{app:collab}}\begin{sloppypar}\hyphenpenalty=5000\widowpenalty=500\clubpenalty=5000
\cmsinstitute{Yerevan Physics Institute, Yerevan, Armenia}
{\tolerance=6000
A.~Hayrapetyan, A.~Tumasyan\cmsAuthorMark{1}\cmsorcid{0009-0000-0684-6742}
\par}
\cmsinstitute{Institut f\"{u}r Hochenergiephysik, Vienna, Austria}
{\tolerance=6000
W.~Adam\cmsorcid{0000-0001-9099-4341}, J.W.~Andrejkovic, T.~Bergauer\cmsorcid{0000-0002-5786-0293}, S.~Chatterjee\cmsorcid{0000-0003-2660-0349}, K.~Damanakis\cmsorcid{0000-0001-5389-2872}, M.~Dragicevic\cmsorcid{0000-0003-1967-6783}, P.S.~Hussain\cmsorcid{0000-0002-4825-5278}, M.~Jeitler\cmsAuthorMark{2}\cmsorcid{0000-0002-5141-9560}, N.~Krammer\cmsorcid{0000-0002-0548-0985}, A.~Li\cmsorcid{0000-0002-4547-116X}, D.~Liko\cmsorcid{0000-0002-3380-473X}, I.~Mikulec\cmsorcid{0000-0003-0385-2746}, J.~Schieck\cmsAuthorMark{2}\cmsorcid{0000-0002-1058-8093}, R.~Sch\"{o}fbeck\cmsorcid{0000-0002-2332-8784}, D.~Schwarz\cmsorcid{0000-0002-3821-7331}, M.~Sonawane\cmsorcid{0000-0003-0510-7010}, S.~Templ\cmsorcid{0000-0003-3137-5692}, W.~Waltenberger\cmsorcid{0000-0002-6215-7228}, C.-E.~Wulz\cmsAuthorMark{2}\cmsorcid{0000-0001-9226-5812}
% ...
\par}
\vskip\cmsinstskip
\dag:~Deceased\\
$^{1}$Also at Yerevan State University, Yerevan, Armenia\\
$^{2}$Also at TU Wien, Vienna, Austria\\
% ...
\end{sloppypar}
\end{document}